\documentclass[10pt,reqno, footinclude=true, headinclude=true, openright]{article}
\pdfoutput=1
\usepackage[utf8]{inputenc}
\usepackage[margin=1.00in]{geometry}
\usepackage{authblk}
\RequirePackage{enumitem}
\setlist[enumerate]{leftmargin=*}
\usepackage{color, xcolor, colortbl}
\usepackage{graphicx,epstopdf}
\usepackage{amsmath,amssymb,amsthm}
\usepackage{algorithm}
\usepackage{algorithmic}
\usepackage{esvect}
\usepackage{bm}
\usepackage[caption=false]{subfig}
\usepackage{appendix}
\usepackage{multirow}
\usepackage{verbatim}
\usepackage{mathtools}
\usepackage{dsfont}
\usepackage{braket}
\usepackage[english]{babel}
\usepackage[qm]{qcircuit}
\usepackage{tikz}
\usetikzlibrary{shapes.geometric}
\usetikzlibrary{arrows.meta}
\usetikzlibrary{positioning}
\usepackage{csquotes}
\usepackage{rotating}
\usepackage{caption}
\numberwithin{equation}{section}

\definecolor{mulberry}{rgb}{0.77, 0.29, 0.55}
\definecolor{lightpink}{rgb}{1.0, 0.71, 0.76}
\definecolor{yellow}{RGB}{240,228,66}
\definecolor{lightblue}{RGB}{0,58,120}
\definecolor{darkblue}{rgb}{0.1,0.1,0.7}
\definecolor{darkred}{rgb}{0.5,0.1,0.1}
\definecolor{darkgreen}{rgb}{0.0,0.72,0.06}
\definecolor{shadecolor}{rgb}{0.85,0.85,0.85}
\definecolor{navyblue}{rgb}{0.0, 0.0, 0.5}
\definecolor{cerisepink}{rgb}{0.93, 0.23, 0.51}
\definecolor{oceanboatblue}{rgb}{0.0, 0.47, 0.75}
\definecolor{red-brown}{rgb}{0.65, 0.16, 0.16}

\usepackage[
    colorlinks=true,
    citecolor=darkgreen,
    linkcolor=blue,
    filecolor=magenta,      
    urlcolor=magenta,
    pdftitle={MPF},
    pdfpagemode=FullScreen,
    ]{hyperref}
    
\usepackage[backend=biber,style=numeric-comp,sorting=none,backref=true,maxbibnames=99]{biblatex}
\DefineBibliographyStrings{english}{%
  backrefpage = {page},
  backrefpages = {pages},
}

\usepackage[capitalize,noabbrev]{cleveref}

\DeclareFontEncoding{FMS}{}{}
\DeclareFontSubstitution{FMS}{futm}{m}{n}
\DeclareFontEncoding{FMX}{}{}
\DeclareFontSubstitution{FMX}{futm}{m}{n}
\DeclareSymbolFont{fouriersymbols}{FMS}{futm}{m}{n}
\DeclareSymbolFont{fourierlargesymbols}{FMX}{futm}{m}{n}
\DeclareMathDelimiter{\VERT}{\mathord}{fouriersymbols}{152}{fourierlargesymbols}{147}

\newcommand{\Z}{{\mathbb{Z}}}

\newtheorem{thm}{\protect\theoremname}
  \theoremstyle{plain}
  \newtheorem{lem}[thm]{\protect\lemmaname}
  \theoremstyle{plain}
  
  \theoremstyle{plain}
  \newtheorem*{lem*}{\protect\lemmaname}
  \theoremstyle{plain}
  \newtheorem{prop}[thm]{\protect\propositionname}
  \theoremstyle{plain}
  \newtheorem{cor}[thm]{\protect\corollaryname}
  \newtheorem{assumption}[thm]{Assumption}

  \newtheorem{result}[thm]{Result}

\PassOptionsToPackage{USenglish}{babel}
\usepackage[USenglish]{babel}
  \providecommand{\corollaryname}{Corollary}
  \providecommand{\lemmaname}{Lemma}
  \providecommand{\propositionname}{Proposition}
  \providecommand{\remarkname}{Remark}
\providecommand{\theoremname}{Theorem}

\newcommand{\norm}[1]{\|#1\|}

\begin{document}

\title{\centering Multi-Product Hamiltonian Simulation with Explicit Commutator Scaling} 

\author[1]{Junaid Aftab}
\author[2]{Dong An}
\author[1,3]{Konstantina Trivisa}

\affil[1]{Department of Mathematics, University of Maryland, College Park}
\affil[2]{Joint Center for Quantum Information and Computer Science, University of Maryland, College Park}
\affil[3]{Institute for Physical Science and Technology, University of Maryland, College Park}


\date{\today}

\maketitle

\begin{abstract}
The well-conditioned multi-product formula (MPF), proposed by [Low-Kliuchnikov-Wiebe (2019)], is a high-order, time-independent Hamiltonian simulation algorithm that implements a linear combination of low-order product formulas. 
Prior work established its well-conditioned algorithmic construction and near-optimal time and precision dependence, but did not simultaneously provide an explicit error bound in terms of nested commutators, which has left its practical advantage uncertain.
In this work, we provide a rigorous complexity analysis of the well-conditioned MPF, explicitly establishing both commutator scaling and near-optimal dependence on time and precision through a rigorous Baker–Campbell–Hausdorff (BCH) and variation-of-parameters-based error analysis.
Using our improved complexity estimates, we present several physically relevant applications where the second-order-based MPF asymptotically outperforms the second-order product formula in all three aspects of system size, time, and precision. 
We further demonstrate that, while MPF yields worse scalings than the best post-Trotter methods in most applications, there exist certain parameter regimes in the nonlocal fast-transform Bogoliubov-de Gennes model where MPF can outperform both leading post-Trotter methods and product formula with fixed or adaptively optimized order.
\end{abstract}

\tableofcontents

\section{Introduction}

The (time-independent) Hamiltonian simulation problem, also referred to as the quantum simulation problem, is widely regarded as one of the most significant applications of quantum computing and a potential route to practical quantum advantage. Its objective is to prepare a quantum state \(\ket{u(T)}\) that evolves according to the time-independent Schr\"odinger equation
\begin{equation}\label{eqn:HamSim}
i \frac{d}{dt} u(t) = H u(t), \quad t \in [0,T],
\end{equation}
where $H \in \mathbb{C}^{N \times N}$ denotes the Hamiltonian of a finite-dimensional quantum system.
The solution to~\cref{eqn:HamSim} is given explicitly by \(\ket{u(T)}=e^{-iHT}\ket{u(0)}\). Accordingly, the central task in quantum simulation is to implement the unitary time-evolution operator \(e^{-iHT}\).

This raises the algorithmic question of how to implement \(e^{-iHT}\) efficiently.
The development of such algorithms dates back to Lloyd's seminal work~\cite{Lloyd1996}, which, in modern terminology, introduced an algorithm based on the first-order Trotter product formula. Specifically, suppose that the Hamiltonian \(H\) admits a decomposition \(H=\sum_{\gamma=1}^{\Gamma}H_\gamma\), where each \(U_\gamma(t)=e^{-itH_\gamma}\) can be implemented efficiently. This assumption is satisfied in many practical applications. Lloyd's method approximates \(e^{-iHT}\) by a carefully chosen product of the unitaries \(U_\gamma(t)\). Since Lloyd's original work, substantial progress has been made in the design of quantum algorithms based on higher-order product formulas, together with rigorous complexity analyses~\cite{BerryAhokasCleveEtAl2007,ChildsMaslovNamEtAl2018,ChildsSu2019,ChildsSuTranEtAl2020,SahinogluSomma2020,SuHuangCampbell2020,AnFangLin2021,YiCrosson2021,morales2022greatly,LowSuTongEtAl2023}. Notably, Childs et al.~\cite{ChildsSuTranEtAl2020} showed that product formulas can exhibit commutator scaling, which reflects the structure of commutators among the Hamiltonian terms. This property allows product formulas to achieve favorable, and in some cases optimal, system-size scaling for physically local Hamiltonians. Despite these advantages, a major limitation of product formulas is their overhead in the evolution time \(T\) and target error \(\epsilon\). A product formula of order \(p\) incurs gate complexity \(\mathcal{O}(T^{1+1/p}\epsilon^{-1/p})\), which is super-linear in \(T\) and super-logarithmic in \(1/\epsilon\) for fixed \(p\). 
Moreover, it is unlikely that this scaling could be improved to the ideal \(\mathcal{O}(T\operatorname{polylog}(T/\epsilon))\) by optimizing \(p\) because, for generic two-operator product formulas, the number of independent order conditions grows as \(\Theta(2^p/p)\)~\cite{Suzuki1992GeneralTheory,ChenChildsHafeziEtAl2022}, which implies the need of exponentially many factors under the usual independence assumption.

To achieve improved dependence on evolution time and target precision, recent work has introduced a class of quantum algorithms, often referred to as post-Trotter methods, that go beyond the traditional product formulas~\cite{BerryChilds2012,ChildsWiebe2012,BerryCleveGharibian2014,BerryChildsCleveEtAl2015,BerryChildsKothari2015,LowChuang2017,LowChuang2019,GilyenSuLowEtAl2019,ChildsOstranderSu2019,Campbell2019,Low2019,ChenHuangKuengEtAl2020,zeng2022simple}. Among these, qubitization-based algorithms developed by Low and Chuang~\cite{LowChuang2019} and quantum singular value transformation (QSVT), introduced by Gily\'en et al.~\cite{GilyenSuLowEtAl2019}, achieve query complexity \(\mathcal{O} (\|H\|T+\frac{\log(1/\epsilon)}{\log\log(1/\epsilon)})\) with respect to black-box access to the Hamiltonian. This dependence on both time \(T\) and precision \(\epsilon\) improves upon that of product formulas of any order. Moreover, it is asymptotically optimal, saturating known lower bounds in both \(T\) and \(\epsilon\)~\cite{BerryChildsCleveEtAl2014}. However, a notable limitation of post-Trotter methods is that their complexity typically scales at least linearly with the spectral norm \(\|H\|\) and does not exploit commutator scaling. As a result, they often exhibit worse system-size dependence than product formulas in physically relevant settings, including local Hamiltonians~\cite{ChildsSuTranEtAl2020} and nuclear effective field theories~\cite{WatsonBringewattShawEtAl2023}. This highlights a fundamental trade-off between favorable dependence on system size and favorable dependence on time and precision, which must be considered when selecting an algorithm for simulating a given quantum system.

Low, Kliuchnikov, and Wiebe~\cite{LowKliuchnikovWiebe2019} present a quantum simulation algorithm based on the multi-product formula (MPF)~\cite{chin2010multi} that aims to combine commutator scaling with near-optimal dependence on time and precision. The MPF approximates the ideal evolution operator by a linear combination of product formulas
\begin{equation}\label{eqn:MPF_intro}
e^{-iHt} \approx \sum_{j=1}^{M} a_j U_p(t/k_j)^{k_j}.
\end{equation}
Here \(U_p(s)\) denotes a \(p\)-th order product formula with time step \(s\) and is referred to as the base sequence of the MPF. The order \(p\) of the base sequence is typically low, such as \(p=2\) or \(p=4\), to avoid the exponential overhead of high-order product formulas. In~\cite{LowKliuchnikovWiebe2019}, Low, Kliuchnikov, and Wiebe construct well-conditioned choices of the coefficients \(a_j\) and powers \(k_j\) satisfying conditions for high-order convergence of the MPF. They show that implementing the MPF using the linear combination of unitaries (LCU) subroutine~\cite{ChildsWiebe2012,GilyenSuLowEtAl2019} has complexity \(\mathcal{O}((\sum_{\gamma}\|H_{\gamma}\|)T\operatorname{polylog}(T/\epsilon))\), which is near-optimal in both \(T\) and \(\epsilon\). Furthermore, when \(T\) and \(\epsilon\) are treated as constants, \cite{LowKliuchnikovWiebe2019} shows that the MPF complexity depends only on Hamiltonian terms appearing in nested commutators.

However, a comprehensive complexity analysis of the MPF remains lacking. First, \cite{LowKliuchnikovWiebe2019} does not provide a uniform complexity estimate that simultaneously exhibits commutator scaling and near-optimal dependence on time and precision. Specifically, \cite{LowKliuchnikovWiebe2019} presents two separate complexity estimates. The first achieves near-optimal dependence on time and precision but contains a Hamiltonian-dependent factor of order \(\mathcal{O}(\sum_{\gamma}\|H_{\gamma}\|)\), while the second captures commutator dependence but does not account for time and precision dependence. More importantly, \cite{LowKliuchnikovWiebe2019} establishes only qualitative commutator dependence and does not provide a rigorous MPF error bound with an explicit nested-commutator expression. Consequently, the system-size dependence remains ambiguous when the MPF is applied to specific systems of practical relevance.

In this work, we establish improved error analysis and complexity estimate for the MPF based on the second-order product formula, which we refer to as the second-order based MPF or 2nd-based MPF. Our analysis rigorously shows that the second-order based MPF simultaneously achieves explicit commutator scaling and near-optimal dependence on evolution time and target precision. 

Using this refined complexity estimate, we identify several application scenarios in which the second-order-based MPF improves upon low and fixed-order product formulas in its dependence on evolution time and precision, while also exhibiting favorable system-size scaling and, in certain parameter regimes, outperforming leading post-Trotter methods.

We further investigate the commutator-scaling behavior of MPFs constructed from higher-order product formulas, showing that increasing the order of the base sequence can yield additional improvements in system-size dependence. 
Our main contribution is a commutator-sensitive error analysis of existing well-conditioned MPF algorithms that yields complexity bounds with near-optimal dependence on evolution time and precision and explicit dependence on nested-commutator quantities.

Besides their significance in quantum algorithm design, MPFs also have a substantial history in classical algorithms and numerical analysis, where they are commonly referred to as multi-product expansions or extrapolated splitting methods.
In fact, the first MPF, though ill-conditioned, was proposed in~\cite{chin2010multi} as a classical Runge--Kutta--Nyström integrator for classical Hamiltonian systems.
Since then, MPFs have been used to construct arbitrary-order imaginary-time and diffusion algorithms for eigenvalue problems \cite{ChinJanecekKrotscheck2009Imaginary,ChinJanecekKrotscheck2009Diffusion}, general integrators for autonomous and non-autonomous equations \cite{ChinGeiser2011}, high-order propagators for path-integral Monte Carlo \cite{ZillichMayrhoferChin2010}, and algorithms for nonlinear ordinary differential equations and semidiscretized parabolic equations \cite{Geiser2013Nonlinear,BeauregardGeiser2015}.
Recent work continues to develop generalized multi-product extrapolation methods and analyze their stability for classical evolution equations \cite{BlanesCasasShaw2024,DuanQuanYangZhu2026}.
Therefore, the operator-level approximation and error analysis studied in this work are conceptually independent of the quantum LCU implementation, and, after spatial discretization, the same bounds apply to the action of the resulting matrix approximation in classical algorithms.
In this sense, our main result on error estimates for MPFs may also provide useful insights for the broader fields of classical algorithms and numerical analysis.

\subsection{Main Results}
Our main result is an explicit commutator dependent error and complexity bound for the well-conditioned second-order-based MPF. 
We first summarize it as follows. 

\begin{result}[Informal version of~\cref{cor:MPF_complexity}]\label{res:complexity}
     Consider using MPF in~\cref{eqn:MPF_intro} based on second-order product formula for simulating the Hamiltonian $H = \sum_{\gamma=1}^{\Gamma} H_{\gamma}$. 
     Then the number of queries to the base product formula is 
     \begin{equation}
        \mathcal{O}\left( \mu T  \mathrm{poly}\log\left(\frac{\mu T}{\epsilon}\right)\right). 
    \end{equation}
    Here $T$ is the evolution time, $\epsilon$ is the tolerated error, and $\mu$ is a parameter with commutator scaling defined as ($\mathbb{Z}^{+}$ is the set of positive integers)
    \begin{align}
        \mu = 2 \sup_{\substack{(j,l) \in 2\mathbb{Z}^+\times \mathbb{Z}^+: \\ j/2 \geq l \geq 1} } \sup_{\substack{j_1,\cdots,j_l \in 2\mathbb{Z}^{+}: \\ j_1+\cdots+j_l=j}} \left(\prod_{\kappa=1}^{l}  \alpha_{\operatorname{comm},j_{\kappa}+1} \right)^{\frac{1}{j+l}} \label{eqn:mu_intro}, 
    \end{align}
where $\alpha_{\operatorname{comm},j}$ is defined as $\alpha_{\operatorname{comm},j} = \sum_{\gamma_1,\cdots,\gamma_j=1}^{\Gamma} \|[H_{\gamma_1}, [H_{\gamma_2}, \cdots [H_{\gamma_j-1}, H_{\gamma_j}]]]\|$.
\end{result}

\cref{res:complexity} reproduces the nearly linear scaling in the evolution time $T$ and polylogarithmic scaling in the target precision $\epsilon$ for the MPF, as originally demonstrated in~\cite{LowKliuchnikovWiebe2019}. The main improvement of \cref{res:complexity} over the analysis in~\cite{LowKliuchnikovWiebe2019} is that it establishes an $\mathcal{O}(\mu)$ query complexity, where the parameter $\mu$ depends solely on the commutators among the constituent Hamiltonians $H_{\gamma}$. The parameter $\mu$ is guaranteed to be no worse than the bound $\mathcal{O}(\sum_{\gamma} \| H_{\gamma} \|)$ used in~\cite{LowKliuchnikovWiebe2019}. This is because the commutator coefficients satisfy $\alpha_{\operatorname{comm},j} \leq \mathcal{O} ( (\sum_{\gamma} \| H_{\gamma} \|)^j )$, and consequently, $\prod_{\kappa=1}^{\ell} \alpha_{\operatorname{comm}, j_{\kappa}+1} \leq \mathcal{O} ( (\sum_{\gamma} \| H_{\gamma} \|)^{j+\ell} )$. Hence, even in the worst-case scenario, the complexity in \cref{res:complexity} is at least comparable to that of~\cite{LowKliuchnikovWiebe2019}. Moreover, in many practical examples, the commutator-based parameter $\mu$ is significantly smaller than $\mathcal{O}(\sum_{\gamma} \| H_{\gamma} \|)$, thereby providing a tangible advantage.

Note that the parameter $\mu$ depends on nested commutators of depth greater than or equal to $3$, and importantly, the depth is allowed to grow arbitrarily large. This contrasts with the commutator scaling established for the standard product formula in~\cite{ChildsSuTranEtAl2020}, where only nested commutators of fixed depth $(p+1)$ appear in a $p$-th order formula. Informally, the commutator parameter $\mu$ in the MPF characterizes the maximal growth rate of nested commutators as a function of their depth. To illustrate this, suppose that the commutator coefficients scale geometrically as $\alpha_{\operatorname{comm},j} = \mathcal{O}(a^j)$ for some $a > 1$ that may depend on the structure of the Hamiltonians. This assumption and geometric growth are observed in some relevant applications we will discuss in~\cref{sec:app}. 
Under this assumption, we have
$\prod_{\kappa=1}^{\ell} \alpha_{\operatorname{comm},j_{\kappa}+1} = \mathcal{O}(a^{j+\ell}),$
and from the definition of $\mu$ in~\cref{eqn:mu_intro}, it follows directly that $\mu = \mathcal{O}(a)$. This confirms that $\mu$ effectively captures the base rate of geometric growth in the nested commutators with respect to depth. In the worst-case scenario, we have $a = \mathcal{O} (\sum_{\gamma} \| H_{\gamma} \| )$. However, for many physical systems of practical interest, the base $a$ and hence $\mu$ can be smaller.

\subsubsection{Applications and Comparison}\label{sec:intro_app}

\begin{table}[t]
    \renewcommand{\arraystretch}{2.2}
    \centering
    \scalebox{0.76}{
    \begin{tabular}{c|cccc}\hline\hline
        \textbf{System} 
        & \textbf{Post-Trotter} 
        & \textbf{$2$nd Product Formula} 
        & \textbf{$p$-th Product Formula} 
        & \textbf{$2$nd-Based MPF} \\\hline 

        Transverse Field Ising Chain
        & $\min\left\{n^2,\,n\Lambda T\operatorname{polylog}\left(\frac{1}{\epsilon}\right)\right\}$
        & $n^{\frac{3}{2}}\Lambda^{\frac{3}{2}}T^{\frac{3}{2}}\epsilon^{-\frac{1}{2}}$
        & $n^{1+\frac{1}{p}}\Lambda^{1+\frac{1}{p}}T^{1+\frac{1}{p}}\epsilon^{-\frac{1}{p}}$
        & $n^{\frac{4}{3}}\Lambda T \operatorname{polylog}\left(\frac{1}{\epsilon}\right)$ \\

        Chiral Potts Spin Chains
        & $n\Lambda T\operatorname{polylog}\left(\frac{1}{\epsilon}\right)$
        & $n^{\frac{3}{2}}\Lambda^{\frac{3}{2}}T^{\frac{3}{2}}\epsilon^{-\frac{1}{2}}$
        & $n^{1+\frac{1}{p}}\Lambda^{1+\frac{1}{p}}T^{1+\frac{1}{p}}\epsilon^{-\frac{1}{p}}$
        & $n^{\frac{4}{3}}\Lambda T \operatorname{polylog}\left(\frac{1}{\epsilon}\right)$ \\

        Nonlocal Fast-Transform BdG 
        & $n^2$
        & $n^{\frac{3}{2}}\Lambda^{\frac{3}{2}}T^{\frac{3}{2}}\epsilon^{-\frac{1}{2}}$
        & $n^{1+\frac{1}{p}}\Lambda^{1+\frac{1}{p}}T^{1+\frac{1}{p}}\epsilon^{-\frac{1}{p}}$
        & $n^{\frac{4}{3}}\Lambda T \operatorname{polylog}\left(\frac{1}{\epsilon}\right)$ \\

        Kitaev's Honeycomb Model
        & $n\Lambda T\operatorname{polylog}\left(\frac{1}{\epsilon}\right)$
        & $n^{\frac{3}{2}}\Lambda^{\frac{3}{2}}T^{\frac{3}{2}}\epsilon^{-\frac{1}{2}}$
        & $n^{1+\frac{1}{p}}\Lambda^{1+\frac{1}{p}}T^{1+\frac{1}{p}}\epsilon^{-\frac{1}{p}}$
        & $n^{\frac{4}{3}}\Lambda T \operatorname{polylog}\left(\frac{1}{\epsilon}\right)$ \\\hline\hline 
    \end{tabular}}
    \caption{Comparison of the multi-product formula based on the second-order product formula ($2$nd MPF), standard product formulas, and representative post-Trotter methods.
    All entries report gate complexities, and we omit the big-\(\widetilde{\mathcal O}\) notation. 
    Here the parameters \(T\) and \(\epsilon\) denote the evolution time and simulation error, respectively. 
    \(\Lambda\) and \(n\) denote the relevant Hamiltonian scale and the system size parameter, respectively, whose concrete meanings vary in different models. 
    For the post-Trotter entry, the \(n\Lambda T\operatorname{polylog}(1/\epsilon)\) branches for the transverse field Ising, chiral Potts and Kitaev's honeycomb use geometrically local lattice simulation~\cite{HaahHastingsKothariLow2021}, the \(n^2\) Ising branch uses exact free-fermion or Cartan compilation~\cite{CerveraLierta2018ExactIsing,KokcuSteckmannWangEtAl2022}, and the \(n^2\) BdG entry uses direct fermionic Gaussian compilation and fast-forwarding~\cite{JiangSungKechedzhiEtAl2018,GuSommaSahinoglu2021,KokcuCampsOftelieEtAl2023}. 
    The chiral Potts entry is stated for fixed local dimension \(d\) and is conditional on~\cref{ass:scp-extensive-onsager-generators}.}
    \label{tab:main}
\end{table}

Using~\cref{res:complexity}, we analyze the gate complexity of the second-order-based MPF for several applications considered in~\cref{sec:app}, including spin Hamiltonians with geometric commutator growth, nonlocal fast-transform Bogoliubov--de Gennes (BdG) Hamiltonians, and Kitaev's honeycomb model.
\Cref{tab:main} summarizes the resulting scalings for these applications and compares them with those of standard product formulas and the best existing post-Trotter methods, as specified for each application in the captions of~\cref{tab:main}.
The product-formula entries use the commutator-scaling bounds of~\cite{ChildsSuTranEtAl2020}.
We remark that the closely related extrapolation methods such as~\cite{endo2019mitigatingalerrorsinasimulation,watkins2023multproductrichardson} are not included in this comparison, as their proven output and resource models differ from the coherent Hamiltonian simulation algorithms being compared here.

Compared to the second-order product formula, our results demonstrate that $2$nd-based MPF can achieve a polynomial speedup in terms of both the system-size parameter $n$ and the evolution time $T$, and an exponential speedup in terms of the error $\epsilon$. 
Similar speedups in $T$ and $\epsilon$ also hold when we compare $2$nd-based MPF with a fixed $p$-th order product formula, but for all the considered applications, the dependence on $n$ of $2$nd-based MPF becomes generally worse than that of the product formula with sufficiently high order, specifically $p>3$. 
This is because a $p$-th order product formula only depends on nested commutators of fixed depth $p+1$, whereas $2$nd-based MPF still depends on commutators beginning at depth $3$. 

The comparison with post-Trotter methods is more subtle and depends on the model.
For most of the applications we consider, the second-order-based MPF is asymptotically worse in at least one parameter than the specialized post-Trotter methods listed in~\cref{tab:main}.
Specifically, for the transverse-field Ising model, chiral Potts model, and Kitaev's honeycomb model, geometrically local lattice simulation~\cite{HaahHastingsKothariLow2021} achieves \(\widetilde{\mathcal O}\left(n\Lambda T\operatorname{polylog}\left(1/\epsilon\right)\right)\), improving on the \(n^{4/3}\) system-size dependence of the second-order-based MPF.
These advantages arise because the post-Trotter baselines exploit additional structure, such as geometric locality or free-fermion integrability, that is not used by the general commutator-based MPF bound.
As a further point of comparison, the MPF may outperform general-purpose post-Trotter methods.
For example, optimal general-purpose qubitization or QSVT~\cite{LowChuang2019,GilyenSuLowEtAl2019} requires \(\mathcal O\left(n \Lambda T+\log(1/\epsilon)\right)\) queries to a block encoding of the Hamiltonian.
A straightforward explicit LCU block encoding of the Ising, chiral Potts, or Kitaev Hamiltonian costs \(\mathcal{O}(n)\) gates, yielding an overall QSVT gate complexity of \(\mathcal O\left(n^2\Lambda T+n\log(1/\epsilon)\right)\).
Thus, the second-order-based MPF has better system-size dependence, improving \(n^2\) to \(n^{4/3}\).

Among all the applications considered, a particularly notable example is the nonlocal fast-transform BdG model, for which the lattice simulation method~\cite{HaahHastingsKothariLow2021} does not apply uniformly due to the nonlocality of the Hamiltonian.
Instead, direct Gaussian compilation and fast-forwarding~\cite{JiangSungKechedzhiEtAl2018,GuSommaSahinoglu2021,KokcuCampsOftelieEtAl2023} yield an \(\mathcal{O}(n^2)\) post-Trotter complexity.
Nevertheless, unlike in the other examples, the MPF can be asymptotically more efficient than both this Gaussian bound and a fixed-order product formula in certain parameter regimes, as derived in~\cref{eqn:fast-bdg-fixed-p-general-regime}.
For example, when \(\Lambda = \mathcal{O}(1)\), choosing \(T = n^{\tau}\) and \(\epsilon = n^{-\kappa}\) with \(\tau<2/3\) and \(\tau+\kappa>p/3-1\) yields a regime in which the second-order-based MPF outperforms both the post-Trotter method and the \(p\)-th order product formula.
Such a regime also exists when comparing against a product formula with adaptively optimized order, although this advantage requires the inverse super-polynomial precision regime described in~\cref{eqn:fast-bdg-optimized-general-regime}.

Thus, the conclusion is that the 2nd-based MPF improves low and fixed-order product formulas in time and precision while retaining commutator-sensitive scaling, and its advantage over post-Trotter methods depends on the Hamiltonian structure and the asymptotic parameter regime.

\subsubsection{Proof Sketch}
The proof of~\cref{res:complexity} provides a rigorous foundation for the formal derivation of the MPF presented in~\cite{LowKliuchnikovWiebe2019}. The main technical contribution is a precise representation of the error associated with the second-order product formula. Specifically, for a fixed short evolution time \(t\) and arbitrary integers \(k\) and \(p\), we establish the following expansion
\begin{equation}\label{eqn:error_representation_intro}
    U_2(t/k)^k = e^{-iHt} + \sum_{j \in 2 \mathbb{Z}^+ }^{\infty} \widetilde{E}_{j+1}(t) \frac{1}{k^j} + \widetilde{F}_p(t,k),
\end{equation}
where the operators \(\widetilde{E}_{j+1}(t)\) are independent of \(k\) and satisfy \(\widetilde{E}_{j+1}(t)=\mathcal{O}(t^{j+1})\), while the remainder term \(\widetilde{F}_p(t,k)\) is of order \(\mathcal{O}(t^{3p})\). Crucially, the constant factors in the bounds for these operators depend only on the sizes of nested commutators among the Hamiltonian components \(H_{\gamma}\). This structure enables the derivation of high-order error bounds for the MPF by applying the MPF order conditions to cancel all lower-order terms. Consequently, the constant factor in the final error bound is determined entirely by the commutator structure of the Hamiltonian.

There are two steps in proving~\cref{eqn:error_representation_intro}. We first use the multi-term Baker--Campbell--Hausdorff (BCH) formula~\cite{ArnalCasasChiralt2020} to express the repeated product \(U_2(t/k)^k\) as a single exponential, \(U_2(t/k)^k=e^{-i(H+\Delta H)t}\), where the perturbation term \(\Delta H\) is given by an infinite series of nested commutators. Unlike many previous works, we make this step mathematically rigorous by identifying a sufficient convergence condition for the infinite BCH series, thereby ensuring the validity of the exponential representation. The second step is to derive an explicit representation of the difference \(e^{-i(H+\Delta H)t}-e^{-iHt}\). This is accomplished using a high-order variation-of-parameters formula, which may be interpreted as a refined version of the truncated Dyson series in the interaction picture. Such a formulation has recently been employed in the design of both classical and quantum algorithms for simulating open quantum dynamics~\cite{CaoLu2021,LiWang2023}.

\subsubsection{MPF Based on High-Order Product Formula}
We have analyzed the complexity of the second-order based MPF and compared it with other quantum simulation algorithms. However, the MPF formalism, as expressed in~\cref{eqn:MPF_intro}, is flexible and permits the use of product formulas of arbitrary order as the base sequence. In the following result, we provide an estimate of the query complexity for the MPF constructed from a \(2p\)-th-order symmetric product formula, referred to as the \(2p\)-th-based MPF, for any positive integer \(p\).

\begin{result}[Informal version of~\cref{cor:MPF_complexity_high_order}]\label{res:complexity_high_order}
    Suppose that there exists a well-conditioned MPF based on $2p$-th order symmetric product formula, and consider using this MPF for simulating the Hamiltonian $H = \sum_{\gamma=1}^{\Gamma} H_{\gamma}$.  
    Then the number of queries to the base product formula is 
        \begin{equation}
            \mathcal{O}\left( \mu^{(2p)} T  \operatorname{polylog}\left(\frac{\mu^{(2p)} T}{\epsilon}\right)\right). 
        \end{equation}
    Here $T$ is the evolution time, $\epsilon$ is the tolerated error, and $\mu^{(2p)}$ is a parameter with commutator scaling defined as ($\mathbb{Z}^{+}$ is the set of positive integers)
    \begin{align}
        \mu^{(2p)} = 2 \sup_{\substack{(j,l) \in 2\mathbb{Z}^+\times \mathbb{Z}^+: \\ j/2 \geq l \geq 1} } \sup_{\substack{j_1,\cdots,j_l \in 2\mathbb{Z}^{+}: \\ j_1+\cdots+j_l=j, \\ j_{\kappa}\geq 2p }} \left(\prod_{\kappa=1}^{l}  \alpha_{\operatorname{comm},j_{\kappa}+1} \right)^{\frac{1}{j+l}} , \label{eqn:mu_intro_high_order}
    \end{align}
where $\alpha_{\operatorname{comm},j}$ is defined as
$\alpha_{\operatorname{comm},j} = \sum_{\gamma_1,\cdots,\gamma_j=1}^{\Gamma} \|[H_{\gamma_1}, [H_{\gamma_2}, \cdots [H_{\gamma_j-1}, H_{\gamma_j}]]]\|$.
\end{result}
Similar to the second-order based MPF, the \(2p\)-th-based MPF also achieves nearly linear scaling in the evolution time \(T\) and polylogarithmic scaling in the error tolerance \(\epsilon\). The key difference lies in the commutator dependence. Specifically, for a general \(2p\)-th-based MPF, the commutator scaling parameter \(\mu^{(2p)}\) depends on nested commutators of depth greater than or equal to \(2p+1\). Thus, MPFs constructed from higher-order product formulas inherently involve more deeply nested commutators. Notably, commutator scalings of greater depth typically correspond to improved system-size dependence in many physical applications, as demonstrated in the analysis of the standard product formula~\cite{ChildsSuTranEtAl2020}. Consequently, one can expect improved asymptotic scaling with respect to system size when employing \(2p\)-th-based MPFs with larger \(p\). However, since the cost of implementing the standard product formula grows exponentially with its order, using very high-order product formulas as the base sequence may be impractical for most applications.

A caveat to~\cref{res:complexity_high_order} is that it assumes the existence of a well-conditioned MPF. By a well-conditioned MPF, we mean that both the sum of the absolute values of the coefficients, \(\sum_j |a_j|\), and the sum of the absolute values of the powers, \(\sum_j |k_j|\), grow at most polynomially with respect to the convergence order of the corresponding MPF. This condition is crucial for ensuring an efficient quantum implementation of the MPF via the LCU subroutine. While~\cite{LowKliuchnikovWiebe2019} rigorously establishes the existence of a well-conditioned second-order based MPF, it remains an open question whether well-conditioned \(2p\)-th-based MPFs can be constructed for general \(p\geq 2\). Nonetheless, the numerical optimization results in~\cite{LowKliuchnikovWiebe2019} provide strong empirical evidence supporting the existence of a well-conditioned fourth-order-based MPF.

\subsection{Related Works}\label{sec:related_work}
Our work focuses on the complexity analysis of the MPF under its standard quantum implementation, namely, using the LCU subroutine to implement the linear combination of product formulas. Several previous works have studied variants of the MPF and their theoretical analysis. In this section, we discuss these works and compare them with ours.

A common variant of the MPF approach performs the linear-combination step classically, particularly when the objective is to estimate observables of the quantum system. For instance, \cite{rendon2023improvedtrottersimulation} proposes a polynomial-interpolation-based algorithm for digital quantum simulation. This method improves accuracy by collecting low-order Trotter simulation results at multiple time-step sizes and subsequently using polynomial interpolation, implemented with Chebyshev nodes and polynomials, to estimate the ideal zero-time-step limit. The motivation of~\cite{rendon2023improvedtrottersimulation} is to overcome limitations of low-order Trotter formulas, such as their low convergence order, closely aligning with the objectives of MPF-based algorithms.

Several extensions of the MPF tailored to near-term and early fault-tolerant quantum devices have also been proposed. For example, \cite{faehrmann2022randomizing} introduces a randomized MPF method for estimating observables using importance sampling. Specifically, the algorithm samples the power parameters in the MPF with probabilities proportional to their coefficients, independently estimates the observable corresponding to each sampled product-formula power on a quantum device, and then classically averages the resulting estimates. By avoiding oblivious amplitude amplification, which is required in the standard quantum LCU subroutine, the randomized MPF reduces circuit depth while maintaining comparable error scaling, at the cost of additional classical repetitions. Similar importance-sampling strategies have been widely used in early fault-tolerant hybrid algorithms for tasks including ground-state energy estimation~\cite{LinTong2022}, phase estimation~\cite{WanBertaCampbell2022}, solving linear systems~\cite{WangMcArdleBerta2023}, linear differential equations~\cite{AnLiuLin2023,AnChildsLin2023}, and general LCU implementations~\cite{Chakraborty2023}. Another related work, \cite{vasquez2023wellharware}, proposes a hybrid quantum--classical implementation of the MPF that is better suited to noisy hardware. Unlike fully quantum MPF implementations based on the LCU algorithm, this approach does not require additional qubits or controlled operations. While these approaches focus on estimating observables on near-term or early fault-tolerant devices, our work addresses a different regime: the complexity of quantum-state preparation for Hamiltonian simulation in the fully fault-tolerant setting, using quantum LCU to implement the MPF. 
We remark that neither output model is universally preferable. If a small collection of expectation values is the final goal, these hybrid approaches could be particularly attractive. However, if coherent state preparation or Hamiltonian simulation as a subroutine is required, Hamiltonian simulation primitive with quantum state output remains necessary. 

To the best of our knowledge, three recent works~\cite{watkins2022timedependentclock,zhuk2023trotter,watkins2023multproductrichardson} attempt to provide error analyses of MPFs exhibiting commutator scaling. However, as we now explain, these works do not fully address the concerns resolved in the present study. The work in~\cite{watkins2022timedependentclock} generalizes the MPF to time-dependent quantum simulation and demonstrates that both time-independent and time-dependent MPFs exhibit commutator scaling. Nevertheless, the resulting commutator scaling remains implicit, reflecting a limitation similar to that of the original well-conditioned MPF formulation in~\cite{LowKliuchnikovWiebe2019}. The analysis in~\cite{zhuk2023trotter} focuses on commutator error scaling for the hybrid quantum--classical MPF algorithm proposed in~\cite{vasquez2023wellharware}. Their results are restricted to the regime in which the number of terms \(M\) in the MPF sum satisfies \(M\leq p+1\) for a \(p\)-th-order base product formula, yielding only quadratic accuracy improvements over the base product formula. Consequently, their analysis does not capture the fast convergence properties of the well-conditioned MPF described in~\cite{LowKliuchnikovWiebe2019}. The recent work~\cite{watkins2023multproductrichardson} provides commutator error scaling for the Richardson-extrapolation-based algorithm introduced in~\cite{endo2019mitigatingalerrorsinasimulation}, which aims to mitigate algorithmic errors in product formulas. Similar to~\cite{zhuk2023trotter}, this work addresses hybrid implementations of the MPF with classical observable outputs rather than fully quantum state preparation. Our work addresses this gap by providing an explicit commutator-scaling error analysis for the fully quantum, well-conditioned MPF of~\cite{LowKliuchnikovWiebe2019}, thereby clarifying its complexity and performance in the fault-tolerant quantum computing regime.

\subsection{Discussions and Open Questions}
We have established a complexity analysis for second-order MPFs that incorporates explicit commutator scaling together with near-optimal dependence on the evolution time and target precision. 

Our results yield advantage of MPF compared to second-order product formula, and identify model and parameter specific regimes in which the resulting complexity bound could be smaller than those of leading post-Trotter methods. 

We now discuss several open questions arising from our work.

Our complexity estimates primarily characterize asymptotic scaling in terms of physical and algorithmic parameters and therefore do not directly imply practical advantages. Specifically, the practical break-even point between MPFs and standard product formulas or other post-Trotter methods depends on implementation-level constants, the overheads associated with LCU and robust oblivious amplitude amplification, and the Hamiltonian family under consideration. A detailed resource estimation comparing these methods in concrete regimes therefore remains an interesting practical question.

 Our commutator-scaling parameter depends on nested commutators of all depths greater than or equal to \(3\). While we demonstrate the effectiveness of this scaling in several applications, the analysis of arbitrarily high-depth nested commutators may remain challenging for certain systems, such as nuclear effective field theories~\cite{WatsonBringewattShawEtAl2023}. It is therefore of practical interest to establish commutator-scaling bounds for second-order MPFs that depend only on low-depth nested commutators. \footnote{As discussed later in the ``Subsequent Work'' subsection, since the release of the first version of our work, Ref.~\cite{mizuta2025commutatorscalinghamiltoniansimulation} has obtained such a result for the special class of \(k\)-local, \(g\)-extensive Hamiltonians. To the best of our knowledge, however, the general case remains open and warrants further investigation. We expect that such a result may be obtained by refining our analysis using a truncated BCH formula together with an appropriate remainder estimate, since the high-depth nested commutators in our current analysis arise solely from the use of the infinite BCH series.}

Our analysis proceeds by establishing an upper bound on the operator-norm distance between the MPF operator \(U_{\mathrm{MP}}\) and the ideal time-evolution operator \(U\), thereby providing a worst-case performance guarantee over all initial states. In specific simulation instances, however, a more relevant error metric may be state-dependent, namely \(\|U_{\mathrm{MP}}\ket{\psi_0}-U\ket{\psi_0}\|\) for a given input state \(\ket{\psi_0}\). Such a vector-norm error can be substantially smaller than the worst-case operator-norm bound when the dynamics are restricted to a subspace of states of interest. Deriving state-dependent vector-norm error bounds may therefore lead to improved complexity estimates in practical simulations. This approach has been applied to standard product formulas and shown to yield certain advantages~\cite{JahnkeLubich2000,SahinogluSomma2020,SuHuangCampbell2020,AnFangLin2021,YiCrosson2021,ZhaoZhouShawEtAl2022,ChenBrandao2023,Burgarth2023,burgarth2023strong}. It would therefore be interesting to investigate whether MPFs can similarly benefit from state-dependent error bounds.

Our analysis shows that MPFs can simultaneously achieve commutator scaling and near-optimal dependence on time and precision. Nevertheless, further improvements in complexity may be possible through a refined analysis of MPFs or the development of improved algorithms. First, although our analysis achieves near-optimal dependence on the evolution time \(T\) and precision \(\epsilon\), it still contains additional polylogarithmic factors in \(T\), and the dependence on \(\log(1/\epsilon)\) has degree greater than one. By comparison, qubitization and quantum singular value transformation algorithms for quantum simulation achieve query complexity \(\mathcal{O}(\|H\|T+\log(1/\epsilon))\), where \(\|H\|\) is the spectral norm of the Hamiltonian. This scaling is exactly linear in both \(T\) and \(\log(1/\epsilon)\) and is optimal, matching known lower bounds. However, current analyses indicate that qubitization and quantum singular value transformation methods do not exploit commutator scaling. Thus, it remains an open problem whether one can design quantum simulation algorithms that combine commutator scaling with exactly optimal dependence on time and precision.

Another notable feature of optimal time-independent quantum simulation algorithms is the additive scaling between the precision dependence and other problem parameters. In contrast, our current MPF analysis achieves only multiplicative scaling among all parameters. It therefore remains an interesting open problem to develop time-independent quantum simulation algorithms that simultaneously exhibit all three desirable features: commutator scaling, optimal time and precision dependence, and additive scaling.

Beyond the second-order MPF, we also provide a complexity estimate for the \(2p\)-th-order MPF that incorporates commutator scaling. Our general result assumes the existence of a well-conditioned \(2p\)-th-order MPF. It is therefore both natural and important to develop well-conditioned MPF schemes whose base product formulas have order higher than two.

\subsection{Subsequent Work}
Since the first version of this paper appeared, there have been notable developments in this area. Here, we discuss some of these developments and highlight the main changes made to the paper since its initial version.

The most closely related work is~\cite{mizuta2025commutatorscalinghamiltoniansimulation}, which identified a technical gap in the previous version of this paper. Specifically,~\cite{mizuta2025commutatorscalinghamiltoniansimulation} emphasized that dependence on high-depth nested commutators can affect applications to certain Hamiltonian families, such as \(k\)-local Hamiltonians, when the constants in the commutator estimates scale poorly with the commutator depth. 
As a result, our general complexity estimate cannot yield the desired MPF efficiency for these applications. 
To address this issue,~\cite{mizuta2025commutatorscalinghamiltoniansimulation} employed a Floquet--Magnus-based approach to establish commutator scaling for MPFs involving only finite-depth commutator words, thereby obtaining a system-size-efficient cost that properly reflects commutator scaling for \(k\)-local Hamiltonians. 
In the present version of the paper, we have replaced all the examples in~\cref{sec:app} by new ones so that the present examples discussed in~\cref{sec:app} can be suitably analyzed within our framework.

We further note that~\cite{mizuta2025commutatorscalinghamiltoniansimulation} and the present work are complementary in scope.
\cite{mizuta2025commutatorscalinghamiltoniansimulation} focuses on \(k\)-local Hamiltonians and establishes stronger finite-depth commutator scaling through a different expansion, whereas the present work applies to general finite-dimensional Hamiltonian decompositions but yields favorable system-size scaling only with the required geometric commutator growth. 
This generality is important in view of the newly written~\cref{sec:app} and our subsequent work~\cite{aftab2026mpf-lchs}, in which commutator scaling for MPFs is extended to more general linear non-homogeneous differential equations using the Linear Combination of Hamiltonian Simulation paradigm~\cite{AnLiuLin2023,low2025optimallchs,AnChildsLin2023}, covering examples of differential equations that do not satisfy the hypotheses of Floquet--Magnus analysis.

Relatedly,~\cite{watkins2023multproductrichardson} developed Richardson extrapolation and polynomial interpolation methods for improving product formula of estimating time-evolved expectation values. 
This work achieves polylogarithmic precision dependence in the Trotter depth and includes an explicit commutator-scaling parameter that is closely related in spirit to the parameter \(\mu_m\) used in the present work. 
The outputs, however, are different. \cite{watkins2023multproductrichardson} analyzes extrapolated estimates of expectation values, leading to scalar observable estimation bounds. 
This approach could be particularly attractive if a small collection of expectation values is the final goal. 
Additionally, in a complementary direction,~\cite{chakraborty2025qsvt} developed a QSVT framework without block encodings, using extrapolation techniques to obtain near-optimal complexity with minimal ancilla overhead in broader circuit settings. This further illustrates the utility of extrapolation-based techniques beyond the Hamiltonian simulation problem.
By contrast, the present work analyzes well-conditioned MPFs implemented coherently through LCU and establishes operator-norm bounds for approximating the full time-evolution operator. 
Such an operator-norm bounds are useful when a coherent quantum state is desired either on its own or as a subroutine of other quantum algorithms.

\subsection{Organization}
The remainder of this paper is organized as follows. In~\cref{sec:prelim}, we review the notation and preliminary results on product formulas and the MPF. In~\cref{sec:results}, we present and prove our main result: a complexity analysis of the \(2\)nd-based MPF with explicit commutator scaling. We then discuss applications in~\cref{sec:app} and extend the analysis to the \(2p\)-th-based MPF in~\cref{sec:general_MPF}.

\section{Preliminaries}\label{sec:prelim}
In this section, we present the preliminary material used throughout the paper. We first introduce the notation and then review product formulas and well-conditioned multi-product formulas (MPFs). Finally, we describe the quantum implementation of MPFs using the LCU technique together with oblivious amplitude amplification.

\subsection{Notation}\label{sec:notation}
We briefly discuss the main notation used throughout this work.

\subsubsection{Standard Notation}
Let \(\mathbb{R}\), \(\mathbb{C}\), and \(\mathbb{N}\) denote the sets of real, complex, and natural numbers, respectively. Let \(\mathbb{Z}^{+}\) denote the set of positive integers, and let \(m+n\mathbb{Z}^{+}\) denote the set of integers of the form \(m+nk\), where \(k\in\mathbb{Z}^{+}\). Let \(S_k\) denote the symmetric group on \(k\) letters. For \(\sigma=(\sigma(1),\ldots,\sigma(k))\in S_k\), let \(d_\sigma\) denote the number of descents of \(\sigma\), namely the number of positions \(i\in\{1,\ldots,k-1\}\) such that \(\sigma(i+1)<\sigma(i)\). If \(f,g:\mathbb{N}\to\mathbb{R}^{+}\) are nonnegative functions, we use the following standard notation from complexity theory:
\begin{enumerate}
  \item  $f(n) = \mathcal{O}(g(n))$ if and only if there exists a constant $C > 0$ and an integer $N \in \mathbb{N}$ such that $f(n) \leq C g(n)$ for all $n \geq N$. The notation $\widetilde{\mathcal O}$ hides polylogarithmic factors.
  \item  $f(n) = \Omega(g(n))$ if and only if there exists a constant $C > 0$ and an integer $N \in \mathbb{N}$ such that $f(n) \geq C g(n)$ for all $n \geq N$.
  \item  $f(n) = \Theta(g(n))$ if and only if $f(n) = \mathcal O (g(n))$ and $f(n) = \Omega(g(n))$.
\end{enumerate}

\subsubsection{Linear Algebra Notation}
All vector spaces considered in this paper are finite-dimensional and defined over \(\mathbb{C}\), and all operators are represented by complex square matrices. Throughout, we assume that all operators are time-independent, as our primary focus is the time-independent quantum simulation problem. We will frequently consider operators of the form
\begin{equation}\label{multi-A}
    A=\sum_{\gamma_1,\ldots,\gamma_k} A_{\gamma_1,\ldots,\gamma_k},
\end{equation}
where each \(A_{\gamma_1,\ldots,\gamma_k}\) acts on the underlying finite-dimensional complex vector space. 
This notation is used because, later in the paper, the indices encode the Hamiltonian terms that appear in a given nested commutator. Consequently, the dependence on the word \((\gamma_1,\ldots,\gamma_k)\) constitutes an essential part of the data to be estimated. Replacing this notation by a single symbol, such as \(A_k\), would therefore obscure the underlying commutator structure.
The linear algebra notation employed throughout the paper is summarized below:

\begin{enumerate}[label={(\alph*)}]
\item $\|A\|$ denotes the spectral norm of $A$, which is the largest singular value of $A$.

    \item  \(\|A\|_1\) denotes the \(1\)-norm of \(A\), defined as follows. If \(A\) admits a decomposition of the form \(A=\sum_{\gamma_1,\ldots,\gamma_k} A_{\gamma_1,\ldots,\gamma_k}\), then the \(1\)-norm of \(A\) is defined by
    \begin{equation}
    \| A \|_1 = \sum_{\gamma_1, \cdots, \gamma_k} \| A_{\gamma_1, \cdots, \gamma_k} \|.
    \end{equation}

    \item \(\VERT A \VERT_1\) denotes the \emph{induced} \(1\)-norm of \(A=\sum_{\gamma_1,\ldots,\gamma_k} A_{\gamma_1,\ldots,\gamma_k}\), defined as
     \begin{equation}
        \VERT A \VERT_1 
        =
        \max_j \max_{\gamma_j} \sum_{\gamma_1,\ldots,\gamma_{j-1}, \gamma_{j+1},\ldots,\gamma_k} \left\| A_{\gamma_1,\ldots,\gamma_k} \right\|. 
    \end{equation}
    Note that $\VERT A  \VERT_1 \leq \left\lVert A \right\rVert_1$. 
    
    \item For two matrices \(A_1\) and \(A_2\), their commutator is denoted by \([A_1,A_2]=A_1A_2-A_2A_1\). More generally, \([A_1A_2\cdots A_\Gamma]\) denotes the nested commutator
\begin{equation}
[A_1,[A_2,\ldots,[A_{\Gamma-1},A_\Gamma]\ldots]].
\end{equation}
When \(A_1=\cdots=A_{\Gamma-1}=A\) and \(A_\Gamma=B\), we abbreviate this nested commutator as \(\operatorname{ad}^{\Gamma-1}_A(B)\).

    \item For a vector $\vec{a} = (a_1, \ldots, a_n)$, $\|\vec{a}\|_1$ denotes its vector 1-norm $\|\vec{a}\|_1 = \sum_{j=1}^n |a_j|$.
    \end{enumerate}
    Note that $\|\vec{a}\|_1 $ should not be confused with the matrix 1-norm $\|A\|_1$ defined above.

\subsection{Product Formulas}\label{sec:product-formulas}
Consider a quantum system governed by the Hamiltonian \(H=\sum_{\gamma=1}^{\Gamma}H_\gamma\). Product formulas assume that the time evolution generated by each individual term \(H_\gamma\), namely \(U_\gamma(t)=e^{-itH_\gamma}\), can be implemented efficiently on a quantum computer. Under this assumption, a generic product formula consisting of \(\Xi\) factors approximates the exact time-evolution operator \(U(t)=e^{-itH}\) by an expression of the form
\begin{equation}\label{generic-prod}
    U_p(t)
    =
    \prod_{\xi=1}^{\Xi}
    \prod_{\gamma=1}^{\Gamma}
    e^{-it a_{\xi,\gamma} H_{\pi_\xi(\gamma)}},
\end{equation}
where \(a_{\xi,\gamma}\in\mathbb{R}\) and \(\pi_\xi\in S_\Gamma\) is a permutation of the Hamiltonian terms. The product formula in \cref{generic-prod} is said to be of order \(p\) if its one-step approximation error satisfies \(\|U_p(t)-U(t)\|=\mathcal{O}(t^{p+1})\). The first-order formula is given by the Lie--Trotter formula~\cite{hall2013lie}, whereas higher-order product formulas can be constructed recursively using the Suzuki recursion. This construction yields the following family of formulas:
\begin{equation}
U_1(t) = \prod_{\gamma=1}^\Gamma e^{-i(t/r)H_\gamma}, \quad 
U_2(t) = \prod_{\gamma=\Gamma}^1 e^{-i(t/2r)H_\gamma} \prod_{\gamma=1}^{\Gamma} e^{-i(t/2r)H_\gamma},
\end{equation}
and higher orders follow the Suzuki recursion
\begin{equation}\label{eqn:Trotter_Suzuki}
U_{2p+2}(t) = U_{2p}(s_pt)^2 U_{2p}((1-4s_p)t) U_{2p}(s_p\Delta)^2,
\quad s_p = (4-4^{1/(2p+1)})^{-1}.
\end{equation}
In practice, the simulation interval \([0,T]\) is partitioned into \(r\) segments, and the evolution over each segment is approximated using a product formula. The resulting \(p\)-th-order approximation to the full time evolution is therefore \(U_p(T/r)^r\). Childs et al.~\cite{ChildsSuTranEtAl2020} showed that, for a \(p\)-th-order product formula, it suffices to choose
\begin{equation}
    r = \mathcal{O}\left( \frac{ \alpha_{\operatorname{comm},p+1}^{1/p} T^{1+1/p}}{\epsilon^{1/p}} \right), 
\end{equation}
where \(\alpha_{\operatorname{comm},p+1}=\sum_{j_1,\ldots,j_{p+1}=1}^{\Gamma}\|[H_{j_1}H_{j_2}\cdots H_{j_{p+1}}]\|\) denotes the sum of the spectral norms of all nested commutators of depth \(p+1\). A key advantage of product formulas is that their error bounds exhibit commutator scaling, which can be substantially smaller than bounds based directly on the spectral norm of the Hamiltonian in many practical settings. However, for any fixed order, their complexity scales superlinearly with \(T\) and superlogarithmically with \(1/\epsilon\).

Moreover, higher-order Trotter--Suzuki formulas incur large order-dependent prefactors. In Suzuki's recursive construction, increasing the order by two multiplies the number of exponentials by a factor of \(5\). Consequently, a recursively constructed formula of order \(2p\) consists of \(\mathcal{O}(5^p)\) second-order blocks and therefore contains a number of exponentials that grows exponentially with the order parameter~\cite{suzuki1991general,suzuki1993general}.

\subsection{Multi-Product Formulas (MPF)}\label{sec:mpf}
The approach of approximating high-order time evolution by a linear combination of low-order product formulas is commonly referred to in the numerical analysis literature as a MPF. For a short time step \(\Delta\), an MPF takes the form
\begin{equation}\label{eq:multi-product}
U_{\operatorname{MP}}(\Delta)=\sum_{j=1}^M a_j U_p(\Delta/k_j)^{k_j},
\end{equation}
where \(a_j\in\mathbb{R}\) and \(k_j\in\mathbb{N}\). Since an MPF is constructed as a linear combination of repeated applications of a \(p\)-th-order product formula, we refer to it as a \(p\)-th-order-based MPF. Despite being constructed from a fixed-order product formula, an MPF can attain arbitrarily high convergence order. The underlying idea is to choose the coefficients \(a_j\) and integers \(k_j\) so that the leading error terms associated with the individual components approximately cancel in the linear combination. In particular, Low, Kliuchnikov, and Wiebe exploit this principle in~\cite{LowKliuchnikovWiebe2019} and, for any integer \(m\), construct a linear combination of second-order product formulas of the form
\begin{equation}\label{eqn:multiproduct}
    U_{\operatorname{MP}} (\Delta) = \sum_{j=1}^m a_j U_{2}(\Delta/k_j)^{k_j} = e^{-iH\Delta} + \mathcal{O}(\Delta^{2m+1}). 
\end{equation} 
Here, \(M\) is chosen to be \(m\). The parameters \(a_j\) and \(k_j\) are selected to satisfy the corresponding order conditions and to ensure that the local truncation error is of order \(2m+1\). The analysis in~\cite{LowKliuchnikovWiebe2019} relies on the observation that any symmetric product formula admits a formal Baker--Campbell--Hausdorff (BCH) expansion
\begin{equation}\label{equation:formal-bch}
    U_{2}(\Delta) = e^{-iH\Delta + E_3\Delta^3 + E_5\Delta^5 + \cdots}, 
\end{equation}
where the \(E_l\) are operators involving the Hamiltonian terms \(H_\gamma\). We note that~\cite{LowKliuchnikovWiebe2019} does not address the convergence of the formal BCH expansion in \cref{equation:formal-bch}. A subsequent formal Taylor expansion then yields an expression for \(U_2^{k_j}(\Delta/k_j)\) of the form
\begin{equation}
    U_{2}^{k_j}(\Delta/k_j) = e^{-iH\Delta} + \frac{\Delta^3}{k_j^2} \widetilde{E}_3(\Delta) + \frac{\Delta^5}{k_j^4} \widetilde{E}_5(\Delta) + \cdots.  
\end{equation}
Here, the operators \(\widetilde{E}_{\ell}(\Delta)\), whose explicit forms are not specified in~\cite{LowKliuchnikovWiebe2019}, are independent of \(k_j\) and uniformly bounded with respect to \(\Delta\). Consequently, satisfying the order conditions requires the cancellation of all lower-order terms, and hence the parameters must satisfy the linear system
\begin{equation}\label{eqn:multiproduct_linear_system}
    \left( \begin{array}{cccc}
        1 & 1 & \cdots & 1 \\
        k_1^{-2} & k_2^{-2} & \cdots & k_m^{-2} \\
        \vdots & \vdots & \ddots & \vdots \\
        k_1^{-2m+2} & k_2^{-2m+2} & \cdots & k_m^{-2m+2}
    \end{array} \right) 
    \left( \begin{array}{c}
        a_1 \\
        a_2 \\
        \vdots \\
        a_m 
    \end{array}\right) 
    = \left(\begin{array}{c}
        1 \\
        0 \\
        \vdots \\
        0 
    \end{array}\right). 
\end{equation}
It is shown in~\cite{LowKliuchnikovWiebe2019} that there exists a solution to~\cref{eqn:multiproduct_linear_system} such that
\begin{equation}
    \|\vec{k}\|_1 = \mathcal{O}(m^2\log m), \quad \|\vec{a}\|_1 = \mathcal{O}(\log m). 
\end{equation}

Intuitively, the MPF is expected to achieve high-order convergence while retaining commutator scaling, since the order conditions in~\cref{eqn:multiproduct_linear_system} cancel all lower-order error terms, and the MPF is exact whenever the Hamiltonian terms \(H_\gamma\) mutually commute. Indeed, \cite{LowKliuchnikovWiebe2019} establishes two results that separately demonstrate these two properties.

\begin{lem}[Theorem 2 in \cite{LowKliuchnikovWiebe2019}]
\label{low-1}
Time evolution operator of a time-independent quantum simulation problem can be approximated by the $2$nd-based MPF with error at most $\epsilon$ and probability at least $1 - \mathcal{O}(\epsilon)$, using $O(T\lambda \log^2(T\lambda/\epsilon))$ controlled-$U_2$ queries, where $\lambda = \sum_{\gamma=1}^{\Gamma} \| H_{\gamma} \|$. 
\end{lem}

\begin{lem}[Theorem 3 in \cite{LowKliuchnikovWiebe2019}]
\label{low-2}
The error of a $p$-th-based MPF depends on Hamiltonian terms that all occur in commutators $[H_{\gamma_q}, [\ldots, [H_{\gamma_2}, H_{\gamma_1}] \ldots ]]$ nested to depth $q > p$. 
\end{lem}
While \cref{low-1} establishes high-order convergence of the MPF constructed from a low-order base sequence, its dependence on the Hamiltonian is characterized in terms of the \(1\)-norm of the Hamiltonian. Commutator scaling is treated separately in~\cref{low-2}. However, that result does not provide an explicit and directly computable expression for this dependence. Consequently, prior to our work, no unified result simultaneously established high-order convergence and explicit commutator scaling, leaving the practical advantages of MPFs only partially characterized.
The purpose of the present analysis is to combine these properties in a single explicit operator norm bound with both near-optimal time and precision dependence and explicit nested-commutator dependence.

\subsection{Quantum Implementation of MPF}\label{sec:prelim_algorithm}
A quantum algorithm for Hamiltonian simulation based on the MPF can be constructed using the LCU framework, following the approach of~\cite{LowKliuchnikovWiebe2019}. As the name suggests, LCU provides an efficient procedure for implementing linear combinations of unitary operators. The LCU technique was introduced in~\cite{ChildsWiebe2012} in the context of simulating Hamiltonians expressed as sums of Pauli operators and has since become a standard tool in quantum linear algebra. Here, we give a concrete LCU implementation of the MPF in~\cref{eq:multi-product}. For a broader discussion of LCU and its general-purpose applications, we refer to~\cite{GilyenSuLowEtAl2019}. We begin by describing the implementation of the MPF over a short time interval \(\Delta\). Suppose that we are given a pair of state-preparation oracles \((O_l,O_r)\) such that
\begin{align}
O_l: \ket{0} \rightarrow \frac{1}{\sqrt{\|\vec{a}\|_1}} \sum_{j=1}^{M} \overline{\sqrt{a_{j}}} \ket{j}, \qquad
O_r: \ket{0} \rightarrow \frac{1}{\sqrt{\|\vec{a}\|_1}} \sum_{j=1}^{M} \sqrt{a_{j}} \ket{j}.
\end{align}
Here, \(\sqrt{z}\) denotes the principal branch square root of \(z\), and \(\overline{z}\) denotes its complex conjugate. We assume access to a select oracle \(\operatorname{SEL}\) satisfying
\begin{equation}
    \text{SEL} = \sum_{j=1}^{M} \ket{j}\bra{j} \otimes U_p(\Delta/k_{j})^{k_{j}}. 
\end{equation}
The operator \((O_l^{\dagger}\otimes I)\operatorname{SEL}(O_r\otimes I)\) then implements the LCU. Specifically, by~\cite[Lemma~52]{GilyenSuLowEtAl2019}, for any input state \(\ket{\psi}\), we have
\begin{equation}
    (O_l^{\dagger}\otimes I) \text{SEL} (O_r\otimes I) \ket{0}\ket{\psi} = \frac{1}{\|\vec{a}\|_1} \ket{0} \left( \sum_{j=1}^M a_j U_{p}(\Delta/k_j)^{k_j} \right)\ket{\psi} + \ket{\perp}, 
\end{equation}
where \(\ket{\perp}\) denotes a possibly unnormalized state satisfying \((\ket{0}\bra{0}\otimes I)\ket{\perp}=0\). The component associated with the ancilla state \(\ket{0}\) encodes the desired MPF, but direct post-selection on \(\ket{0}\) can have success probability as small as \(\mathcal{O}(1/\|\vec{a}\|_1^2)\).
For long-time simulation up to time \(T\), we divide the interval \([0,T]\) into \(r\) equally spaced segments. On each segment, we implement the MPF via LCU for time \(\Delta=T/r\), as discussed above. To prevent the success probability from decaying exponentially, each step uses robust oblivious amplitude amplification~\cite{BerryChildsCleveEtAl2015} to increase the stepwise success probability to \(1-\mathcal{O}(\epsilon/r)\). The overall success probability after \(r\) steps is then \(1-\mathcal{O}(\epsilon)\).

The complexity of the LCU implementation is governed by two principal contributions. First, prior to post-selection, each LCU circuit requires one query to each of the oracles \(O_l\), \(O_r\), and \(\operatorname{SEL}\). The state-preparation oracles \(O_l\) and \(O_r\) can be implemented using \(\mathcal{O}(m)\) gates. This cost is typically subdominant, since it is independent of the Hamiltonian dimension and \(m\) remains small for well-conditioned MPFs. The dominant cost arises from implementing the select oracle \(\operatorname{SEL}\), which requires \(\mathcal{O}(\|\vec{k}\|_1)\) queries to controlled applications of \(U_p\). Second, amplifying the success probability requires \(\mathcal{O}(\|\vec{a}\|_1)\) rounds of robust oblivious amplitude amplification, with each round invoking the LCU circuit a constant number of times. Consequently, each time step requires \(\mathcal{O}(\|\vec{a}\|_1\|\vec{k}\|_1)\) queries to the product formula \(U_p\), and this procedure is repeated over all \(r\) time steps. A more detailed analysis of the resulting complexity is presented later in the paper.
\section{MPF Based on Second-Order Product Formula}\label{sec:results}
In this section, we establish a complexity bound for MPFs based on second-order product formulas. We first apply the multi-term Baker--Campbell--Hausdorff (BCH) formula to represent \(U_2\) as the exponential of a single effective Hamiltonian. To rigorously justify this representation, we derive a sufficient condition for convergence of the BCH expansion under commutator scaling. We then combine the resulting BCH representation with a high-order variation-of-parameters formula to obtain a commutator-scaling error bound for the corresponding MPF and subsequently analyze its overall computational complexity.

\subsection{BCH Formula and Convergence}
We begin with the general form of the Baker--Campbell--Hausdorff (BCH) formula. Let \(X_1,\ldots,X_{\Gamma}\) be arbitrary matrices. Following~\cite{ArnalCasasChiralt2020}, we write \(e^{X_1}e^{X_2}\cdots e^{X_{\Gamma}}=e^{Z}\) for some matrix \(Z\). Formally, \(Z\) admits the expansion
    \begin{align}
        Z &= \sum_{k = 1}^{\infty} \sum_{ \substack{i_1+\cdots+i_{\Gamma }= k,\\ i_j\geq 0}}  \frac{1}{i_1!\cdots i_{\Gamma}!} \phi_k(\underbrace{X_1,\cdots,X_1}_{i_1},\cdots, \underbrace{X_{\Gamma},\cdots,X_{\Gamma}}_{i_{\Gamma}}) \\
        & = X_1+\cdots + X_{\Gamma} + \sum_{k = 2}^{\infty} \sum_{ \substack{i_1+\cdots+i_{\Gamma}=k,\\ i_j\geq 0}}  \frac{1}{i_1!\cdots i_{\Gamma}!} \phi_k(\underbrace{X_1,\cdots,X_1}_{i_1},\cdots, \underbrace{X_{\Gamma},\cdots,X_{\Gamma}}_{i_{\Gamma}}) \label{eqn:BCH_formal}
    \end{align}
    where, for matrices $Y_1,\cdots,Y_k$, 
    \begin{equation}\label{eqn:def_phi}
        \phi_k(Y_1,\cdots,Y_k) = \frac{1}{k^2} \sum_{\sigma \in S_k} (-1)^{d_{\sigma}} \frac{1}{\binom{k-1}{d_{\sigma}}} [Y_{\sigma(1)}Y_{\sigma(2)}\cdots Y_{\sigma(k)}]. 
    \end{equation} 
Note that~\cref{eqn:BCH_formal} is, in general, only a formal expansion and is not guaranteed to converge. However, convergence can be established provided that nested commutators of increasing depth do not grow too rapidly. Moreover, since this section considers symmetric product formulas as the base sequences, with the second-order Trotter--Suzuki formula as the primary example, all even-order terms in the corresponding BCH expansion vanish. We summarize these two properties in \cref{prop:BCH_symmetric}, which provides a rigorous convergence guarantee.

    \begin{prop}\label{prop:BCH_symmetric}
        Let $X_1,\cdots,X_{\Gamma}$ be matrices. 
        Suppose that there exists an integer $J$ such that for any odd positive integer $k \geq J$, we have $\sum_{\gamma_1,\cdots,\gamma_k=1}^{\Gamma} \|[X_{\gamma_1}X_{\gamma_2}\cdots X_{\gamma_k}]\| \leq 1$. 
        Then we have 
        \begin{equation}
            e^{X_1/2}e^{X_2/2}\cdots e^{X_{\Gamma}/2}e^{X_{\Gamma}/2} \cdots e^{X_2/2}e^{X_1/2} = e^Z, 
        \end{equation}
        where $Z = X_1+\cdots + X_{\Gamma} + \sum_{k \in 1+2\mathbb{Z}^+} \Phi_k$ and 
        \begin{align}
            \Phi_k  = \sum_{ \substack{\sum i_j + i_j' = k,\\ i_j,i_j'\geq 0}}  
            \prod_{i_j,i_j'=1}^\Gamma \frac{1}{i_j! i_j'!}
            \phi_k(\underbrace{X_1/2,\cdots,X_1/2}_{i_1},\cdots, \underbrace{X_{\Gamma}/2,\cdots,X_{\Gamma}/2}_{i_{\Gamma}},\underbrace{X_{\Gamma}/2,\cdots,X_{\Gamma}/2}_{i_{\Gamma}'},\cdots, \underbrace{X_1/2,\cdots,X_1/2}_{i_1'}) \nonumber. 
        \end{align}
        Furthermore, we have
        \begin{equation}
            \|\Phi_k\| \leq \frac{1}{k^2} \sum_{\gamma_1,\cdots,\gamma_k=1}^{\Gamma} \|[X_{\gamma_1}X_{\gamma_2}\cdots X_{\gamma_k}]\|. 
        \end{equation} 
    \end{prop}

    \begin{proof}
Let \(X_{\Gamma+j}=X_{\Gamma+1-j}\) and \(i_{\Gamma+j}=i_{\Gamma+1-j}'\) for \(j=1,\ldots,\Gamma\). We write \(\widetilde{X}_j=X_j/2\) for \(j=1,\ldots,2\Gamma\). 
Let \(s\in\mathbb{C}\) and define
\begin{equation}\label{eqn:proof_BCH_Os}
    O(s)=e^{s\widetilde{X}_1}e^{s\widetilde{X}_2}\cdots e^{s\widetilde{X}_{2\Gamma}}.
\end{equation}
The function \(O(s)\) is entire because it is a finite product of matrix exponentials. Consider the matrix-valued formal power series
\begin{equation}\label{eqn:proof_BCH_formal_Z}
    \widehat{Z}(s)=s\widetilde{X}_1+\cdots+s\widetilde{X}_{2\Gamma}+\sum_{k\geq2}\Phi_k(s),
\end{equation}
where
\begin{equation}\label{eqn:proof_BCH_eq_1}
\Phi_k(s)=\sum_{\substack{i_1+\cdots+i_{2\Gamma}=k\\i_j\geq0}}\frac{1}{i_1!\cdots i_{2\Gamma}!}\phi_k(\underbrace{s\widetilde{X}_1,\ldots,s\widetilde{X}_1}_{i_1},\ldots,\underbrace{s\widetilde{X}_{2\Gamma},\ldots,s\widetilde{X}_{2\Gamma}}_{i_{2\Gamma}})=s^k\Phi_k, \qquad \Phi_k:=\Phi_k(1).
\end{equation}
The second equality follows because \(\phi_k\) is homogeneous of degree \(k\). By the formal multi-term BCH formula in~\cref{eqn:BCH_formal}, we have 
\begin{equation}\label{eqn:proof_BCH_formal_identity}
    e^{\widehat{Z}(s)}=e^{s\widetilde{X}_1}\cdots e^{s\widetilde{X}_{2\Gamma}}
\end{equation}
as an identity of formal power series in \(s\). The convention \(i_{\Gamma+j}=i_{\Gamma+1-j}'\) shows that the resulting expression for \(\Phi_k\) agrees with the expression in the statement of~\cref{prop:BCH_symmetric}. 

We first establish the claimed bound on \(\Phi_k\). Fix \(k\) and an ordered symbolic word \((\gamma_1,\ldots,\gamma_k)\in\{1,\ldots,2\Gamma\}^k\). There is a unique multiplicity vector \((i_1,\ldots,i_{2\Gamma})\) appearing in~\cref{eqn:proof_BCH_eq_1} that corresponds to this word. In the associated term \(\phi_k\), exactly \(i_1!\cdots i_{2\Gamma}!\) permutations yield the ordered symbolic word \((\gamma_1,\ldots,\gamma_k)\). Let \(\widetilde{S}\) denote the set of such permutations, so that \(|\widetilde{S}|=i_1!\cdots i_{2\Gamma}!\). Consequently, prior to evaluating the symbolic labels as matrices, the sum of the absolute values of the scalar coefficients of all contributions associated with \((\gamma_1,\ldots,\gamma_k)\) is bounded by
\begin{equation}\label{eqn:proof_BCH_coefficient_bound}
    \frac{1}{i_1!\cdots i_{2\Gamma}!}\frac{1}{k^2}\sum_{\sigma\in\widetilde{S}}\frac{1}{\binom{k-1}{d_\sigma}}\leq\frac{1}{k^2}.
\end{equation}
Here, the indices \(1,\ldots,2\Gamma\) are regarded as distinct symbolic labels for the purpose of counting, even when the corresponding matrices coincide. Applying the triangle inequality therefore yields
\begin{equation}\label{eqn:proof_BCH_bound_2Gamma}
    \|\Phi_k\|\leq\frac{1}{k^2}\sum_{\gamma_1,\ldots,\gamma_k=1}^{2\Gamma}\|[\widetilde{X}_{\gamma_1}\widetilde{X}_{\gamma_2}\cdots\widetilde{X}_{\gamma_k}]\|=\frac{1}{k^22^k}\sum_{\gamma_1,\ldots,\gamma_k=1}^{2\Gamma}\|[X_{\gamma_1}X_{\gamma_2}\cdots X_{\gamma_k}]\|.
\end{equation}
Since each index in \(\{1,\ldots,\Gamma\}\) has exactly two preimages in \(\{1,\ldots,2\Gamma\}\), every indexed word in \(X_1,\ldots,X_\Gamma\) appears exactly \(2^k\) times in the resulting sum. Therefore,
\begin{equation}\label{eqn:proof_BCH_bound_Gamma}
    \|\Phi_k\|\leq\frac{1}{k^2}\sum_{\gamma_1,\ldots,\gamma_k=1}^{\Gamma}\|[X_{\gamma_1}X_{\gamma_2}\cdots X_{\gamma_k}]\|.
\end{equation}
The remaining proof proceeds in four steps:
\begin{enumerate}
    \item We first prove that the formal series \(\widehat{Z}(s)\) converges in a neighborhood of \(s=0\).
    \item We then identify the resulting analytic function \(Z(s)\) as a local analytic logarithm of \(O(s)\) and show that \(Z(s)\) is odd.
    \item We next show that the resulting odd-power series for \(Z(s)\) converges absolutely and uniformly on \(|s|\leq1\).
    \item Finally, we use the identity theorem to extend \(e^{Z(s)}=O(s)\) from a neighborhood of \(0\) to the open unit disk, and use continuity to permit evaluation at \(s=1\).
\end{enumerate}

We first show that \(\widehat{Z}(s)\) converges in a neighborhood of \(s=0\). Set \(\Lambda=\sum_{\gamma=1}^{\Gamma}\|X_\gamma\|\). If \(\Lambda=0\), then \(X_\gamma=0\) for every \(\gamma\), and the result is immediate. We therefore assume that \(\Lambda>0\). The elementary estimate \(\|[A,B]\|\leq2\|A\|\|B\|\), applied recursively through the \(k-1\) commutator brackets, gives \(\|[X_{\gamma_1}X_{\gamma_2}\cdots X_{\gamma_k}]\|\leq2^{k-1}\prod_{j=1}^{k}\|X_{\gamma_j}\|\). Using~\cref{eqn:proof_BCH_bound_Gamma}, we obtain
\begin{equation}\label{eqn:proof_BCH_crude_Phi_bound}
    \|\Phi_k\|\leq\frac{2^{k-1}}{k^2}\sum_{\gamma_1,\ldots,\gamma_k=1}^{\Gamma}\prod_{j=1}^{k}\|X_{\gamma_j}\|=\frac{2^{k-1}}{k^2}\left(\sum_{\gamma=1}^{\Gamma}\|X_\gamma\|\right)^k=\frac{2^{k-1}}{k^2}\Lambda^k.
\end{equation}
Consequently, for every \(r<1/(2\Lambda)\) we have
\begin{equation}\label{eqn:proof_BCH_local_absolute_convergence}
    \sum_{k\geq2}\sup_{|s|\leq r}\|s^k\Phi_k\|\leq\frac{1}{2}\sum_{k\geq2}\frac{(2\Lambda r)^k}{k^2}<\infty.
\end{equation}
Thus, the formal series in~\cref{eqn:proof_BCH_formal_Z} converges absolutely and locally uniformly for \(|s|<1/(2\Lambda)\). On this disk, let \(Z(s)\) denote the analytic matrix-valued function defined by the convergent series
\begin{equation}\label{eqn:proof_BCH_local_Z}
    Z(s)=s\widetilde{X}_1+\cdots+s\widetilde{X}_{2\Gamma}+\sum_{k\geq2}s^k\Phi_k.
\end{equation}
Since this series converges locally uniformly and satisfies \(Z(0)=0\), its substitution into the matrix exponential is legitimate, and the Taylor series of \(e^{Z(s)}\) at \(s=0\) is the formal exponential of \(\widehat{Z}(s)\). By~\cref{eqn:proof_BCH_formal_identity}, this Taylor series agrees coefficientwise with the Taylor series of \(O(s)\). Since both \(e^{Z(s)}\) and \(O(s)\) are analytic for \(|s|<1/(2\Lambda)\), uniqueness of analytic power-series expansions gives 
\begin{equation}\label{eqn:proof_BCH_local_exponential_identity}
    e^{Z(s)}=O(s), \qquad |s|<\frac{1}{2\Lambda}.
\end{equation}

We now identify \(Z(s)\) with a local analytic logarithm of \(O(s)\). Since \(O(0)=I\) and \(O(s)\) is continuous, we may choose \(\rho\in\left(0,\min\left\{1,\frac{1}{2\Lambda}\right\}\right)\) such that
$\sup_{|s|\leq\rho}\|O(s)-I\|\leq\frac{1}{2}.$
For \(|s|\leq\rho\), define
\begin{equation}\label{eqn:proof_BCH_Mercator}
    L(s)=\sum_{n=1}^{\infty}\frac{(-1)^{n+1}}{n}\bigl(O(s)-I\bigr)^n.
\end{equation}
This series converges absolutely and uniformly in operator norm on \(|s|\leq\rho\), since
\begin{equation}\label{eqn:proof_BCH_Mercator_convergence}
    \sum_{n=1}^{\infty}\sup_{|s|\leq\rho}\frac{\|(O(s)-I)^n\|}{n}\leq\sum_{n=1}^{\infty}\frac{2^{-n}}{n}<\infty.
\end{equation}
Consequently, \(L(s)\) is analytic for \(|s|<\rho\), satisfies \(L(0)=0\), and obeys \(e^{L(s)}=O(s)\) for \(|s|<\rho\). The derivative of the matrix exponential at the zero matrix is the identity map. The analytic inverse function theorem therefore gives a neighborhood of the zero matrix on which the matrix exponential is injective. We may take this neighborhood to be symmetric about the origin. Since \(Z(0)=L(0)=0\), after decreasing \(\rho\), if necessary, both \(Z(s)\) and \(L(s)\) belong to this neighborhood for \(|s|<\rho\). We have
$e^{Z(s)}=O(s)=e^{L(s)}$ for $|s|<\rho$. 
Local injectivity of the exponential map therefore gives \(Z(s)=L(s)\) for \(|s|<\rho\). 

The palindromic relation \(\widetilde{X}_{2\Gamma+1-j}=\widetilde{X}_j\) gives
\begin{equation}\label{eqn:proof_BCH_palindromic_inverse}
    O(s)^{-1}=e^{-s\widetilde{X}_{2\Gamma}}\cdots e^{-s\widetilde{X}_1}=e^{-s\widetilde{X}_1}\cdots e^{-s\widetilde{X}_{2\Gamma}}=O(-s).
\end{equation}
After decreasing \(\rho\) once more, if necessary, both \(L(-s)\) and \(-L(s)\) belong to the neighborhood on which the matrix exponential is injective. Hence,
\begin{equation}\label{eqn:proof_BCH_logarithm_odd}
    e^{L(-s)}=O(-s)=O(s)^{-1}=e^{-L(s)}, \qquad |s|<\rho,
\end{equation}
and local injectivity gives \(L(-s)=-L(s)\) for \(|s|<\rho\). Since \(Z(s)=L(s)\) locally, we have \(Z(-s)=-Z(s)\) for \(|s|<\rho\). Comparing coefficients in the convergent power series~\cref{eqn:proof_BCH_local_Z} yields $\Phi_k=0$ for $k\in2\mathbb{Z}^+$.

It remains to extend the convergence of \(Z(s)\) to \(s=1\). By~\cref{eqn:proof_BCH_bound_Gamma} and the assumption in~\cref{prop:BCH_symmetric}, for every sufficiently large odd \(k\), \(\|\Phi_k\|\leq1/k^2\). Hence,
\begin{equation}\label{eqn:proof_BCH_uniform_sum}
    \sum_{\substack{k\geq3\\k\ \mathrm{odd}}}\sup_{|s|\leq1}\|s^k\Phi_k\|<\infty
\end{equation}
because the sufficiently large odd terms are bounded by the convergent series \(\sum_k k^{-2}\), while the remaining terms form a finite sum. The Weierstrass \(M\)-test therefore shows that
\begin{equation}\label{eqn:proof_BCH_global_Z}
    Z_{\mathrm{ext}}(s)=s(X_1+\cdots+X_\Gamma)+\sum_{\substack{k\geq3\\k\ \mathrm{odd}}}s^k\Phi_k
\end{equation}
converges absolutely and uniformly for \(|s|\leq1\). It is therefore analytic for \(|s|<1\) and continuous for \(|s|\leq1\). For sufficiently small \(s\), this series agrees with the locally defined function \(Z(s)\) in~\cref{eqn:proof_BCH_local_Z}, since all even coefficients vanish. Thus, \(Z_{\mathrm{ext}}(s)\) extends the locally defined function \(Z(s)\). We henceforth denote this extension again by \(Z(s)\). The functions \(e^{Z(s)}\) and \(O(s)\) are analytic on the connected open unit disk and agree near \(s=0\).  The identity theorem, applied entrywise, therefore gives
\begin{equation}\label{eqn:proof_BCH_global_exponential_identity}
    e^{Z(s)}=O(s), \qquad |s|<1.
\end{equation}
Finally, letting \(s\uparrow1\) along the real interval and using continuity yields \(O(1)=e^{Z(1)}\). Setting \(Z=Z(1)\), we conclude that
\begin{equation}
    e^{X_1/2}e^{X_2/2}\cdots e^{X_\Gamma/2}e^{X_\Gamma/2}\cdots e^{X_2/2}e^{X_1/2}=e^Z.
\end{equation}
This completes the proof.
\end{proof}

    \subsection{Representation of Powers of the Second-Order Product Formula}
    We now derive a polynomial expansion of \(\bigl(U_2(\Delta/k)\bigr)^k\) in powers of \(1/k\). We first apply the Baker--Campbell--Hausdorff (BCH) formula to represent the product formula as the exponential of a single matrix, thereby facilitating the analysis of its powers. We then expand these powers using a high-order variation-of-parameters formula, which may be regarded as an analogue of the truncated Dyson series in the interaction picture. A variant of this formula has also been employed in recent work on classical and quantum algorithms for simulating open quantum dynamics~\cite{CaoLu2021,LiWang2023}. For completeness, we state the formula in \cref{lem:VoP_high_order} and provide a proof.

    \begin{lem}\label{lem:VoP_high_order}
        Let $A$ and $B$ be two anti-Hermitian matrices. 
        Then for any positive integer $p$, we have 
        \begin{align}\label{eqn:VoP_high_order}
        e^{A+B} &= e^{A} + \sum_{l=1}^{p-1} \int_0^1 \int_0^{s_{1}}\cdots \int_0^{s_{l-1}} e^{A(1-s_1)} B  e^{A (s_1-s_2)}B \cdots  e^{A (s_{l-1}-s_l)}B e^{As_l} ds_l \cdots ds_2 ds_1 + R_p(A,B), 
    \end{align}
    where $R_p(A,B)$ is an operator depending on $A,B$ and $p$ such that 
    \begin{equation}
        \|R_p(A,B)\| \leq \frac{\|B\|^p}{p!}. 
    \end{equation}
    \end{lem}
    
    \begin{proof}
    We begin with the variation-of-parameters formula for two matrices \(A\) and \(B\), which states that
    \begin{equation}\label{eqn:VoP_op}
        e^{(A+B)t} = e^{At} + \int_0^t e^{A(t-s)} B  e^{(A+B)s} ds. 
    \end{equation}
    This formula can be derived as follows. Let \(V_1(t)=e^{At}\) and \(V_2(t)=e^{(A+B)t}\). Then \(V_1\) and \(V_2\) solve the ODEs
    \begin{align}
        \frac{d V_1(t)}{dt} & = A V_1(t), \quad V_1(0) = I, \\
        \frac{d V_2(t)}{dt} & = A V_2(t) + BV_2(t), \quad V_2(0) = I. 
    \end{align}
    We regard \(V_2\) as a perturbation of \(V_1\), with perturbation term \(BV_2(t)\). The variation-of-parameters formula then gives 
    \begin{equation}
        V_2(t) = V_1(t) + \int_0^t e^{A(t-s)} BV_2(s) ds, 
    \end{equation}
    which is precisely~\cref{eqn:VoP_op}. To derive a high-order formula, we apply~\cref{eqn:VoP_op} with \(t=1\), obtaining
    \begin{equation}
        e^{A+B} = e^{A} + \int_0^1 e^{A(1-s)} B  e^{(A+B)s} ds. 
    \end{equation}
    Applying~\cref{eqn:VoP_op} again to replace \(e^{(A+B)s}\) in the preceding equation yields
    \begin{equation}
        e^{A+B} = e^{A} + \int_0^1 e^{A(1-s_1)} B  e^{A s_1} ds_1 + \int_0^1 \int_0^{s_1} e^{A(1-s_1)} B  e^{A (s_1-s_2)}B e^{(A+B)s_2} ds_2 ds_1.
    \end{equation}
    Repeating this substitution once more gives 
    \begin{align}
        e^{(A+B)} &= e^{A} + \int_0^1 e^{A(1-s_1)} B  e^{A s_1} ds_1 + \int_0^1 \int_0^{s_1} e^{A(1-s_1)} B  e^{A (s_1-s_2)}B e^{A s_2} ds_2 ds_1 \\
        & + \int_0^1 \int_0^{s_1} \int_0^{s_2} e^{A(1-s_1)} B  e^{A (s_1-s_2)}B e^{A (s_2-s_3)} B e^{(A+B)s_3} ds_3 ds_2 ds_1. 
    \end{align}
    Iterating this formula \(p\) times yields
     \begin{align}
        e^{(A+B)} &= e^{A} + \sum_{l=1}^{p-1} \int_0^1 \int_0^{s_{1}}\cdots \int_0^{s_{l-1}} e^{A(1-s_1)} B  e^{A (s_1-s_2)}B \cdots  e^{A (s_{l-1}-s_l)}B e^{As_l} ds_l \cdots ds_2 ds_1 \\
        & \quad\quad + \int_0^1 \int_0^{s_1} \cdots \int_0^{s_{p-1}} e^{A(1-s_1)} B  e^{A (s_1-s_2)}B \cdots  e^{A (s_{p-1}-s_{p})}B e^{(A+B)s_{p}} ds_{p} \cdots ds_2 ds_1, 
    \end{align}
   which is precisely the form of~\cref{eqn:VoP_high_order}, with 
    \begin{equation}
        R_p(A,B) = \int_0^1 \int_0^{s_1} \cdots \int_0^{s_{p-1}} e^{A(1-s_1)} B  e^{A (s_1-s_2)}B \cdots  e^{A (s_{p-1}-s_{p})}B e^{(A+B)s_{p}} ds_{p} \cdots ds_2 ds_1. 
    \end{equation}
        Since \(A\) and \(B\) are anti-Hermitian, we can bound \(R_p\) as
        \begin{equation}
            \|R_p(A,B)\| \leq \int_0^1 \int_0^{s_1} \cdots \int_0^{s_{p-1}} \|B\|^p ds_{p} \cdots ds_2 ds_1 = \frac{\|B\|^p}{p!}. 
        \end{equation}
    This completes the proof.
    \end{proof}
    
    We now establish an expansion for powers of the second-order Trotter formula.   
    \begin{lem}\label{lem:Trotter_2nd_representation}
        Let $U_{2}(s)$ be the second-order Trotter-Suzuki formula, and $U(s) = e^{-iHs}$ be the exact evolution operator. 
        Define $\alpha_{\operatorname{comm},j}$ to be the summation of the norms of all the $j$-th nested commutators, i.e.,  $\alpha_{\operatorname{comm},j} = \sum_{\gamma_1,\cdots,\gamma_j=1}^{\Gamma} \|[H_{\gamma_1}H_{\gamma_2}\cdots H_{\gamma_j}]\|. $
        Suppose that there exists an integer $J$ such that 
        \begin{equation}\label{eqn:lem_assumption_delta}
            \Delta \leq \inf_{j \geq J,\text{ odd} } \alpha_{\operatorname{comm},j}^{-1/j}. 
        \end{equation}
        Then, for any positive integer $k$ and $p$, we have 
        \begin{equation}
            \left(U_{2}(\Delta/k)\right)^k = U(\Delta) + \sum_{j \in 2\mathbb{Z}^+}^{\infty} \widetilde{E}_{j+1,p}(\Delta) \frac{1}{k^j} + \widetilde{F}_p(\Delta,k).
        \end{equation}
        Here $\widetilde{E}_{j+1,p}$'s are operators independent of $k$ and can be bounded as 
        \begin{equation}
            \|\widetilde{E}_{j+1,p}(\Delta) \| \leq \Delta^{j+1} \left(\sum_{l=1}^{\min\left\{j/2,p-1\right\}}  \frac{\Delta^{l-1}}{l!} \left(\sum_{\substack{j_1,\cdots,j_l \in 2\mathbb{Z}^+,\\ j_1+\cdots+j_l=j}} \left( \prod_{\kappa=1}^{l}  \alpha_{\operatorname{comm},j_{\kappa}+1}\right) \right) \right), 
        \end{equation}
        and $\widetilde{F}_p(\Delta,k)$ is an operator such that
        \begin{equation}
            \|\widetilde{F}_p(\Delta,k)\| \leq \frac{\Delta^{3p}}{p!} \sum_{j \in 2\mathbb{Z}^+, j\geq 2p} \Delta^{j-2p} \left(\sum_{\substack{j_1,\cdots,j_p \in 2\mathbb{Z}^+,\\ j_1+\cdots+j_p=j}} \left(\prod_{\kappa=1}^p \alpha_{\operatorname{comm},j_{\kappa}+1} \right) \right). 
        \end{equation}
    \end{lem}
    \begin{proof}
    Using~\cref{prop:BCH_symmetric}, we have
    \begin{align}   
    U_{2}(s) &= e^{-i H_1 s/2} \cdots e^{-i H_{\Gamma} s/2} e^{-i H_{\Gamma} s/2} \cdots e^{-i H_1 s/2} = \exp\left( -iHs + \sum_{j \in 1+2\mathbb{Z}^+} E_j s^j \right),
    \end{align}
    where
    \begin{equation}
        E_j = \sum_{ \substack{\sum (i_{\gamma}+i_{\gamma}') = j,\\ i_{\gamma},i_{\gamma}'\geq 0}}  \frac{(-i)^j }{i_1!\cdots i_{\Gamma}!i_1'!\cdots i_{\Gamma}'!} \phi_j\Big(\underbrace{\frac{H_1}{2},\cdots, \frac{H_1}{2}}_{i_1},\cdots, \underbrace{\frac{H_{\Gamma}}{2},\cdots,\frac{H_{\Gamma}}{2}}_{i_{\Gamma}},\underbrace{\frac{H_{\Gamma}}{2},\cdots,\frac{H_{\Gamma}}{2}}_{i_{\Gamma}'},\cdots, \underbrace{\frac{H_1}{2},\cdots,\frac{H_1}{2}}_{i_1'}\Big). 
    \end{equation}
    By~\cref{prop:BCH_symmetric}, this expansion converges if \(s^j\alpha_{\operatorname{comm},j}\leq 1\) for sufficiently large odd \(j\). This condition is guaranteed for all \(s\leq \Delta\) by \cref{eqn:lem_assumption_delta}. Moreover, \cref{prop:BCH_symmetric} gives the bound
    $
        \|E_j\| \leq \frac{1}{j^2} \alpha_{\operatorname{comm},j} \leq \alpha_{\operatorname{comm},j}. 
    $
    Then 
    \begin{equation}\label{eqn:proof_error_representation_eq1}
        \left( U_{2}(\Delta/k) \right)^k = \exp\left( -iH \Delta + \sum_{j \in 1+2\mathbb{Z}^+} E_j \frac{\Delta^j}{k^{j-1}} \right). 
    \end{equation}
    We expand \(\bigl(U_2(\Delta/k)\bigr)^k\) as a polynomial in \(1/k\). 
    Let \(A=-iH\Delta\) and \(B=\sum_{j\in 2\mathbb{Z}^{+}} E_{j+1}\frac{\Delta^{j+1}}{k^{j}}\). The matrix \(A\) is anti-Hermitian. Moreover, \(B\) is also anti-Hermitian, as follows from~\cref{eqn:proof_error_representation_eq1}. Indeed, the second-order product formula \(U_2\) is unitary, and hence both sides of~\cref{eqn:proof_error_representation_eq1} are unitary. Taking the logarithm therefore yields an anti-Hermitian matrix. Since \(-iH\Delta\) is anti-Hermitian, the remaining term \(B\) must also be anti-Hermitian. We may therefore apply~\cref{lem:VoP_high_order} with the above choices of \(A\) and \(B\) to obtain
    \begin{align}
         \left( U_{2}(\Delta/k) \right)^k &= U(\Delta) + \sum_{l=1}^{p-1} \int_0^1 \int_0^{s_{1}}\cdots \int_0^{s_{l-1}} e^{A(1-s_1)} B  e^{A (s_1-s_2)}B \cdots  e^{A (s_{l-1}-s_l)}B e^{As_l} ds_l \cdots ds_2 ds_1  \nonumber \\
         & + R_p(A,B). 
    \end{align}
    First, consider the summation term. Since \(B\) is a polynomial in \(1/k\), this term can likewise be expressed as a polynomial in \(1/k\). For notational convenience, let \(s_0=1\), and write \(ds=ds_l\cdots ds_1\) whenever the meaning is clear from the context. 
Substituting the expression for \(B\) and expanding the product, we have
    \begin{align}
        & \quad \sum_{l=1}^{p-1} \int_0^1 \int_0^{s_{1}}\cdots \int_0^{s_{l-1}} e^{A(1-s_1)} B  e^{A (s_1-s_2)}B \cdots  e^{A (s_{l-1}-s_l)}B e^{As_l} ds_l \cdots ds_2 ds_1   \\
        & = \sum_{l=1}^{p-1} \int_0^1 \int_0^{s_{1}}\cdots \int_0^{s_{l-1}} ds \left( \prod_{\kappa=l}^{1} \left(\sum_{j \in 2\mathbb{Z}^+ } e^{A(s_{\kappa-1 }-s_{\kappa})}E_{j+1} \frac{\Delta^{j+1}}{k^{j}}\right)\right)e^{As_l} \\
        & = \sum_{l=1}^{p-1} \int_0^1 \int_0^{s_{1}}\cdots \int_0^{s_{l-1}} ds \left( \sum_{j_1,\cdots,j_l \in 2\mathbb{Z}^+ } \left( \prod_{\kappa=l}^{1} e^{A(s_{\kappa-1}-s_{\kappa})}E_{j_{\kappa}+1}\right) \frac{\Delta^{l+\sum_{\kappa}j_{\kappa} }}{k^{\sum_{\kappa}j_{\kappa}}}\right)e^{As_l}. 
    \end{align}
We then collect terms with the same powers of \(1/k\) to obtain
\begin{align}
        & \quad \sum_{l=1}^{p-1} \int_0^1 \int_0^{s_{1}}\cdots \int_0^{s_{l-1}} e^{A(1-s_1)} B  e^{A (s_1-s_2)}B \cdots  e^{A (s_{l-1}-s_l)}B e^{As_l} ds_l \cdots ds_2 ds_1   \\
         & = \sum_{l=1}^{p-1} \int_0^1 \int_0^{s_{1}}\cdots \int_0^{s_{l-1}} ds \left( \sum_{j \in 2\mathbb{Z}^+, j \geq 2l} \sum_{\substack{j_1,\cdots,j_l \in 2\mathbb{Z}^+, \\ j_1+\cdots+j_l=j}}  \left( \prod_{\kappa=l}^{1} e^{A(s_{\kappa-1}-s_{\kappa})}E_{j_{\kappa}+1}\right) \frac{\Delta^{l+j }}{k^{j}}\right)e^{As_l} \\
        & = \sum_{j \in 2\mathbb{Z}^+} \widetilde{E}_{j+1,p}(\Delta) \frac{1}{k^j} \label{label-2}. 
    \end{align}
Here, the operator \(\widetilde{E}_{j+1,p}(\Delta)\) is independent of \(k\) and admits the representation
    \begin{equation}
        \widetilde{E}_{j+1,p}(\Delta) = \sum_{l=1}^{\min\left\{j/2,p-1\right\}} \sum_{\substack{j_1,\cdots,j_l \in 2\mathbb{Z}^+, \\ j_1+\cdots+j_l=j}} \Delta^{l+j} \int_0^1 \int_0^{s_{1}}\cdots \int_0^{s_{l-1}} ds  \left( \prod_{\kappa=l}^{1} e^{A(s_{\kappa-1}-s_{\kappa})}E_{j_{\kappa}+1}\right) e^{As_l}. 
    \end{equation}
    Moreover, by the matrix norm inequality and the fact that \(A=-iH\Delta\) is anti-Hermitian, the norm of \(\widetilde{E}_{j+1,p}(\Delta)\) can be bounded above by
    \begin{align}
        \|\widetilde{E}_{j+1,p}(\Delta) \| & \leq \sum_{l=1}^{\min\left\{j/2,p-1\right\}} \sum_{\substack{j_1,\cdots,j_l \in 2\mathbb{Z}^+, \\ j_1+\cdots+j_l=j}} \Delta^{l+j} \int_0^1 \int_0^{s_{1}}\cdots \int_0^{s_{l-1}} ds  \left( \prod_{\kappa=l}^{1} \|E_{j_{\kappa}+1}\|\right) \label{label-3}.     \end{align}
    Since the integrand is independent of the integration variables \(s_1,\ldots,s_l\), the nested integral can be evaluated explicitly as \(\int_0^1 \int_0^{s_1}\cdots \int_0^{s_{l-1}} ds=1/l!\). Furthermore, using~\cref{prop:BCH_symmetric}, we bound \(\|E_{j_{\kappa}+1}\|\leq \alpha_{\operatorname{comm},j_{\kappa}+1}\), which yields
    \begin{align}
        \|\widetilde{E}_{j+1,p}(\Delta) \| 
        & \leq \Delta^{j+1} \left(\sum_{l=1}^{\min\left\{j/2,p-1\right\}}  \frac{\Delta^{l-1}}{l!} \left(\sum_{\substack{j_1,\cdots,j_l \in 2\mathbb{Z}^+, \\ j_1+\cdots+j_l=j}} \prod_{\kappa=1}^{l} \|E_{j_{\kappa}+1}\| \right) \right) \\
        & \leq \Delta^{j+1} \left(\sum_{l=1}^{\min\left\{j/2,p-1\right\}}  \frac{\Delta^{l-1}}{l!} \left(\sum_{\substack{j_1,\cdots,j_l \in 2\mathbb{Z}^+, \\ j_1+\cdots+j_l=j}} \prod_{\kappa=1}^{l}  \alpha_{\operatorname{comm},j_{\kappa}+1} \right) \right) \label{label-4}. 
    \end{align} 
    Next, we analyze the remainder term \(R_p(A,B)\). Since $\|R_p(A,B)\| \leq \frac{\|B\|^p}{p!}$ from \cref{lem:VoP_high_order}, we have
    \begin{align}
        \|R_p(A,B)\| 
        & \leq \frac{1}{p!} \left( \sum_{j\in 2\mathbb{Z}^+} \|E_{j+1}\| \frac{\Delta^{j+1}}{k^{j}} \right)^p \label{label-5} \\
        & = \frac{\Delta^p}{p!} \sum_{j_1,\cdots,j_p \in 2\mathbb{Z}^+ } \left(\prod_{\kappa=1}^p \|E_{j_\kappa+1}\| \right) \frac{\Delta^{j_1+\cdots+j_p}}{k^{j_1+\cdots+j_p}} \\
        & \leq \frac{\Delta^{3p}}{p!} \sum_{j \in 2\mathbb{Z}^+, j\geq 2p} \Delta^{j-2p} \left(\sum_{\substack{j_1,\cdots,j_p \in 2\mathbb{Z}^+, \\ j_1+\cdots+j_p = j}} \left(\prod_{\kappa=1}^p \alpha_{\operatorname{comm},j_{\kappa}+1} \right) \right) \label{label-6}, 
    \end{align}
    where, in the second line, we again expand the power and collect terms according to their powers of \(\Delta\). In the third line, we drop the factor involving the power of \(1/k\) and apply the bound \(\|E_{j_{\kappa}+1}\|\leq \alpha_{\operatorname{comm},j_{\kappa}+1}\).
    Note that we use the notation \(\widetilde{F}_p(\Delta,k)\) in the statement to emphasize its dependence on \(\Delta\) and \(k\). This completes the proof.
    \end{proof}

    \subsection{Error Bound for Short-Time MPF}
    Combining~\cref{lem:Trotter_2nd_representation} with the order conditions in~\cref{eqn:multiproduct_linear_system} yields the single-step error bound for the MPF in \cref{thm:MP_error_bound}.
    
    \begin{thm}\label{thm:MP_error_bound}
        Let $U_{\operatorname{MP}}(\Delta)$ denote the $2$nd-based MPF as 
        $U_{\operatorname{MP}} (\Delta) = \sum_{l=1}^M a_l U_{2}(\Delta/k_l)^{k_l}. $
        Suppose that there exists an integer $J$ such that 
            $\Delta \leq  \inf_{j \geq J,\text{ odd}} \alpha_{\operatorname{comm},j}^{-1/j}$.  
        Then we have 
        \begin{equation}
            \|U_{\operatorname{MP}} (\Delta) - U(\Delta)\| \leq \|\vec{a}\|_1 \sum_{j \in 2\mathbb{Z}^+, j\geq 2m} \sum_{l=1}^{m}  \frac{\Delta^{j+l}}{l!} \left(\sum_{\substack{j_1,\cdots,j_l \in 2\mathbb{Z}^+, \\ j_1+\cdots+j_l=j}} \left(\prod_{\kappa=1}^{l}  \alpha_{\operatorname{comm},j_{\kappa}+1}\right) \right). 
        \end{equation}
    \end{thm}
    \begin{proof}
        For any \(p\geq 1\), combining \cref{lem:Trotter_2nd_representation} with the MPF order conditions in~\cref{eqn:multiproduct_linear_system} yields
        \begin{equation}
            U_{\operatorname{MP}}(\Delta) = U(\Delta) + \sum_{j \in 2\mathbb{Z}^+, j\geq 2m} \sum_{l=1}^M a_l \widetilde{E}_{j+1,p}(\Delta) \frac{1}{k_l^j} + \sum_{l=1}^M a_l \widetilde{F}_p(\Delta,k_l). 
        \end{equation}
        Using the bounds on \(\widetilde{E}\) and \(\widetilde{F}\), together with the crude estimate \(k_l\geq 1\), we obtain
        \begin{align}
            \|U_{\operatorname{MP}}(\Delta) - U(\Delta)\| & \leq \|\vec{a}\|_1 \sum_{j \in 2\mathbb{Z}^+, j\geq 2m} \Delta^{j+1} \left(\sum_{l=1}^{\min \{j/2,p-1\}}  \frac{\Delta^{l-1}}{l!} \left(\sum_{\substack{j_1,\cdots,j_l \in 2\mathbb{Z}^+, \\ j_1+\cdots+j_l=j}} \left(\prod_{\kappa=1}^{l}  \alpha_{\operatorname{comm},j_{\kappa}+1} \right) \right) \right) \label{eqn:proof_error_eq1}\\
            & + \|\vec{a}\|_1 \frac{\Delta^{3p}}{p!} \sum_{j \in 2\mathbb{Z}^+, j\geq 2p} \Delta^{j-2p} \left(\sum_{\substack{j_1,\cdots,j_p \in 2\mathbb{Z}^+, \\ j_1+\cdots+j_p=j}} \left(\prod_{\kappa=1}^p \alpha_{\operatorname{comm},j_{\kappa}+1} \right) \right).\label{eqn:proof_error_eq2} 
        \end{align}
        Choosing \(p=m\), the summation over \(l\) in~\cref{eqn:proof_error_eq1} ranges from \(1\) to \(m-1\). Moreover, \cref{eqn:proof_error_eq2} has the same form as the summand in~\cref{eqn:proof_error_eq1} corresponding to \(l=m\). Therefore,
        \begin{equation}
            \|U_{\operatorname{MP}}(\Delta) - U(\Delta)\| \leq \|\vec{a}\|_1 \sum_{j \in 2\mathbb{Z}^+, j\geq 2m} \sum_{l=1}^{m}  \frac{\Delta^{j+l}}{l!} \left(\sum_{\substack{j_1,\cdots,j_l \in 2\mathbb{Z}^+, \\ j_1+\cdots+j_l=j}} \left(\prod_{\kappa=1}^{l}  \alpha_{\operatorname{comm},j_{\kappa}+1} \right)  \right). 
        \end{equation}
    This completes the proof.
    \end{proof}

    \subsection{Complexity Analysis}
    We are now prepared to estimate the overall complexity of the second-order based MPF for long-time simulation, following the implementation described in~\cref{sec:prelim_algorithm}. The resulting complexity bound is stated in~\cref{thm:MPF_complexity}. Its proof follows the general strategy of~\cite[Theorem~2]{LowKliuchnikovWiebe2019}, adapted to incorporate the commutator-scaling representation for powers of the second-order Trotter formula.

    \begin{thm}\label{thm:MPF_complexity}
        Let $\epsilon \in (0,1)$. 
        Consider the quantum simulation problem up to time $T$ using $2$nd-based MPF of convergence order $2m$. 
        Then, in order to obtain an $\epsilon$-approximation with probability at least $1-\epsilon$, it suffices to use 
        \begin{equation}
            \mathcal{O}\left( m^2 (\log m)^2 \mu_m T \left( \frac{\mu_m T }{\epsilon}\right)^{1/(2m)} \right) 
        \end{equation}
        queries to the controlled version of $U_{2}$. 
        Here 
        \begin{align}
            \mu_m := \sup_{\substack{j \in 2\mathbb{Z}^+, j\geq 2m, \\ 1\leq l \leq m}} \lambda_{j,l}, \quad \quad 
            \lambda_{j,l} = \left(\sum_{\substack{j_1,\cdots,j_l \in 2\mathbb{Z}^+, \\ j_1+\cdots+j_l=j}} \left(\prod_{\kappa=1}^{l}  \alpha_{\operatorname{comm},j_{\kappa}+1} \right)\right)^{\frac{1}{j+l}}, \label{eqn:def_mu_m}
        \end{align}
    where $\alpha_{\operatorname{comm},j} = \sum_{\gamma_1,\cdots,\gamma_j=1}^{\Gamma} \|[H_{\gamma_1}H_{\gamma_2}\cdots H_{\gamma_j}]\|. $
    \end{thm}
    \begin{proof}
        Suppose that the time interval \([0,T]\) is partitioned into \(r\) equal-length segments, and let \(\Delta=T/r\) denote the length of each segment. Then~\cref{thm:MP_error_bound} yields
        \begin{equation}\label{eqn:proof_complexity_eq1}
            \|U_{\operatorname{MP}} (\Delta) - U(\Delta)\| \leq \|\vec{a}\|_1 \sum_{j \in 2\mathbb{Z}^+, j\geq 2m} \sum_{l=1}^{m}  \frac{( \mu_m \Delta)^{j+l}}{l!} := \epsilon_{\Delta}. 
        \end{equation}
        Choose $r$ such that
        \begin{equation}\label{r-choice}
    r=
    \left\lceil
    2\mu_m T
    \left(
    1+
    \frac{8\pi\mu_m T\|\vec a\|_1}{\epsilon}
    \right)^{1/(2m)}
    \right\rceil .
    \end{equation}
    We first verify that the choice of \(r\) in~\cref{r-choice} satisfies the step-size condition in~\cref{lem:Trotter_2nd_representation}. It suffices to consider commutators of depth \(q=j+1\), where \(j\in 2\mathbb{Z}^{+}\) and \(j\geq 2m\). For such \(q\), choosing \(l=1\) in the definition of \(\mu_m\) yields
        \begin{equation}
        \alpha_{\operatorname{comm},q}^{1/q}
        =
        \lambda_{q-1,1}
        \leq \mu_m .
        \end{equation}
        Since $r\geq 2\mu_m T$, then $\Delta=T/r$ satisfies
        \begin{equation}
        \Delta \leq \frac{1}{2\mu_m}
        \leq \frac{1}{2}\alpha_{\operatorname{comm},q}^{-1/q}
        \end{equation}
        for every $q\geq 2m+1$. Hence, the step-size condition in \cref{lem:Trotter_2nd_representation} is satisfied. For any $j \geq 2m$ and $1 \leq l \leq m$, we have $j+l-1\geq 2m$. By the choice of \(r\) in~\cref{r-choice}, we have
    \begin{equation}
    r\geq
    2\mu_m T
    \left(
    1+\frac{8\pi\mu_m T\|\vec a\|_1}{\epsilon}
    \right)^{1/(2m)}.
    \end{equation}
    Raising both sides to the power \(j+l-1\) gives
    \begin{equation}
    r^{j+l-1}\geq
    (2\mu_m T)^{j+l-1}
    \left(
    1+\frac{8\pi\mu_m T\|\vec a\|_1}{\epsilon}
    \right)^{(j+l-1)/(2m)}.
    \end{equation}
    Since \((j+l-1)/(2m)\geq 1\), we obtain
    \begin{equation}
    r^{j+l-1}\geq
    \frac{(2\mu_m T)^{j+l-1}8\pi\mu_m T\|\vec a\|_1}{\epsilon}.
    \end{equation}
        Therefore, we have
        \begin{align}
        r\frac{(\mu_m\Delta)^{j+l}}{l!}
        &=
        \frac{(\mu_m T)^{j+l}}{l!r^{j+l-1}}\leq
        \frac{1}{l!}
        \frac{(\mu_m T)^{j+l}}{(2\mu_m T)^{j+l-1}}
        \frac{\epsilon}{8\pi\mu_m T\|\vec a\|_1} =
        \frac{\epsilon}{4\pi\,2^{j+l}l!\|\vec a\|_1}.
    \end{align}
    Hence, we have
    \begin{align}
        r\epsilon_{\Delta}=
        \|\vec{a}\|_1
        \sum_{\substack{j\in 2\mathbb{Z}^{+}\\ j\geq 2m}}
        \sum_{l=1}^{m}
        \frac{(\mu_m T)^{j+l}}{l!r^{j+l-1}} \leq
        \|\vec{a}\|_1
        \sum_{\substack{j\in 2\mathbb{Z}^{+}\\ j\geq 2m}}
        \sum_{l=1}^{m}
        \frac{\epsilon}{4\pi\,2^{j+l}l!\|\vec a\|_1} =
        \frac{\epsilon}{4\pi}
        \sum_{\substack{j\in 2\mathbb{Z}^{+}\\ j\geq 2m}}
        \sum_{l=1}^{m}
        \frac{1}{2^{j+l}l!}<
        \frac{\epsilon}{4\pi}.
    \end{align}
    The global error can be bounded using the triangle inequality. Specifically,~\cref{eqn:proof_complexity_eq1} implies \(\|U_{\operatorname{MP}}(\Delta)\|\leq \|U(\Delta)\|+\epsilon_{\Delta}=1+\epsilon_{\Delta}\). Hence, we have
    \begin{align}
        \|U_{\operatorname{MP}}(T/r)^r - U(T)\|
        &=
        \left\|
        \sum_{j=0}^{r-1}
        U_{\operatorname{MP}}(\Delta)^{r-j-1}
        \left(U_{\operatorname{MP}}(\Delta)-U(\Delta)\right)
        U(\Delta)^j
        \right\| \\
        &\leq
        \sum_{j=0}^{r-1}
        \|U_{\operatorname{MP}}(\Delta)\|^{r-j-1}
        \|U_{\operatorname{MP}}(\Delta)-U(\Delta)\| \\
        &\leq
        r\epsilon_{\Delta}(1+\epsilon_{\Delta})^{r-1}
        <
        \frac{\epsilon}{4\pi}
        \left(1+\frac{\epsilon}{4\pi r}\right)^r\leq
        \frac{\epsilon}{4\pi}e^{\epsilon/(4\pi)}<
        \frac{\epsilon}{4},
    \end{align}
    where the last inequality holds because $\epsilon \in  (0,1)$. 
    Since \(\|\vec{a}\|_1=\mathcal{O}(\log m)\), we have \(\|\vec{a}\|_1^{1/(2m)}=\mathcal{O}(1)\). Therefore, the choice of \(r\) is given by
        \begin{equation}
            r = \mathcal{O}\left( \mu_m T \left( \frac{\mu_m T }{\epsilon}\right)^{1/(2m)}\right). 
        \end{equation}
        
         We now count the overall query complexity. For each short-time MPF \(U_{\operatorname{MP}}(\Delta)\), the LCU step requires \(\|\vec{k}\|_1\) queries to \(U_2\).
         However, LCU introduces a normalization factor of \(\|\vec{a}\|_1\), so the success probability at each step must be amplified to enable efficient long-time simulation. To this end, we choose the smallest odd integer \(q\) such that \(\alpha_q:=\csc\!\left(\pi/(2q)\right)\geq\|\vec{a}\|_1\). By appending a standard normalization-padding ancilla and performing a single-qubit rotation, we slightly increase the normalization factor to \(\alpha_q\). We may then apply robust oblivious amplitude amplification~\cite{BerryChildsCleveEtAl2015} for \((q-1)/2\) rounds. Since \(\|U_{\operatorname{MP}}(\Delta)-U(\Delta)\|\leq \epsilon_{\Delta}<\epsilon/(4\pi r)\), robust OAA yields an amplified-block error of \(2q\sin\!\left(\pi/(2q)\right)\epsilon_\Delta\leq \pi\epsilon_\Delta<\epsilon/(4r)\). Therefore, the overall failure after \(r\) segments is \(\mathcal{O}(\epsilon)\). 
         Since \(q=\Theta(\|\vec{a}\|_1)\), exact-angle OAA amplifies the ideal block exactly using \(\mathcal{O}(\|\vec{a}\|_1)\) calls to \(U_{\operatorname{MP}}\). Notice that we use the exact-angle version of robust oblivious amplitude amplification rather than fixed-point amplitude amplification, and therefore no polylogarithmic overhead is introduced into the complexity.
         Therefore, the number of queries to \(U_2\) per time step is \(\mathcal{O}(\|\vec{a}\|_1\|\vec{k}\|_1)\), and the total query complexity for long-time simulation is \(\mathcal{O}(r\|\vec{a}\|_1\|\vec{k}\|_1)\). For the well-conditioned \(2\)nd-based MPF, we have \(\|\vec{a}\|_1=\mathcal{O}(\log m)\) and \(\|\vec{k}\|_1=\mathcal{O}(m^2\log m)\). Hence, the total number of queries to \(U_2\) is
        \begin{equation}\mathcal{O}(r\|\vec{a}\|_1\|\vec{k}\|_1) = \mathcal{O}\left( m^2 (\log m)^2 \mu_m T \left( \frac{\mu_m T }{\epsilon}\right)^{1/(2m)} \right). 
        \end{equation}
    This completes the proof.
    \end{proof}
    
    \cref{thm:MPF_complexity} provides an \(m\)-specific complexity estimate. In practice, one may choose \(m\) to minimize the query complexity, provided that the growth rate of high-order nested commutators remains bounded independently of \(m\). This is formalized in~\cref{cor:MPF_complexity}.

    \begin{cor}\label{cor:MPF_complexity}
        Consider Hamiltonian simulation problem up to time $T$ using $2$nd-based MPF. 
        Let $\mu_m$ be a commutator-related factor as defined in~\cref{thm:MPF_complexity}, and suppose that there exists an integer $M$ such that $\sup_{m\geq M} \mu_m \leq \mu$. 
        Then, in order to obtain an $\epsilon$-approximation with probability at least $1-\epsilon$, it suffices to use 
        \begin{equation}
            \mathcal{O}\left( \mu T  \operatorname{polylog}\left(\frac{\mu T}{\epsilon}\right)\right)
        \end{equation}
        queries to the controlled version of $U_2$. 
    \end{cor}
    \begin{proof}
        By~\cref{thm:MPF_complexity}, the query complexity for \(m\geq M\) is given by 
        \begin{equation}
            \mathcal{O}\left( m^2 (\log m)^2 \mu T \left( \frac{\mu T }{\epsilon}\right)^{1/(2m)} \right). 
        \end{equation}
        Choose \(m=\left\lceil \frac{1}{2}\log \left(\frac{\mu T}{\epsilon}\right)\right\rceil\). For sufficiently small \(\epsilon\) or sufficiently large \(T\), this choice satisfies \(m\geq M\). Then
        \begin{equation}
            \left( \frac{\mu T }{\epsilon}\right)^{1/(2m)} \leq \left( \frac{\mu T }{\epsilon}\right)^{1/\log\left(\frac{\mu T}{\epsilon}\right)} = e. 
        \end{equation}
        Therefore, the overall complexity becomes 
        \begin{equation}
            \mathcal{O}\left( m^2 (\log m)^2 \mu T  \right) = \mathcal{O}\left( \mu T  \left(\log\left(\frac{\mu T}{\epsilon}\right)\right)^2 \left(\log \log\left(\frac{\mu T}{\epsilon}\right)\right)^2 \right). 
        \end{equation}
    This completes the proof.
    \end{proof}
    By~\cref{thm:MPF_complexity} and~\cref{cor:MPF_complexity}, the query complexity of the second-order based MPF depends crucially on the commutator parameters \(\mu_m\) and \(\mu\), which involve sums over all indices \(j_1,\ldots,j_l\) satisfying \(j_1+\cdots+j_l=j\). This summation can be further simplified by observing that the number of positive integer solutions to \(j_1+\cdots+j_l=j\) is \(\binom{j-1}{l-1}<2^{j-1}\). Hence,
    \begin{equation}\label{eqn:improved_mu_m}
        \mu_m = \sup_{\substack{j \in 2\mathbb{Z}^+, j\geq 2m\\ 1\leq l \leq m}} \left(\sum_{\substack{j_1,\cdots,j_l \in 2\mathbb{Z}^+, \\ j_1+\cdots+j_l=j}} \left(\prod_{\kappa=1}^{l}  \alpha_{\operatorname{comm},j_{\kappa}+1} \right)\right)^{\frac{1}{j+l}} \leq 2 \sup_{\substack{j \in 2\mathbb{Z}^+, j\geq 2m, \\ 1\leq l \leq m}}  \max_{\substack{j_1,\cdots,j_l \in 2\mathbb{Z}^+, \\ j_1+\cdots+j_l=j}} \left(\prod_{\kappa=1}^{l}  \alpha_{\operatorname{comm},j_{\kappa}+1} \right)^{\frac{1}{j+l}}. 
    \end{equation}
    \cref{eqn:improved_mu_m} is useful for computing commutator scaling in specific applications.

\section{Applications}\label{sec:app}
In this section, we analyze the computational complexity of the second-order MPF in several applications, including the simulation of the transverse field Ising model, chiral Potts model, nonlocal fast-transform Bogoliubov–de Gennes Hamiltonians and Kitaev's honeycomb model.

\subsection{Spin Hamiltonians with Geometric Commutator Growth}\label{subsec:spin-onsager}
We consider spin Hamiltonians for which high-depth nested commutators can be controlled geometrically in the commutator depth. This property does not hold for arbitrary spin Hamiltonians. We therefore restrict our attention to spin Hamiltonians whose generating Hamiltonians satisfy strong constraints on their commutator structure. The examples below are based on the Onsager algebra~\cite{onsager1944crystal,DolanGrady1982,davies1991onsager}. Let \(A_0,A_1\) be two Hermitian operators satisfying the Dolan--Grady commutation relations
\begin{equation}\label{eqn:dolan-grady}
    [A_0,[A_0,[A_0,A_1]]]=16[A_0,A_1],
    \qquad
    [A_1,[A_1,[A_1,A_0]]]=16[A_1,A_0].
\end{equation}
These relations generate a representation of the Onsager algebra. More explicitly, starting from \(A_0\) and \(A_1\), define \(G_1:=\frac{1}{4}[A_1,A_0]\). The remaining Onsager generators are then generated recursively by
\begin{align}\label{eqn:onsager-recursion-positive}
    A_{m+1}
    &:=
    A_{m-1}
    +
    \frac{1}{2}[G_1,A_m],
    \\
    G_m:&=\frac{1}{4}[A_m,A_0].
\end{align}
for \(m\in\mathbb Z\). In particular, \(G_0=0\), and the resulting generators satisfy \(G_{-m}=-G_m\). By the Dolan--Grady/Onsager theorem~\cite{davies1990onsager,Roan1991Onsager}, the operators generated in this manner satisfy the Onsager commutation relations
\begin{equation}\label{eqn:onsager-relations}
    [A_m,A_\ell]=4G_{m-\ell},
    \qquad
    [G_m,A_\ell]=2(A_{\ell+m}-A_{\ell-m}),
    \qquad
    [G_m,G_\ell]=0 .
\end{equation}
\cref{lem:onsager-coefficient-growth} shows that every depth-\(q\) nested commutator of \(A_0\) and \(A_1\) has coefficient size at most \(4^{q-1}\) in the \(\ell^1\) semi-norm on the Onsager algebra. 
We state the result here and give its proof in~\cref{app:spin_onsager}. 

\begin{lem}\label{lem:onsager-coefficient-growth}
Let \(A_0,A_1\) generate an Onsager algebra satisfying \cref{eqn:onsager-relations}. For \(q\geq 2\) and \(\gamma_1,\ldots,\gamma_q\in\{0,1\}\), define
\begin{equation}
    C_q(\gamma_1,\ldots,\gamma_q)
    :=
    [A_{\gamma_1},[A_{\gamma_2},\cdots,[A_{\gamma_{q-1}},A_{\gamma_q}]\cdots]] .
\end{equation}
We have $\|C_q(\gamma_1,\ldots,\gamma_q)\|_{\ell^1(\operatorname{Ons})} \leq 4^{q-1}$. 
\end{lem}

In the \(n\)-spin-chain representations of the Onsager algebra considered below, the generators satisfy a uniform norm bound
\begin{equation}\label{eqn:onsager-generator-norm-bound}
    \|A_m\|\leq C_0 n,
    \qquad
    \|G_m\|\leq C_0 n,
    \qquad
    m\in\mathbb Z,
\end{equation}
where \(C_0>0\) is independent of \(m\) and \(n\). Combining 
\cref{lem:onsager-coefficient-growth} with \cref{eqn:onsager-generator-norm-bound}, we obtain
\begin{equation}\label{eqn:onsager-word-bound}
    \left\|
    [A_{\gamma_1},[A_{\gamma_2},\cdots,[A_{\gamma_{q-1}},A_{\gamma_q}]\cdots]]
    \right\|
    \leq
    C_0 n\,4^{q-1},
\end{equation}
for every \(q\geq 2\) and every word \(\gamma_1,\ldots,\gamma_q\in\{0,1\}\). Equivalently, there exists a \(C_{\operatorname{Ons}}>0\), independent of \(n\) and \(q\), such that
\begin{equation}\label{eqn:onsager-word-bound-simplified}
    \left\|
    [A_{\gamma_1},[A_{\gamma_2},\cdots,[A_{\gamma_{q-1}},A_{\gamma_q}]\cdots]]
    \right\|
    \leq
    n C_{\operatorname{Ons}}^{q}.
\end{equation}

In the examples below, we consider a Hamiltonian of the form \(H=H_0+H_1\), where \(H_0:=\lambda A_0\), \(H_1:=hA_1\), and \(A_0\) and \(A_1\) satisfy \cref{eqn:dolan-grady}. Summing \cref{eqn:onsager-word-bound} over all words \(\gamma_1,\ldots,\gamma_q\in\{0,1\}\), we obtain
\begin{align}
    \alpha_{\operatorname{comm},q}
    &=
    \sum_{\gamma_1,\ldots,\gamma_q\in\{0,1\}}
    \left\|
    [H_{\gamma_1},[H_{\gamma_2},\cdots,[H_{\gamma_{q-1}},H_{\gamma_q}]\cdots]]
    \right\| \leq
    n C_{\operatorname{Ons}}^q
    \sum_{\gamma_1,\ldots,\gamma_q\in\{0,1\}}
    \prod_{r=1}^q |\theta_{\gamma_r}|,
\end{align}
where \(\theta_0=\lambda\) and \(\theta_1=h\). Hence, we have
\begin{equation}\label{eqn:onsager-alpha-bound}
    \alpha_{\operatorname{comm},q}
    \leq
    n\left(C_{\operatorname{Ons}}(|\lambda|+|h|)\right)^q .
\end{equation}
Using \cref{eqn:improved_mu_m}, we get, for any positive even integers \(j_1,\ldots,j_l\) satisfying \(j_1+\cdots+j_l=j\),
\begin{align}
    \left(
    \prod_{\kappa=1}^{l}
    \alpha_{\operatorname{comm},j_\kappa+1}
    \right)^{1/(j+l)}
    &\leq
    \left(
    \prod_{\kappa=1}^{l}
    n\left(C_{\operatorname{Ons}}(|\lambda|+|h|)\right)^{j_\kappa+1}
    \right)^{1/(j+l)}=
    n^{l/(j+l)}
    C_{\operatorname{Ons}}(|\lambda|+|h|).
\end{align}
Since \(j_\kappa\geq 2\), we have \(l\leq j/2\), and therefore \(l/(j+l)\leq 1/3\). Thus,
\begin{equation}\label{eqn:onsager-mu-bound}
    \mu_m
    =
    \mathcal O (
    n^{1/3}(|\lambda|+|h|)
    ),
    \qquad
    \mu
    =
    \mathcal O (
    n^{1/3}(|\lambda|+|h|)
     ).
\end{equation}
Consequently, \cref{cor:MPF_complexity} implies that the number of controlled applications of the second-order product formula is
\begin{equation}\label{eqn:onsager-query-complexity}
    \mathcal O\left(
    n^{1/3}(|\lambda|+|h|)T
    \operatorname{polylog}\left(
    \frac{n(|\lambda|+|h|)T}{\epsilon}
    \right)
    \right).
\end{equation}

\subsubsection{The Transverse Field Ising Model}\label{subsubsec:tfim-onsager}
Consider the one-dimensional transverse field Ising model with periodic boundary conditions, 
\begin{equation}\label{eqn:tfim-hamiltonian}
    H
    =
    \lambda \sum_{j=1}^{n} Z_jZ_{j+1}
    +
    h\sum_{j=1}^{n}X_j,
\end{equation}
where $Z_{n+1}:=Z_1$. This model is a standard spin-chain model and a common target for quantum simulation~\cite{Pfeuty1970,KimEtAl2010FrustratedIsing}. Define \(A_0:=\sum_{j=1}^{n} Z_jZ_{j+1}\) and \(A_1:=\sum_{j=1}^{n}X_j\). \Cref{lem:tfim-onsager-verification} shows that this pair satisfies the assumptions introduced above. We state the result here and provide its proof in~\cref{app:Ising}.

\begin{lem}\label{lem:tfim-onsager-verification}
Let $A_0:=\sum_{j=1}^{n}Z_jZ_{j+1}$ and $A_1:=\sum_{j=1}^{n}X_j$, where $Z_{n+1}:=Z_1$. Then \(A_0,A_1\) satisfy the commutation relations in \cref{eqn:dolan-grady}. Moreover, in the Onsager representation generated by this pair, the auxiliary Onsager generators \(A_m,G_m\) satisfy
\begin{equation}\label{eqn:tfim-onsager-generator-bound}
    \|A_m\|\leq C_{\operatorname{I}}n,
    \qquad
    \|G_m\|\leq C_{\operatorname{I}}n,
    \qquad
    m\in\mathbb Z,
\end{equation}
for an absolute constant \(C_{\operatorname{I}}>0\) independent of \(m\) and \(n\).
\end{lem}

A single second-order product-formula step for the decomposition \(H=\lambda A_0+hA_1\) requires implementing \(e^{-isA_0}\) and \(e^{-isA_1}\). Since the terms in \(A_0\) commute with one another, as do the terms in \(A_1\), each such exponential can be implemented using \(\mathcal O(n)\) elementary one-qubit and two-qubit gates. Therefore, \cref{eqn:onsager-query-complexity} implies that the total gate complexity is
\begin{equation}\label{eqn:tfim-gate-complexity}
    \mathcal O\left(
    n^{4/3}(|\lambda|+|h|)T
    \operatorname{polylog}\left(
    \frac{n(|\lambda|+|h|)T}{\epsilon}
    \right)
    \right).
\end{equation}
Note that this example is \(2\)-local and \(g\)-extensive, with \(g=\mathcal O(|\lambda|+|h|)\). Thus, it lies within the locality/extensivity framework~\cite{mizuta2025commutatorscalinghamiltoniansimulation}. 
It is useful here because the arbitrary-depth commutator growth can be computed directly from the commutation relations in \cref{eqn:dolan-grady}, showing that our analysis applies to such examples without invoking the special finite-commutator-depth analysis of~\cite{mizuta2025commutatorscalinghamiltoniansimulation}.

\subsubsection{A Non-Local Onsager Chain}

The example in \cref{subsubsec:tfim-onsager} is local. We now construct a spin Hamiltonian that exhibits the same high-depth commutator growth while having nonlocal Pauli summands. Let \(U_{\operatorname{Cl}}\) be a Clifford unitary and define
\begin{equation}
B_0:=U_{\operatorname{Cl}}A_0U_{\operatorname{Cl}}^\dagger, \qquad B_1:=U_{\operatorname{Cl}}A_1U_{\operatorname{Cl}}^\dagger .
\end{equation}
We consider the Hamiltonian \(H=\lambda B_0+hB_1\). Since Clifford unitaries map Pauli strings to Pauli strings, \(B_0\) and \(B_1\) are again sums of Pauli strings. However, for a nonlocal Clifford unitary \(U_{\operatorname{Cl}}\), the Pauli strings appearing in \(B_0\) and \(B_1\) may have weight \(\Theta(n)\). Thus, the Hamiltonian need not be \(k\)-local for constant \(k\), even though it is unitarily equivalent to the transverse-field Ising model. For example, let \(U_{\operatorname{Cl}}=\prod_{j=2}^{n}\operatorname{CNOT}_{1,j}\), where qubit \(1\) is the control and qubit \(j\) is the target. Under conjugation by \(U_{\operatorname{Cl}}\), the Pauli operators transform as
\begin{align}
    U_{\operatorname{Cl}}X_1U_{\operatorname{Cl}}^\dagger
    &=
    X_1X_2\cdots X_n,
    \\
    U_{\operatorname{Cl}}X_jU_{\operatorname{Cl}}^\dagger
    &=
    X_j,
    \\
    U_{\operatorname{Cl}}Z_1U_{\operatorname{Cl}}^\dagger
    &=
    Z_1,
    \\
    U_{\operatorname{Cl}}Z_jU_{\operatorname{Cl}}^\dagger
    &=
    Z_1Z_j,
\end{align}
where $j \geq 2$. Therefore, for the transverse field term transforms 
\begin{equation}
    U_{\operatorname{Cl}}
    \left(\sum_{j=1}^{n}X_j\right)
    U_{\operatorname{Cl}}^\dagger
    =
    X_1X_2\cdots X_n
    +
    \sum_{j=2}^{n}X_j .
\end{equation}
Thus, \(B_1=U_{\operatorname{Cl}}A_1U_{\operatorname{Cl}}^\dagger\) contains a Pauli string of weight \(n\). Consequently, for \(h\neq 0\), the Hamiltonian \(H=\lambda B_0+hB_1\) is not \(k\)-local for any fixed \(k\) independent of \(n\).\footnote{
For completeness, the term $\sum_{j=1}^{n}Z_jZ_{j+1}$ transforms as
\(U_{\operatorname{Cl}} (\sum_{j=1}^{n}Z_jZ_{j+1} )U_{\operatorname{Cl}}^\dagger
=
Z_2+\sum_{j=2}^{n-1}Z_jZ_{j+1}+Z_n\)
where periodic boundary conditions are used.  
}
The commutator analysis is unchanged from \cref{subsubsec:tfim-onsager}. If \(g_{\operatorname{Cl}}\) denotes the cost of implementing a controlled \(U_{\operatorname{Cl}}\), then exponentials of \(B_0\) and \(B_1\) can be implemented as $e^{-isB_\nu} = U_{\operatorname{Cl}}e^{-isA_\nu}U_{\operatorname{Cl}}^\dagger$ for $\nu\in\{0,1\}$. Thus, a single second-order product formula step costs \(\mathcal O(n+g_{\operatorname{Cl}})\) gates. \cref{eqn:onsager-query-complexity} implies the  total gate complexity is
\begin{equation}\label{eqn:clifford-scrambled-gate-complexity}
    \mathcal O\left(
    (n+g_{\operatorname{Cl}})
    n^{1/3}(|\lambda|+|h|)T
    \operatorname{polylog}\left(
    \frac{n(|\lambda|+|h|)T}{\epsilon}
    \right)
    \right).
\end{equation}

This example illustrates a simple setting in which the present commutator-scaling analysis applies beyond a constant-locality representation of the Hamiltonian. The high-depth commutator structure is inherited from the Onsager algebra, while the Pauli strings appearing in the Hamiltonian may be highly nonlocal.

\subsubsection{Chiral Potts Model}\label{subsubsec:superintegrable-chiral-potts}
We next consider the \(d\)-state chiral Potts model~\cite{VonGehlenRittenberg1985,Baxter1989SICP,Lychkovskiy2021Closed,Minami2021Onsager}.
Let \(d\geq 2\) and let \(\omega:=e^{2\pi i/d}\). Let \(X\) and \(Z\) denote the \(d\)-dimensional qudit Pauli operators satisfying
\begin{equation}
    X^d=Z^d=I,
    \qquad
    ZX=\omega XZ .
\end{equation}
We write \(X_j,Z_j\) for the corresponding operators on site \(j\), and impose periodic boundary conditions \(Z_{n+1}=Z_1\). Define
\begin{equation}
    \eta_{2j-1}:=X_j,
    \qquad
    \eta_{2j}:=Z_j^\dagger Z_{j+1}
\end{equation}
for $j=1,\ldots,n$. 
Then
\begin{equation}\label{eqn:clock-eta-relations}
    \eta_r^d=I,
    \qquad
    \eta_r\eta_{r+1}
    =
    \omega \eta_{r+1}\eta_r,
    \qquad
    \eta_r\eta_s=\eta_s\eta_r
    \quad
    \text{if } r-s\not\equiv \pm 1 \pmod{2n}.
\end{equation}
Indeed, the only nontrivial adjacent relations follow from $X_jZ_j^\dagger=\omega Z_j^\dagger X_j$ and $Z_{j+1}X_{j+1}=\omega X_{j+1}Z_{j+1}$ for $j=1,\ldots,n$.  Define 
\begin{align}\label{eqn:scp-unnormalized-generators}
    \mathcal A_0
    &:=
    \sum_{j=1}^{n}\sum_{r=1}^{d-1}
    \frac{\eta_{2j-1}^{r}}{1-\omega^{-r}}
    =
    \sum_{j=1}^{n}\sum_{r=1}^{d-1}
    \frac{X_j^r}{1-\omega^{-r}},
    \\
    \mathcal A_1
    &:=
    \sum_{j=1}^{n}\sum_{r=1}^{d-1}
    \frac{\eta_{2j}^{r}}{1-\omega^{-r}}
    =
    \sum_{j=1}^{n}\sum_{r=1}^{d-1}
    \frac{(Z_j^\dagger Z_{j+1})^r}{1-\omega^{-r}}.
\end{align}
These operators are Hermitian, because \(\overline{\frac{1}{1-\omega^{-r}}} = \frac{1}{1-\omega^{r}} = \frac{1}{1-\omega^{-(d-r)}}\), and \((\eta_j^r)^\dagger = \eta_j^{d-r}\) for $j=1,\ldots,n$. 
Let $A_0:=\frac{4}{d}\mathcal A_0$ and $A_1:=\frac{4}{d}\mathcal A_1$. 
Consider the two-term Hamiltonian \(H=\lambda A_0+hA_1\). For \(d=2\), one has \(\omega=-1\), \(A_0=\sum_{j=1}^{n}X_j\), and \(A_1=\sum_{j=1}^{n}Z_jZ_{j+1}\). Thus, the \(d=2\) case recovers the transverse-field Ising chain, up to the harmless interchange of \(A_0\) and \(A_1\). \cref{lem:chiral_potts_commutator} computes the commutator relation. We state the result here and provide its proof in \cref{app:chiral_potts}.

\begin{lem}\label{lem:chiral_potts_commutator}
Let $A_0:=\frac{4}{d}\mathcal A_0$ and $A_1:=\frac{4}{d}\mathcal A_1$, where $\mathcal{A}_0$ and $\mathcal{A}_1$ are defined in  \cref{eqn:scp-unnormalized-generators}.  Then \(A_0,A_1\) satisfy the commutation relations in \cref{eqn:dolan-grady}. 
\end{lem}

To apply the analysis stated at the beginning of this section, it remains to bound the norms of arbitrary \(A_m\) and \(G_m\). To the best of our knowledge, there is no elementary derivation of this fact analogous to the one available for \(d=2\). From a theoretical perspective, the finite-dimensional representation theory of the Onsager algebra can be described through its realization as a fixed-point subalgebra of the \(\mathfrak{sl}_2\) loop algebra. The desired norm estimate can then be derived as a consequence of this representation-theoretic framework. Since we do not carry out this derivation in detail here, we state the required estimate as an assumption below.

\begin{assumption}\label{ass:scp-extensive-onsager-generators}
For the \(d\)-dimensional chiral Potts model, \(A_0,A_1\), the associated Onsager generators satisfy the bound
\begin{equation}
    \|A_m\|\leq K_d n,
    \qquad
    \|G_m\|\leq K_d n,
    \qquad
    m\in\mathbb Z,
\end{equation}
where \(K_d>0\) depends only on \(d\), and not on \(m,n\), or \(q\).
\end{assumption}

Let \(g_d\) denote the cost of implementing one such single-site or two-site \(d\)-level gate in the chosen elementary gate set. Combining these observations with an argument analogous to that given at the end of \cref{subsubsec:tfim-onsager}, we obtain the following total gate complexity:
\begin{equation}\label{eqn:scp-gate-complexity-assumption}
    \mathcal O\left(
    g_d K_d^{1/3}
    n^{4/3}(|\lambda|+|h|)T
    \operatorname{polylog}\left(
    \frac{
    K_d n (|\lambda|+|h|)T
    }{\epsilon}
    \right)
    \right).
\end{equation}
If \(d\) is fixed and single-site and two-site \(d\)-level gates are treated as elementary, then \(g_d\) and \(K_d\) are constants, and the total gate complexity simplifies to
\begin{equation}\label{eqn:scp-gate-complexity-fixed-d-assumption}
    \mathcal O\left(
    n^{4/3}(|\lambda|+|h|)T
    \operatorname{polylog}\left(
    \frac{
    n^{1/3}(|\lambda|+|h|)T
    }{\epsilon}
    \right)
    \right).
\end{equation}
Thus, the chiral Potts model exhibits the same \(n^{1/3}\)-type commutator parameter as the transverse-field Ising model.

\subsection{Fermionic Hamiltonains}\label{sparse-fermion-ham}
We now consider fermionic Hamiltonians, which provide standard examples in the theory of noninteracting and interacting fermions, matchgates, and fermionic linear optics~\cite{TerhalDiVincenzo2002,BravyiKitaev2002}. 

\subsubsection{Quadratic Majorana Hamiltonians}
Let \(c_1,\ldots,c_{2N}\) be Majorana operators satisfying the relations
\begin{equation}\label{majorana-rels}
c_i=c_i^\dagger, \qquad \{c_i,c_j\}=2\delta_{ij}I, \qquad i=1,\ldots,2N.
\end{equation}
Note that \cref{majorana-rels} implies \(c_i^2=I\) for each \(i=1,\ldots,2N\). In what follows, let \(\Gamma_{ij}:=ic_ic_j\). For \(i\neq j\), note that \(\Gamma_{ji}=-\Gamma_{ij}\). Quadratic Majorana Hamiltonians are of the form
\begin{equation}\label{eqn:majorana-hamiltonian}
H = \frac{i}{2} \sum_{\{i,j\} \in E } A_{ij}c_ic_j:=  \frac{1}{2} \sum_{\{i,j\} \in E } A_{ij} \Gamma_{ij}, 
\end{equation}
where \(A=(A_{ij})\) is a real antisymmetric matrix and \(E\) is the edge set of a graph on the Majorana modes. Such Hamiltonians describe noninteracting fermionic systems, with important examples including mean-field superconducting models such as Bogoliubov--de Gennes Hamiltonians and the Kitaev chain~\cite{deGennes1999,Kitaev2001}. Note that \(\|\Gamma_{ij}\|=1\), since each \(\Gamma_{ij}\) is Hermitian and squares to the identity. Indeed, for \(i\neq j\), \(\Gamma_{ij}^{\dagger}=(ic_ic_j)^\dagger=-ic_jc_i=ic_ic_j=\Gamma_{ij}\). Moreover, \(\Gamma_{ij}^2=(ic_ic_j)^2=-c_ic_jc_ic_j=c_i^2c_j^2=I\). Thus, \(\Gamma_{ij}\) is a Hermitian unitary, and hence \(\norm{\Gamma_{ij}}=1\), since its spectrum is contained in \(\{\pm1\}\). We also record the elementary commutation relation for the quadratic Majorana operators. If the two unordered pairs \(\{i,j\}\) and \(\{k,l\}\) are disjoint, then each of \(c_i,c_j\) anticommutes with each of \(c_k,c_l\). Since moving \(c_ic_j\) past \(c_kc_l\) requires four sign changes, we have \(c_ic_jc_kc_l=c_kc_lc_ic_j\), and therefore \([\Gamma_{ij},\Gamma_{kl}]=0\). If the two pairs share exactly one index, the commutator reduces to another quadratic Majorana generator. For instance, when \(i,j,l\) are distinct, we have
\begin{align}
[\Gamma_{ij},\Gamma_{jl}]&= [ic_ic_j,ic_jc_l] =
    -c_ic_jc_jc_l+c_jc_lc_ic_j  =
    -c_ic_l+c_jc_lc_ic_j=
    -c_ic_l-c_ic_l=
    -2c_ic_l =
    2i\Gamma_{il}.
\end{align}
The remaining one-index-overlap cases are identical up to signs. Thus, for \(i\neq j\) and \(k\neq l\), we have
\begin{equation}\label{eqn:maj-comm}
    [\Gamma_{ij},\Gamma_{kl}]
    =
    2i\left(
    \delta_{jk}\Gamma_{il}
    -\delta_{ik}\Gamma_{jl}
    -\delta_{jl}\Gamma_{ik}
    +\delta_{il}\Gamma_{jk}
    \right).
\end{equation}
In particular, the commutator vanishes unless the two unordered pairs \(\{i,j\}\) and \(\{k,l\}\) intersect in exactly one index. This estimate will be used below in the discussion of Kitaev's honeycomb model in \cref{kitaev}.

\subsubsection{Nonlocal Fast-Transform Bogoliubov--de Gennes Hamiltonians}
\label{subsubsec:fast-transform-bdg}
We first consider a structured class of Bogoliubov--de Gennes (BdG) Hamiltonians whose individual components are diagonalized by fast fermionic Gaussian transforms. This example is useful because an individual component may contain a dense collection of long-range hopping and pairing terms in the physical-mode basis, while its exponential remains efficiently implementable, where fermionic Fourier and Bogoliubov transforms are standard ingredients in Gaussian-state and free-fermion circuits~\cite{Ferris2014FermionicFourier,JiangSungKechedzhiEtAl2018,HacklBianchi2021}. We remark that controlled and long-range free-fermion systems admit algebraic circuit compression methods~\cite{KokcuCampsOftelieEtAl2023}, and that their dynamics is efficiently classically simulable for the usual Gaussian inputs and measurements~\cite{TerhalDiVincenzo2002}. Accordingly, this BdG example is meaningful as a quantum subroutine when acting on non-Gaussian quantum data, under coherent control, or as the free part of a larger interacting calculation, rather than as an unconditional quantum-advantage application. As shown below, this example demonstrates that a Hamiltonian may be dense and nonlocal in its physical presentation while still exhibiting genuinely geometric arbitrary-depth commutator growth and admitting inexpensive grouped exponentials.

Let \(\vec{c}=(c_1,\ldots,c_{2N})^T\) denote the Majorana operators in~\cref{majorana-rels}. 
For a real anti-symmetric matrix \(K\in\mathbb{R}^{2N\times 2N}\), define
\begin{equation}\label{eqn:fast-bdg-quadratic-map}
    H(K)
    :=
    \frac{i}{4}\vec{c}^{\,T}K\vec{c}
    =
    \frac{1}{2}\sum_{1\leq r<s\leq 2N}K_{rs}\Gamma_{rs}.
\end{equation}
Up to an irrelevant scalar multiple of the identity, every quadratic BdG Hamiltonian on \(N\) complex fermionic modes can be written in this form. Consider the case in which \(K\) admits the decomposition \(K=\sum_{\gamma=1}^L K_{\gamma}\), and
\begin{equation}\label{eqn:fast-bdg-decomposition}
    H=\sum_{\gamma=1}^{L}H_\gamma,
    \qquad
    H_\gamma:=H(K_\gamma),
\end{equation}
where \(L=\mathcal{O}(1)\). 
We further assume that each component admits a fast orthogonal diagonalization
\begin{equation}\label{eqn:fast-bdg-diagonalization}
    K_\gamma
    =
    O_\gamma^{T}D_\gamma O_\gamma,
    \qquad
    D_\gamma
    =
    \bigoplus_{a=1}^{N}
    \varepsilon_{\gamma,a}\mathcal{J},
    \qquad
    \mathcal{J}:=
    \begin{pmatrix}
        0&1\\
        -1&0
    \end{pmatrix},
\end{equation}
with \(O_\gamma\in\mathrm{SO}(2N)\).  Let \(V_\gamma\) be a fermionic Gaussian circuit satisfying \(V_\gamma\vec{c} \; V_\gamma^\dagger=O_\gamma\vec{c}\). We further assume that \(V_\gamma\) can be implemented using \(g_{\mathrm{FT}}(N)\) elementary gates, where \(g_{\mathrm{FT}}(N)\) is substantially smaller than the generic \(\mathcal{O}(N^2)\) Gaussian synthesis cost. For example, a fermionic fast Fourier network followed by parallel \((k,-k)\) Bogoliubov rotations requires \(\mathcal{O}(N\log N)\) two-mode Gaussian gates in the corresponding native gate model~\cite{Ferris2014FermionicFourier}. The factorization in~\cref{eqn:fast-bdg-diagonalization} then yields
\begin{equation}\label{eqn:fast-bdg-exponential}
    e^{-isH_\gamma}
    =
    V_\gamma
    \left(
        \prod_{a=1}^{N}
        e^{-is\varepsilon_{\gamma,a}\Gamma_{2a-1,2a}/2}
    \right)
    V_\gamma^\dagger.
\end{equation}
The rotations in the product commute and act on disjoint Majorana pairs. Thus, each grouped exponential has gate complexity \(\mathcal{O}(g_{\mathrm{FT}}(N)+N)\).
A concrete nonlocal instance is obtained by taking one \(O_\gamma\) to be a fermionic Fourier transform followed by parallel momentum-space Bogoliubov rotations, and another \(O_\gamma\) to be a product of real-space Bogoliubov rotations.
\cref{lem:fast-bdg-commutator-bound} computes the fixed-depth commutator scale. We state the result here and provide its proof in \cref{app:BdG}.
\begin{lem}\label{lem:fast-bdg-commutator-bound}
Let $J_\gamma:=\|K_\gamma\|$ and  $\Lambda_{\mathrm{BdG}}:=\sum_{\gamma=1}^{L}J_\gamma$. 
Then, for every \(q\geq2\) and every word \(\gamma_1,\ldots,\gamma_q\in\{1,\ldots,L\}\),
\begin{equation}\label{eqn:fast-bdg-word-bound}
    \left\|
    [H_{\gamma_1},[H_{\gamma_2},\ldots,
    [H_{\gamma_{q-1}},H_{\gamma_q}]]\cdots]]
    \right\|
    \leq
    \frac{N}{2}\,2^{q-1}
    \prod_{r=1}^{q}J_{\gamma_r}.
\end{equation}
Consequently,
$\alpha_{\operatorname{comm},q}
    \leq
    \frac{N}{2}\,2^{q-1}
    \Lambda_{\mathrm{BdG}}^q$. 
\end{lem}

For positive even integers \(j_1,\ldots,j_l\) satisfying \(j_1+\cdots+j_l=j\), \cref{lem:fast-bdg-commutator-bound} gives
\begin{align}
    \left(
        \prod_{\kappa=1}^{l}
        \alpha_{\operatorname{comm},j_\kappa+1}
    \right)^{1/(j+l)}
    &\leq
    \Lambda_{\mathrm{BdG}}
    2^{j/(j+l)}
    \left(\frac{N}{2}\right)^{l/(j+l)}.
\end{align}
Since every \(j_\kappa\geq 2\), we have \(l/(j_\kappa+l)\leq 1/3\). Thus,~\cref{eqn:improved_mu_m} gives \(\mu_m=\mathcal{O}(N^{1/3}\Lambda_{\mathrm{BdG}})\) and \(\mu=\mathcal{O}(N^{1/3}\Lambda_{\mathrm{BdG}})\). Consequently, the number of controlled applications of the second-order product formula is
\begin{equation}\label{eqn:fast-bdg-query-complexity}
    \mathcal{O}\left(
        N^{1/3}\Lambda_{\mathrm{BdG}}T
        \operatorname{polylog}\left(
            \frac{N \Lambda_{\mathrm{BdG}}T}{\epsilon}
        \right)
    \right), 
\end{equation} 
and the total gate complexity is
\begin{equation}\label{eqn:fast-bdg-gate-complexity}
    \mathcal{O}\left(
        (g_{\mathrm{FT}}(N)+N)
        N^{1/3}\Lambda_{\mathrm{BdG}}T
        \operatorname{polylog}\left(
            \frac{N \Lambda_{\mathrm{BdG}}T}{\epsilon}
        \right)
    \right).
\end{equation}
For a fermionic fast transform with \(g_{\mathrm{FT}}(N)=\mathcal{O}(N\log N)\), this becomes
\begin{equation}\label{eqn:fast-bdg-gate-complexity-fft}
    \mathcal{O}\left(
        N^{4/3}\log N\,
        \Lambda_{\mathrm{BdG}}T
        \operatorname{polylog}\left(
            \frac{N \Lambda_{\mathrm{BdG}}T}{\epsilon}
        \right)
    \right).
\end{equation}

We have \(\alpha_{\operatorname{comm},3}=\mathcal{O}(N\Lambda_{\mathrm{BdG}}^3)\). The second-order product formula requires \(\mathcal{O}((\Lambda_{\mathrm{BdG}}T)^{3/2}\sqrt{N/\epsilon})\) steps, and the gate complexity for a fermionic fast transform is \(\mathcal{O}\!\left((\Lambda_{\mathrm{BdG}}T)^{3/2}N^{3/2}\log N/\sqrt{\epsilon}\right)\), illustrating the improved dependence on both \(N\) and \(\epsilon\) provided by the MPF bound. More generally, for a fixed \(p\)-th-order product formula, we have \(\alpha_{\operatorname{comm},p+1}=\mathcal{O}(N\Lambda_{\mathrm{BdG}}^{p+1})\). Then,~\cite{ChildsSuTranEtAl2020} shows that the number of Trotter steps is \(\mathcal{O}\!\left(N^{1/p}(\Lambda_{\mathrm{BdG}}T)^{1+1/p}\epsilon^{-1/p}\right)\), and the gate complexity becomes
\begin{equation}
\mathcal{O} ((g_{\mathrm{FT}}(N)+N)N^{1/p}(\Lambda_{\mathrm{BdG}}T)^{1+1/p}\epsilon^{-1/p} ).
\end{equation}
Thus, for fixed \(p>3\), the system-size dependence \(N^{1+1/p}\) is smaller than that of the second-order-based MPF, whereas the MPF retains nearly linear dependence on \(T\) and only polylogarithmic dependence on \(1/\epsilon\).

For this strictly quadratic model, specialized Gaussian methods provide the strongest model-specific post-Trotter approaches, while qubitization and QSVT yield worse gate complexity, as discussed in~\cref{app:BdG}. A generic fermionic Gaussian unitary can be implemented using \(\mathcal O(N^2)\) gates and \(\mathcal O(N)\) depth~\cite{JiangSungKechedzhiEtAl2018}. Quadratic fermionic dynamics can be exponentially fast-forwarded~\cite{GuSommaSahinoglu2021}, while controlled long-range free-fermion evolution admits implementations using \(\mathcal O(N^2)\) gates and \(\mathcal O(N)\) depth~\cite{KokcuCampsOftelieEtAl2023}.

We now explore asymptotic regimes in which the second-order-based MPF may outperform both post-Trotter methods and the \(p\)-th-order product formula. Here, we state the conclusion directly, while the detailed calculations are given in \cref{app:BdG}. Assume \(\Lambda_{\mathrm{BdG}}=\Theta(1)\), set \(R:=\log(NT/\epsilon)\), and assume that the \(p\)-th-order product formula uses \(\mathcal{O}(a^p)\) exponentials for some constant \(a>1\) (note that \(a=\sqrt{5}\) for the Trotter--Suzuki formula). For fixed \(p\), a direct comparison, obtained by requiring the MPF scaling to be asymptotically smaller than both the Trotter and post-Trotter scalings, yields the MPF-favorable regime
\begin{equation}\label{eqn:fast-bdg-fixed-p-general-regime}
    \epsilon a^{-p^2}N^{p/3-1}R^{4p}
    <
    T
    <
    \frac{N^{2/3}}{R^4 \log N}.
\end{equation}
If \(T=N^\tau\) and \(\epsilon=N^{-\kappa}\), then, away from logarithmic boundary cases, this reduces to \(\tau<2/3\) and \(\tau+\kappa>p/3-1\). Thus, at constant accuracy, such a regime exists for fixed \(p<5\). 
For example, \(p=4\) admits \(1/3<\tau<2/3\). 
A corresponding regime with polynomial accuracy also exists for \(p\geq5\).

In practice, the convergence order \(p\) of the product formula may be chosen adaptively to improve the asymptotic scaling. Minimizing the asymptotic scaling of the product formula gives \(p_{\mathrm{opt}}=\sqrt{\frac{R}{\log a}}\). Compared with this product formula at the optimized convergence order, as well as with post-Trotter methods, the corresponding general MPF-favorable regime becomes
\begin{equation}\label{eqn:fast-bdg-optimized-general-regime}
    \frac13\log N+4\log R
    <
    2\sqrt{R\log a},
    \qquad
    T<
    \frac{N^{2/3}}{R^4 \log N}.
\end{equation}
The first condition cannot hold asymptotically when both \(T\) and \(1/\epsilon\) are polynomial in \(N\). However, if \(T=N^\tau\) and \(\epsilon=\exp[-c(\log N)^2]\), then the two conditions reduce to \(c>1/(36\log a)\) and \(\tau<2/3\). Hence, within this cost model, an advantage over optimized-order Trotterization arises in a sufficiently high, inverse-quasi-polynomial accuracy regime.

\subsubsection{Kitaev's Honeycomb Model}\label{kitaev}
We now consider Kitaev's honeycomb model~\cite{Kitaev2006Anyons}. In contrast to the quadratic Hamiltonians considered in~\cref{subsubsec:fast-transform-bdg}, Kitaev's honeycomb model is an interacting spin system whose Majorana representation is quartic. Nevertheless, as we show below, its conserved \(\mathbb{Z}_2\) gauge variables decompose the Hamiltonian into invariant sectors on which the dynamics is quadratic. We show that this additional structure is sufficient to obtain geometric commutator growth and hence to apply our MPF complexity analysis. Let \(\Lambda\) be a finite honeycomb lattice with \(n\) vertices.  The edge set is partitioned into three classes \(E_x,E_y,E_z\).  The Hamiltonian is
\begin{equation}\label{eqn:kitaev-honeycomb-spin-hamiltonian}
H = -\sum_{\alpha\in\{x,y,z\}}J_\alpha \sum_{\{j,k\}\in E_\alpha} \sigma_j^\alpha\sigma_k^\alpha :=H_x+H_y+H_z, \qquad
H_\alpha = - J_\alpha \sum_{\{j,k\}\in E_\alpha} \sigma_j^\alpha\sigma_k^\alpha
\end{equation}
\Cref{kitaev-majorana} shows that \cref{eqn:kitaev-honeycomb-spin-hamiltonian} can be expressed as a quartic Hamiltonian in the standard Majorana representation. We state the result here and give its proof in~\cref{app:Kitaev}. 

\begin{lem}\label{kitaev-majorana}
The Hamiltonian of Kitaev's honeycomb model can be represented on an enlarged Hilbert space by the Hamiltonian
 \begin{equation}\label{eqn:kitaev-honeycomb-enlarged}
\widetilde H=\sum_{\alpha\in\{x,y,z\}}J_\alpha\sum_{\{j,k\}\in E_\alpha} u_{jk} i c_j c_k
=-\sum_{\alpha\in\{x,y,z\}}J_\alpha\sum_{\{j,k\}\in E_\alpha}  b_j^\alpha b_k^\alpha c_j c_k,
\end{equation}
where $u_{jk}=\pm 1$, and $b_j^x$, $b_j^y$, $b_j^z$, and $c_j$ are Majorana operators acting on an enlarged local four-dimensional Hilbert space at each site $j\in\Lambda$.
\end{lem}

 For norm upper bounds, it suffices to work on the enlarged four-dimensional Hilbert space, since restriction to the physical two-dimensional subspace cannot increase the operator norm. Accordingly, the estimates below are carried out prior to projection onto the physical subspace. Although \cref{eqn:kitaev-honeycomb-enlarged} is quartic in the Majorana operators, \cref{lem:kitaev-sectorwise-quadratic} provides the key estimate that allows us to reduce the analysis to commutation relations for quadratic Majorana Hamiltonians.

\begin{lem}
\label{lem:kitaev-sectorwise-quadratic}
The operators \(u_{jk}\) are mutually commuting and commute with \(\widetilde H_\alpha=-J_\alpha\sum_{\{j,k\}\in E_\alpha}  b_j^\alpha b_k^\alpha c_j c_k\) for $\alpha=x,y,z$. 
Hence the enlarged Majorana Hilbert space decomposes into joint eigenspaces labelled by choices of signs $u=\{u_{jk}=\pm1\}$. On each such sector, the Hamiltonian \(\widetilde H \) reduces to a quadratic Majorana Hamiltonian in \(c_j\)'s:
\begin{equation}\label{eqn:kitaev-sector-hamiltonian}
\widetilde H^{(u)}
=
\sum_{\alpha\in\{x,y,z\}}J_\alpha
\sum_{\langle jk\rangle\in E_\alpha}
u_{jk}\, i c_jc_k=
\widetilde H_x^{(u)}+\widetilde H_y^{(u)}+\widetilde H_z^{(u)},
\qquad
\widetilde H_\alpha^{(u)} = J_\alpha
\sum_{\langle jk\rangle\in E_\alpha}
u_{jk}\Gamma_{jk},
\end{equation}
where $\Gamma_{jk}:=ic_jc_k$. 
\end{lem}

\begin{proof}
Let \(e=\{j,k\}\) and \(e'=\{\ell,m\}\) be distinct edges.  If the edges are disjoint, then $u_{jk}=i b_j^\alpha b_k^\alpha$ and $u_{\ell m}=i b_\ell^\beta b_m^\beta$ are products of disjoint pairs of Majorana operators. Moving \(b_j^\alpha b_k^\alpha\) past \(b_\ell^\beta b_m^\beta\) requires four transpositions. Hence, \(u_{jk}u_{\ell m}=u_{\ell m}u_{jk}\). 
If the edges share a vertex, then $\alpha \neq \beta$ necessarily.
If \(j=\ell\), then $u_{jk}=i b_j^\alpha b_k^\alpha$ and $u_{jm}=i b_j^\beta b_m^\beta$. Once again, the four Majorana operators appearing in \(u_{jk}\) and \(u_{jm}\) are distinct. By the same argument as above, \(u_{jk}u_{jm}=u_{jm}u_{jk}\). The other one vertex overlap cases are identical. Hence,  the operators \(u_{jk}\) are mutually commuting. We now show that each \(u_{jk}\) commutes with the Hamiltonian.  Consider a single Hamiltonian term of the form $J_\beta u_{\ell m}\ i c_\ell c_m$ for $\{\ell, m\}\in E_\beta$. Since we have already shown that \(u_{jk}\) commutes with \(u_{\ell m}\), it suffices to show that \(u_{jk}\) commutes with \(ic_\ell c_m\). This is clear, since the two operators involve distinct Majorana operators. Summing over all Hamiltonian terms, we obtain
\begin{equation}
[u_{jk},\widetilde H_\beta]=0
\end{equation}
for every \(\beta\in\{x,y,z\}\). Hence, $[u_{jk},\widetilde{H}]=0$. Since the \(u_{jk}\)'s are mutually commuting Hermitian involutions, the Majorana Hilbert space decomposes into their joint eigenspaces.
On a joint eigenspace labelled by signs \(u=\{u_{jk}=\pm1\}\), each \(u_{jk}\) acts as a scalar.  Therefore,
\begin{equation}
\widetilde H 
=
\sum_{\alpha\in\{x,y,z\}}J_\alpha
\sum_{\langle jk\rangle\in E_\alpha}
u_{jk}\, i c_jc_k
\end{equation}
reduces on each sector to a quadratic Majorana Hamiltonian. This completes the proof. 
\end{proof}

\cref{lem:kitaev-sectorwise-quadratic} is useful for the commutator estimates below. This is because, even though \(\widetilde H\) is quartic on the Majorana Hilbert space, the operators \(u_{jk}\)'s commute with each \(\widetilde H_\alpha\). Hence, every \(\widetilde H_\alpha\) is block diagonal with respect to the joint eigenspace decomposition of the \(u_{jk}\)'s. Hence
\begin{equation}
\widetilde H_\alpha
=
\bigoplus_u
\widetilde H_\alpha^{(u)}.
\end{equation}
Consequently, every nested commutator is also block diagonal. More precisely, for every word \(\gamma_1,\ldots,\gamma_q\in\{x,y,z\}\), we have
\begin{equation}
[\widetilde H_{\gamma_1},[\cdots,[\widetilde H_{\gamma_{q-1}},\widetilde H_{\gamma_q}]\cdots]] = \bigoplus_u [\widetilde H_{\gamma_1}^{(u)},[\cdots,[\widetilde H_{\gamma_{q-1}}^{(u)},\widetilde H_{\gamma_q}^{(u)}]\cdots]].
\end{equation}
Taking operator norms gives
\begin{equation}\label{lift-sector-comms}
 \| [\widetilde H_{\gamma_1},[\cdots,[\widetilde H_{\gamma_{q-1}},\widetilde H_{\gamma_q}]\cdots]]  \| = \max_u  \| [\widetilde H_{\gamma_1}^{(u)},[\cdots,[\widetilde H_{\gamma_{q-1}}^{(u)},\widetilde H_{\gamma_q}^{(u)}]\cdots]]  \|.
\end{equation}
Thus, to bound commutators of the full quartic Majorana Hamiltonian, it suffices to bound the corresponding commutators within each fixed $u$-sector. In such a sector, each $u_{jk}\in\{\pm1\}$ is a scalar sign, and $\widetilde H_\alpha^{(u)}$ is a quadratic Majorana Hamiltonian. \Cref{lem:kitaev-sectorwise-commutator-bound} provides the corresponding commutator bound. We state the result here and give its proof in~\cref{app:Kitaev}.

\begin{lem}
\label{lem:kitaev-sectorwise-commutator-bound}
Fix a joint eigensector \(u=\{u_{jk}=\pm1\}\).  For every word \(\gamma_1,\ldots,\gamma_q\in\{x,y,z\}\), we have
\begin{equation}\label{eqn:kitaev-sector-word-bound}
\| [\widetilde H_{\gamma_1}^{(u)},[\cdots,[\widetilde H_{\gamma_{q-1}}^{(u)},\widetilde H_{\gamma_q}^{(u)}]\cdots]] \| \le n4^{q-1} \prod_{r=1}^{q}|J_{\gamma_r}|.
\end{equation}
Consequently, the same bound holds for the corresponding nested commutator of the full Hamiltonian \(\widetilde H\), and hence also for the original spin Hamiltonian \(H\).
\end{lem}

By \cref{lem:kitaev-sectorwise-commutator-bound}, we have
\begin{align}
\alpha_{\operatorname{comm},q} &= \sum_{\gamma_1,\ldots,\gamma_q\in\{x,y,z\}} \left\| [H_{\gamma_1},[\cdots,[H_{\gamma_{q-1}},H_{\gamma_q}]\cdots]] \right\| \\
&\le n4^{q-1} \sum_{\gamma_1,\ldots,\gamma_q\in\{x,y,z\}} \prod_{r=1}^{q}|J_{\gamma_r}| \\
&= n4^{q-1} \left( |J_x|+|J_y|+|J_z| \right)^q 
\leq C_{\operatorname{KH}}^q  n \left( |J_x|+|J_y|+|J_z| \right)^q,
\end{align}
where $C_{\operatorname{KH}}$ absorbs all constant factors. Thus, Kitaev's honeycomb model has geometric commutator growth and our algorithm can be used to simulate it efficiently. We now plug this estimate into our commutator scaling analysis. For any positive even integers \(j_1,\ldots,j_l\) satisfying \(j_1+\cdots+j_l=j\), we have
\begin{equation}
\left( \prod_{\kappa=1}^{l} \alpha_{\operatorname{comm},j_\kappa+1} \right)^{1/(j+l)} \leq  C_{\operatorname{KH}} n^{l/(j+l)} (|J_x|+|J_y|+|J_z|). 
\end{equation}
Since each \(j_\kappa\ge2\), we have \(l\le j/2\), and hence $\frac{l}{j+l}\le\frac13$.  Therefore, $\mu_m=\mu=\mathcal O ( n^{1/3}  (|J_x|+|J_y|+|J_z|)  )$. Consequently, \cref{cor:MPF_complexity} implies that the number of controlled applications of the second-order product formula is
\begin{equation} 
\mathcal O\left( n^{1/3} (|J_x|+|J_y|+|J_z|) T \operatorname{polylog} \left( \frac{n^{1/3}(|J_x|+|J_y|+|J_z|)T}{\epsilon} \right) \right).
\end{equation}
Each \(E_\alpha\) is an independent edge set. Equivalently, each vertex of the honeycomb lattice is incident to exactly one bond of each type \(x,y,z\), so two distinct \(\alpha\)-bonds cannot touch the same vertex. Consequently, the terms in \(H_\alpha\) act on disjoint pairs of spins and hence mutually commute. Therefore, one product-formula step for the grouped decomposition \(H=H_x+H_y+H_z\) has gate complexity \(\mathcal O(n)\). Hence, the total gate complexity is
\begin{equation} 
\mathcal O\left( n^{4/3} (|J_x|+|J_y|+|J_z|) T \operatorname{polylog} \left( \frac{n^{1/3}(|J_x|+|J_y|+|J_z|)T}{\epsilon} \right) \right).
\end{equation}

The main takeaway is that Kitaev's honeycomb model circumvents the factorial growth of nonzero nested-commutator words that can occur for generic local quartic Majorana Hamiltonians. This simplification arises because the quartic variables $u_{jk}$ are conserved quantities. As a result, the full commutator problem decomposes into independent invariant $u$-sectors, within each of which the Hamiltonian is quadratic in the Majorana operators. The resulting sectorwise quadratic structure makes the model directly amenable to our algorithm and commutator-scaling analysis.

\section{MPF Based on Other Base Sequences}\label{sec:general_MPF}
We have established a complexity estimate for the second-order based MPF. We now examine the effect of using product formulas of different orders as the base sequence. Our main observation is that the convergence order of the base sequence influences the commutator factor appearing in the overall algorithmic complexity.

\subsection{First-Order Trotter}
Let us first consider the case \(p=1\) as an illustrative example. Since the first-order Trotter formula is not symmetric, its BCH expansion contains both odd- and even-order terms.
\begin{equation}
    \left(U_{1}(\Delta/k) \right)^k = \exp \left( -i H \Delta + \sum_{j\geq 2} E_{j}^{(1)} \frac{\Delta^j}{k^{j-1}} \right). 
\end{equation}
Here, we add the superscript \((1)\) to the operators \(E_j\) to emphasize that the \(E_j^{(1)}\) arise from the expansion of the first-order Trotter formula and may differ from the operators \(E_j\) used in~\cref{sec:results}. Then, a Taylor expansion, or more rigorously~\cref{lem:VoP_high_order}, implies that the order condition for an \(m\)th-order MPF is
\begin{equation}\label{eqn:multiproduct_linear_system_1st}
    \left( \begin{array}{cccc}
        1 & 1 & \cdots & 1 \\
        (k_{1}^{(1)})^{-1} & (k_{2}^{(1)})^{-1} & \cdots & (k_{m}^{(1)})^{-1} \\
        \vdots & \vdots & \ddots & \vdots \\
        (k_{1}^{(1)})^{-m+1} & (k_{2}^{(1)})^{-m+1} & \cdots & (k_{m}^{(1)})^{-m+1}
    \end{array} \right) 
    \left( \begin{array}{c}
        a_{1}^{(1)} \\
        a_{2}^{(1)} \\
        \vdots \\
        a_{m}^{(1)} 
    \end{array}\right) 
    = \left(\begin{array}{c}
        1 \\
        0 \\
        \vdots \\
        0 
    \end{array}\right). 
\end{equation}

\begin{equation}
    \|\vec{a}^{(1)}\|_1 = \mathcal{O}(\log m), \quad \|\vec{k}^{(1)}\|_1 = \mathcal{O}(m^4 \left(\log m\right)^2). 
\end{equation}
The query complexity of the first-order-based MPF can be established by following the same steps as in~\cref{sec:results}. The only difference is that the BCH expansion also contains all even-order terms. Consequently, the commutator factor \(\mu\) must also include commutators involving an even number of Hamiltonians, allowing the indices \((j_1,\ldots,j_l)\) in~\cref{eqn:def_mu_m} to be arbitrary positive integers.

\begin{cor}\label{cor:MPF_complexity_1st}
        Consider the Hamiltonian simulation problem up to time $T$ using $1$st-based MPF.
        Then, in order to obtain an $\epsilon$-approximation with probability at least $1-\epsilon$, it suffices to use 
        \begin{equation}
            \mathcal{O}\left( \mu^{(1)} T  \operatorname{polylog}\left(\frac{\mu^{(1)} T}{\epsilon}\right)\right)
        \end{equation}
        queries to controlled version of $U_1$. 
        Here, $\mu^{(1)} = \sup_{m\geq M} \mu_m^{(1)}$ for a sufficiently large $M$, and 
        \begin{equation}
            \mu_m^{(1)} = \sup_{(j,l): j \geq m, 1 \leq l \leq m } \lambda_{j,l}^{(1)}, \quad \quad 
            \lambda_{j,l}^{(1)} \geq \left(\sum_{\substack{j_1,\cdots,j_l\in \mathbb{Z}^+, \\ j_1+\cdots+j_l=j}} \prod_{\kappa=1}^{l}  \alpha_{\operatorname{comm},j_{\kappa}+1}\right)^{\frac{1}{j+l}}. 
        \end{equation}
    \end{cor}

\subsection{High-Order Symmetric Product Formula}
We now consider the \(2p\)th-order-based MPF for an arbitrary positive integer \(p\). We assume that the underlying product formula is symmetric, in the sense that \(U_{2p}(s)U_{2p}(-s)=U_{2p}(-s)U_{2p}(s)=I\). Notice that the Trotter--Suzuki formula constructed in~\cref{eqn:Trotter_Suzuki} satisfies this symmetry condition. The BCH expansion then takes the form

\begin{equation}
    \left(U_{2p}(\Delta/k) \right)^k = \exp \left( -i H \Delta + \sum_{j \in 2\mathbb{Z}^+, j \geq 2p} E_{j+1}^{(2p)} \frac{\Delta^{j+1}}{k^{j}} \right). 
\end{equation}
Notice that, by symmetry, the expansion contains only odd-order terms, and the perturbation Hamiltonian begins at order \(2p+1\) due to the convergence order of the product formula. Therefore, the order condition for the corresponding \(2m\)th-order MPF is
\begin{equation}\label{eqn:multiproduct_linear_system_high_order}
    \left( \begin{array}{cccc}
        1 & 1 & \cdots & 1 \\
        (k_{1}^{(2p)})^{-2p} & (k_{2}^{(2p)})^{-2p} & \cdots & (k_{m}^{(2p)})^{-2p} \\
        (k_{1}^{(2p)})^{-2p-2} & (k_{2}^{(2p)})^{-2p-2} & \cdots & (k_{m}^{(2p)})^{-2p-2} \\
        \vdots & \vdots & \ddots & \vdots \\
        (k_{1}^{(2p)})^{-2m+2} & (k_{2}^{(2p)})^{-2m+2} & \cdots & (k_{m}^{(2p)})^{-2m+2}
    \end{array} \right) 
    \left( \begin{array}{c}
        a_{1}^{(2p)} \\
        a_{2}^{(2p)} \\
        a_{3}^{(2p)} \\
        \vdots \\
        a_{m}^{(2p)} 
    \end{array}\right) 
    = \left(\begin{array}{c}
        1 \\
        0 \\
        0 \\
        \vdots \\
        0 
    \end{array}\right). 
\end{equation}
Unlike the first-order and second-order based MPFs, it is not clear whether~\cref{eqn:multiproduct_linear_system_high_order} admits a well-conditioned solution, in the sense that both \(\|\vec{a}^{(2p)}\|\) and \(\|\vec{k}^{(2p)}\|\) are at most polynomial in \(m\). Nevertheless, the numerical results in~\cite{LowKliuchnikovWiebe2019} suggest the existence of well-conditioned solutions at least for \(p=2\), corresponding to the fourth-order-based MPF. Assuming the existence of well-conditioned coefficients and powers, the following complexity estimate can be established using the same approach as in~\cref{sec:results}.

\begin{cor}\label{cor:MPF_complexity_high_order}
        Consider the quantum simulation problem up to time $T$ using MPF based on $2p$-th order symmetric product formula. 
        Suppose that there exist $\vec{k}^{(2p)}$ and $\vec{a}^{(2p)}$ satisfying~\cref{eqn:multiproduct_linear_system_high_order} and $\|\vec{k}^{(2p)}\|, \|\vec{a}^{(2p)}\| = \mathcal{O}(\operatorname{poly}(m))$.  
        Then, in order to obtain an $\epsilon$-approximation with probability at least $1-\epsilon$, it suffices to use 
        \begin{equation}
            \mathcal{O}\left( \mu^{(2p)} T  \operatorname{polylog}\left(\frac{\mu^{(2p)} T}{\epsilon}\right)\right)
        \end{equation}
        queries to controlled version of $U_{2p}$. 
        Here, $\mu^{(2p)} = \sup_{m\geq M} \mu_m^{(2p)}$ for a sufficiently large $M$, and 
        \begin{equation}
            \mu_m^{(2p)} = \sup_{(j,l): j \geq 2m, 1 \leq l \leq m} \lambda_{j,l}^{(2p)}, \quad \quad 
            \lambda_{j,l}^{(2p)} \geq \left(\sum_{\substack{j_1,\cdots,j_l \in 2\mathbb{Z}^+,\\j_1,\cdots,j_l \geq 2p,\\ j_1+\cdots+j_l=j}} \prod_{\kappa=1}^{l} \alpha_{\operatorname{comm},j_{\kappa}+1}\right)^{\frac{1}{j+l}}. 
        \end{equation}
    \end{cor}

\subsection{Effects and Comparison}
We now compare MPFs based on different base sequences. A caveat is that we assume the existence of well-conditioned MPFs, which remains to be further explored. From~\cref{cor:MPF_complexity}, \cref{cor:MPF_complexity_1st}, and~\cref{cor:MPF_complexity_high_order}, we observe that the scalings in the evolution time \(T\) and precision parameter \(\epsilon\) are essentially the same, up to logarithmic factors: they are nearly linear in \(T\) and polylogarithmic in \(1/\epsilon\), thereby achieving near-optimal dependence on time and precision. The only difference lies in the commutator factor \(\mu^{(p)}\), which is the supremum of another commutator factor \(\lambda_{j,l}^{(p)}\). We observe that MPFs based on higher-order product formulas depend on nested commutators of greater depth. Specifically, for a \(p\)th-order-based MPF, the commutator factor \(\mu^{(p)}\) depends on \(\alpha_{\operatorname{comm},k}\) for all \(k\geq p+1\). Therefore, MPFs based on higher-order product formulas may be asymptotically advantageous in applications where the homogenized quantity \(\alpha_{\operatorname{comm},k}^{1/k}\) decreases as \(k\) increases.

\section*{Acknowledgements}
We thank Yuan Su for helpful discussions and for pointing out the multi-term BCH formula. 
We also thank Andrew Childs, James Watson, and Nathan Wiebe for valuable discussions.
Dong An acknowledges support from the Department of Defense through the Hartree Postdoctoral Fellowship at QuICS, and from the seed grant at the NSF Quantum Leap Challenge Institute for Robust Quantum Simulation (QLCI grant OMA-2120757). 
Konstantina Trivisa acknowledges support from the National Science Foundation under grants DMS-2008568 and DMS-2231533.

\section*{Competing Interests}

The authors have no competing interests to declare that are relevant to the content of this article. 

\section*{Data Availability Statement} 
Data sharing not applicable to this article as no datasets were generated or analyzed during the current study.

\printbibliography
\appendix

\section{Detailed Calculations for the Applications}\label{app:commutator_calculation}

Here we provide complete calculations of the commutator scalings used in~\cref{sec:app}, as well as other technical details omitted from the main text.

\subsection{General Spin Hamiltonians with the Onsager Algebra}\label{app:spin_onsager}

\begin{proof}[Proof of~\cref{lem:onsager-coefficient-growth}]
Let $X=\sum_{m \in \Z} a_m A_m+\sum_m b_mG_m$ be a finite linear combination of Onsager generators.
The coefficient \(\ell^1\) semi-norm is defined by
\begin{equation}
    \|X\|_{\ell^1(\operatorname{Ons})}
    :=
    \inf
    \left\{
    \sum_m |a_m|+\sum_m |b_m|
    :
    X=\sum_m a_mA_m+\sum_m b_mG_m
    \right\},
\end{equation}
where the infimum is included to account for the non-uniqueness of representations of \(X\). We first prove the one-step estimate. For each \(m\) and \(s\in\{0,1\}\), \([A_s,G_m]=2(A_{s-m}-A_{s+m})\) is a linear combination of at most two \(A\)-generators, and the sum of the absolute values of its coefficients is at most \(4\). Similar remarks apply to \([A_s,A_m]=4G_{s-m} \) for each $m$. 
By linearity and by taking the infimum over all Onsager expansions of \(X\), we obtain
\begin{equation}\label{eqn:onsager-ad-bound-proof}
    \|[A_s,X]\|_{\ell^1(\operatorname{Ons})}
    \leq
    4\|X\|_{\ell^1(\operatorname{Ons})},
    \qquad
\end{equation}
for $s\in\{0,1\}$. We now argue by induction on \(q\). For \(q=2\), if \(\gamma_1=\gamma_2\), then $C_2(\gamma_1,\gamma_2)=0$.  If \(\gamma_1\neq\gamma_2\), then
\begin{equation}
    C_2(\gamma_1,\gamma_2)
    =
    [A_{\gamma_1},A_{\gamma_2}]
    =
    4G_{\gamma_1-\gamma_2}.
\end{equation}
Hence, $\|C_2(\gamma_1,\gamma_2)\|_{\ell^1(\operatorname{Ons})} \leq 4$. Assume now that the estimate holds for depth \(q-1\). Using the identity $C_q(\gamma_1,\ldots,\gamma_q) = [A_{\gamma_1},C_{q-1}(\gamma_2,\ldots,\gamma_q)]$, and applying \cref{eqn:onsager-ad-bound-proof}, we get
\begin{align}
\|C_q(\gamma_1,\ldots,\gamma_q)\|_{\ell^1(\operatorname{Ons})}
    &\leq
    4
    \|C_{q-1}(\gamma_2,\ldots,\gamma_q)\|_{\ell^1(\operatorname{Ons})} \leq
    4\cdot 4^{q-2}
    =
    4^{q-1}.
\end{align}
This completes the proof.
\end{proof}

\subsection{The Transverse Field Ising Model}\label{app:Ising}

\begin{proof}[Proof of~\cref{lem:tfim-onsager-verification}]

Write $B_j:=Z_jZ_{j+1}$ for $j=1,\ldots,n$ with periodic indices. Then \(A_0=\sum_{j=1}^{n} B_j\). For each fixed \(j\), the only terms in \(A_0\) that fail to commute with \(X_j\) are \(B_{j-1}\) and \(B_j\). Moreover, \(B_{j-1}\) and \(B_j\) commute with each other, both square to the identity, and both anti-commute with \(X_j\). Hence, if $K_j:=B_{j-1}+B_j$ then we have $[A_0,X_j]=[K_j,X_j]$ for $r=1,2,3$. Since \(K_jX_j=-X_jK_j\), we have $[K_j,X_j]=2K_jX_j$. Furthermore, we have
\begin{equation}
    K_j^2=(B_{j-1}+B_j)^2=2(I+B_{j-1}B_j),
\end{equation}
Therefore, we have $K_j^3=4K_j$. It follows that
\begin{align}
    \operatorname{ad}_{K_j}^{3}(X_j)
    &=[K_j,[K_j,[K_j,X_j]]]   \\
    &=[K_j,[K_j,2K_jX_j]]   \\
    &=[K_j,4K_j^2X_j]  \\
    &=8K_j^3X_j=32K_jX_j=16[K_j,X_j]=16 \operatorname{ad}_{K_j}(X_j).
\end{align}
Equivalently, $[A_0,[A_0,[A_0,X_j]]]=16[A_0,X_j]$. Summing over \(j\) gives the first commutation relation in \cref{eqn:dolan-grady}. The second commutation relation follows similarly. Explicit expressions for \(A_m\) and \(G_m\) are given in~\cite{Minami2021Onsager}, which shows that each \(A_m\) and \(G_m\) is a sum of \(\mathcal O(n)\) Pauli strings with bounded coefficients. Since every Pauli string has spectral norm one, the triangle inequality gives
\begin{equation}
    \|A_m\|\leq C_{\operatorname{I}}n,
    \qquad
    \|G_m\|\leq C_{\operatorname{I}}n,
\end{equation}
for an absolute constant \(C_{\operatorname{I}}>0\), independent of \(m\) and \(n\). This completes the proof.
\end{proof}

\subsection{Chiral Potts Model}\label{app:chiral_potts}

\begin{proof}[Proof of~\cref{lem:chiral_potts_commutator}]
Let $h_r:=\sum_{m=1}^{d-1} \frac{\eta_r^m}{1-\omega^{-m}}$ for $r=0,1$. We first prove the commutation relation for $\mathcal{A}_0$ and $\mathcal{A}_1$. The identity $\sum_{m=1}^{d-1} \frac{\omega^{km}}{1-\omega^{-m}} = \frac{d-1}{2}-k$ for $k=0,\ldots,d-1,$
shows that each \(h_r\) has spectrum
\begin{equation}
    \operatorname{spec}(h_r)
    =
    \left\{
    \frac{d-1}{2},
    \frac{d-3}{2},
    \ldots,
    -\frac{d-1}{2}
    \right\}.
\end{equation}
Indeed, if \(\eta_r\ket{k}=\omega^k\ket{k}\), then
\begin{equation}
    h_r\ket{k}
    =
    \left(\frac{d-1}{2}-k\right)\ket{k}.
\end{equation}
Fix \(j\). Since \(h_{2j}\) fails to commute only with
\(h_{2j-1}\) and \(h_{2j+1}\), set $K_j:=h_{2j-1}+h_{2j+1}$.  
We have $\operatorname{ad}_{\mathcal A_0}^m(h_{2j})=\operatorname{ad}_{K_j}^m(h_{2j})$ for \(m=1,2,3\). We now prove the identity
\begin{equation}\label{eqn:local-dg-raw}
    \operatorname{ad}_{K_j}^3(h_{2j})
    =
    d^2 \operatorname{ad}_{K_j}(h_{2j}).
\end{equation}
We work in a joint eigenbasis of \(\eta_{2j-1}\) and \(\eta_{2j+1}\).
Write the corresponding eigenvalue labels as \(p,q\in\{0,\ldots,d-1\}\).
We have
\begin{equation}
    K_j\ket{p,q}
    =
    \left(
        \frac{d-1}{2}-p
        +
        \frac{d-1}{2}-q
    \right)\ket{p,q}.
\end{equation}
The operator \(\eta_{2j}^m\) maps \(\ket{p,q}\) to a scalar multiple of
$    \ket{p+m,q-m},$
where the labels are read modulo \(d\). Hence each nonzero matrix element of \(h_{2j}\) connects two \(K_j\)-eigenvectors whose eigenvalue difference is one of $0$, $d$ or $-d$. Therefore, on every matrix element of \(h_{2j}\) we have $    \operatorname{ad}_{K_j}^3= d^2 \operatorname{ad}_{K_j},$ which proves \eqref{eqn:local-dg-raw}. Summing \eqref{eqn:local-dg-raw} over \(j\) gives
\begin{equation}
    [\mathcal A_0,[\mathcal A_0,[\mathcal A_0,\mathcal A_1]]]
    =
    d^2[\mathcal A_0,\mathcal A_1].
\end{equation}
Similarly, we have  $[\mathcal A_1,[\mathcal A_1,[\mathcal A_1,\mathcal A_0]]] = d^2[\mathcal A_1,\mathcal A_0]$. We have $A_i=\alpha \mathcal A_i$ for $i=0,1$, where $\alpha:=4/d$. We have
\begin{align}
[A_0,[A_0,[A_0,A_1]]] =  \alpha^4 [\mathcal A_0,[\mathcal A_0,[\mathcal A_0,\mathcal A_1]]]  = \alpha^4 d^2[\mathcal A_0,\mathcal A_1] = \alpha^2 d^2[A_0,A_1] = 16[A_0,A_1],
\end{align}
since \([A_0,A_1]=\alpha^2[\mathcal A_0,\mathcal A_1]\) and $\alpha^2 d^2 = \left(\frac{4}{d}\right)^2d^2 = 16$.  The same argument gives $[A_1,[A_1,[A_1,A_0]]]=16[A_1,A_0]$. 
Since $\|h_r\|=\frac{d-1}{2}$, we have $\|\mathcal A_0\|,\|\mathcal A_1\| \leq \frac{n(d-1)}{2}$. Hence, we have
\begin{equation}
\|A_0\|,\|A_1\|\leq \frac{4}{d}\frac{n(d-1)}{2}  = \frac{2n(d-1)}{d} := C_d n.
\end{equation} 
This completes the proof.
\end{proof}

\subsection{Nonlocal Fast-Transform Bogoliubov--de Gennes Hamiltonians}\label{app:BdG}

\subsubsection{Commutator scaling}

\begin{proof}[Proof of \cref{lem:fast-bdg-commutator-bound}]
The Majorana commutation relation in~\cref{eqn:maj-comm} implies the Lie algebra identity
\begin{equation}\label{eqn:fast-bdg-lie-homomorphism}
    [H(A),H(B)]
    =
    iH([A,B])
\end{equation}
for real anti-symmetric matrices \(A,B\). 
Hence, every nested commutator in~\cref{eqn:fast-bdg-word-bound}, up to a phase of unit modulus, equals \(H(C_q)\), where
\begin{equation}
    C_q
    =
    [K_{\gamma_1},[K_{\gamma_2},\ldots,
    [K_{\gamma_{q-1}},K_{\gamma_q}]]\cdots]].
\end{equation}
Using \(\|[A,B]\|\leq2\|A\|\|B\|\) recursively gives
$\|C_q\|
    \leq
    2^{q-1}\prod_{r=1}^{q}J_{\gamma_r}.$
To convert this matrix bound into a many-body operator norm bound, let \(\kappa_1(C),\ldots,\kappa_N(C)\geq0\) denote the canonical singular values of a real anti-symmetric matrix \(C\). 
Orthogonal block diagonalization shows that \(H(C)\) is unitarily equivalent to \(\frac{1}{2}\sum_{a=1}^{N}\kappa_a(C)\Gamma_{2a-1,2a}\). The operators in this sum are commuting Hermitians, and therefore
\begin{equation}\label{eqn:fast-bdg-many-body-norm}
    \|H(C)\|
    =
    \frac{1}{2}\sum_{a=1}^{N}\kappa_a(C)
    \leq
    \frac{N}{2}\|C\|.
\end{equation}
Note that nested commutators are essential because product formula errors arise from the noncommutativity of the grouped Hamiltonians and therefore cannot be characterized by \(\|H\|\) alone. 
Although we bound the resulting matrix commutators using \(\|[A,B]\|\leq2\|A\|\|B\|\) earlier, doing so before lifting to Fock space incurs the extensive factor \(N\) only once, rather than at every commutator level shown in~\cref{eqn:fast-bdg-many-body-norm}. 
Summing over the indices and using \(\sum_{\gamma_1,\ldots,\gamma_q}\prod_rJ_{\gamma_r}=(\sum_\gamma J_\gamma)^q\) proves the claim.
\end{proof}

\subsubsection{Comparison}
In the main text, we discussed the complexity of specialized Gaussian methods as the post-Trotter candidate. On the other hand, generic qubitization and QSVT require \(\mathcal O\!\left(\alpha_H T+\log(1/\epsilon)\right)\) queries to an \(\alpha_H\)-normalized block encoding~\cite{LowChuang2017,LowChuang2019}. The estimate \(\|H\|\leq (N/2)\|K\|\leq N\Lambda_{\mathrm{BdG}}/2\) gives the generic normalization scale \(\alpha_H=\mathcal{O}(N\Lambda_{\mathrm{BdG}})\). If a block encoding achieving this normalization is available with gate cost \(g_{\mathrm{BE}}(N)\), then the overall gate complexity of qubitization becomes \(\mathcal{O}\!\left(g_{\mathrm{BE}}(N)\left(N\Lambda_{\mathrm{BdG}}T+\log(1/\epsilon)\right)\right)\). A natural LCU-based construction gives \(g_{\mathrm{BE}}(N)=\mathcal{O}(g_{\mathrm{FT}}(N)+N)\). Thus, the \(\mathcal O(N^2)\) Gaussian complexity remains the relevant best existing model-specific upper bound. Next, we derive the regimes stated in \cref{eqn:fast-bdg-fixed-p-general-regime,eqn:fast-bdg-optimized-general-regime}.
Assume \(\Lambda_{\mathrm{BdG}}=\Theta(1)\) and define
\begin{equation}\label{eqn:app-fast-bdg-XR}
    X:=\frac{NT}{\epsilon},
    \qquad
    R:=\log X.
\end{equation}
We compare the following scalings for the second-order-based MPF, the \(p\)-th-order product formula, and the post-Trotter method,
\begin{equation}\label{eqn:app-fast-bdg-costs}
\begin{split}
    G_{\mathrm{MPF}}
    &=
    N^{4/3}\log N\,T R^4,\\
    G_p
    &=
    a^pN^{1+1/p}\log N\,
    T^{1+1/p}\epsilon^{-1/p}
    =
    NT\log N\,a^pX^{1/p},\\
    G_{\mathrm{post}}
    &=
    N^2,
\end{split}
\end{equation}
where \(a>1\) is a constant independent of \(N,T,\epsilon\) that characterizes the growth in the number of exponentials in the product formula (for example, \(a=\sqrt{5}\) for the Trotter--Suzuki formula). Here, we consider the case of a fermionic fast transform and assume that the degree of the polylogarithmic factor in the MPF is \(4\), which is a conservative estimate based on the proof of~\cref{cor:MPF_complexity}. Comparison with the post-Trotter Gaussian method requires solving \(G_{\mathrm{MPF}}<G_{\mathrm{post}}\), which gives
\begin{equation}\label{eqn:app-fast-bdg-fixed-post-condition}
    T<
    \frac{N^{2/3}}{\log N\,R^4}.
\end{equation}
For the fixed-\(p\) product formula, we solve $G_{\mathrm{MPF}}<G_p$ and obtain 
\begin{align}
    T>
    \epsilon a^{-p^2}N^{p/3-1}R^{4p}.
    \label{eqn:app-fast-bdg-fixed-p-condition}
\end{align} 
Combining the two comparisons gives the general simultaneous advantage window
\begin{equation}\label{eqn:app-fast-bdg-fixed-p-window}
    \epsilon a^{-p^2}N^{p/3-1}R^{4p} < T < \frac{N^{2/3}}{\log N\,R^4}.
\end{equation}
Although this condition is implicit through \(R=\log(NT/\epsilon)\), it applies without requiring a prior specification of how \(T\) and \(\epsilon\) depend on \(N\). In particular, the interval can be nonempty only if
\begin{equation}\label{eqn:app-fast-bdg-fixed-p-existence}
    \epsilon\log N\,R^{4p+4} < a^{p^2}N^{(5-p)/3}.
\end{equation}

For a clearer interpretation, we now consider the case in which \(T\) and \(\epsilon\) depend on \(N\). Let $ T=N^\tau$, $\epsilon=N^{-\kappa}$ for  $\tau, \kappa\geq0$ Then
$R=(1+\tau+\kappa)\log N,$
and the three costs become, up to parameter-independent constants,
\begin{equation}\label{eqn:app-fast-bdg-fixed-p-polynomial-costs}
\begin{split}
    G_{\mathrm{MPF}}
    &=
    {\mathcal{O}}\!\left(
        N^{4/3+\tau}\log^5N
    \right),\\
    G_p
    &=
    {\mathcal{O}}\!\left(
        a^p
        N^{1+\tau+(1+\tau+\kappa)/p}
        \log N
    \right),\\
    G_{\mathrm{post}}
    &=
    {\mathcal{O}}(N^2).
\end{split}
\end{equation}
Then, the MPF power of \(N\) is smaller than the post-Trotter power when
$\tau<\frac23.$
Comparison with \(G_p\) instead gives
\begin{equation}
    \frac43+\tau < 1+\tau+\frac{1+\tau+\kappa}{p},
\end{equation}
which is equivalent to
\begin{equation}\label{eqn:app-fast-bdg-fixed-p-power-condition}
    \tau+\kappa>\frac p3-1.
\end{equation}
These inequalities must be strict because, at \(\tau+\kappa=p/3-1\), the two costs have the same power of \(N\), while the MPF incurs an additional factor of \(\Theta(\log^4 N)\). Similarly, when \(\tau=2/3\), the MPF exceeds \(N^2\) by a logarithmic factor. We now consider several more concrete scenarios. At constant accuracy, \(\kappa=0\), a polynomially wide advantage interval exists when

\begin{equation}\label{eqn:app-fast-bdg-fixed-p-constant-accuracy}
    \frac p3-1<\tau<\frac23,
\end{equation}
which requires \(p<5\). 
For the usual even-order formulas, this includes
\(p=2\) and \(p=4\). 
For example, \(p=4\), \(\epsilon=\Theta(1)\), and
\(T=N^{1/2}\) give
\begin{equation}
        G_{\mathrm{MPF}}
    =
    \Theta\ (N^{11/6}\log^5N ),
    \qquad
    G_4
    =
    \Theta (a^4N^{15/8}\log N ),
    \qquad
    G_{\mathrm{post}}
    =
    \Theta(N^2).
\end{equation}
Since \(11/6<15/8<2\), the MPF scaling is the smallest. More generally, for any fixed \(p\), there exists some \(\tau<2/3\) satisfying both comparisons whenever
$ \kappa>\frac{p-5}{3}.$

So far, we have fixed the convergence order \(p\) of the product formula. In practice, however, \(p\) may be chosen adaptively as a function of the other parameters, including \(N,T,\epsilon\). In this case, we must optimize the product-formula order and compare the MPF with the resulting optimized formula. Specifically, we now allow \(p\) to depend on \(N,T,\epsilon\). The logarithm of the \(p\)-dependent part of \(G_p\) is
\begin{equation}\label{eqn:app-fast-bdg-order-objective}
    f(p):=\log(a^pX^{1/p})=p\log a+\frac{R}{p}.
\end{equation}
Treating \(p\) as continuous, we can minimize $f(p)$ at 
$p_{\mathrm{opt}} = \sqrt{\frac{R}{\log a}},$
and
$f(p_{\mathrm{opt}}) = 2\sqrt{R\log a}.$
Rounding \(p_{\mathrm{opt}}\) to the nearest allowed order, such as the nearest positive even integer, changes the cost by at most an asymptotically irrelevant constant factor when \(p_{\mathrm{opt}}\to\infty\). For simplicity, we therefore proceed with \(p_{\mathrm{opt}}\). The optimized cost is
\begin{equation}\label{eqn:app-fast-bdg-optimized-cost}
    G_{\mathrm{PF,opt}} = NT\log N\, \exp\!\left(2\sqrt{R\log a}\right).
\end{equation}
The MPF is asymptotically smaller than this optimized product formula expression precisely when $G_{\mathrm{MPF}}<G_{\mathrm{PF,opt}}$, i.e., 
\begin{align}
    \frac13\log N+4\log R < 2\sqrt{R\log a}. 
    \label{eqn:app-fast-bdg-optimized-condition}
\end{align}
Combining this condition with the post-Trotter comparison yields the general regime
\begin{equation}\label{eqn:app-fast-bdg-optimized-general-regime}
    \frac13\log N+4\log R < 2\sqrt{R\log a}, \qquad 
    T < \frac{N^{2/3}}{\log N\,R^4}.
\end{equation}

We now again consider the special case in which \(T\) and \(\epsilon\) depend on \(N\).
First suppose $T=N^\tau$ and 
$\epsilon=N^{-\kappa}.$
Then \(R=\mathcal{O}(\log N)\). 
Notice that the left-hand side of~\cref{eqn:app-fast-bdg-optimized-condition} grows as \(\frac13\log N+O(\log\log N)\), whereas the right-hand side grows only as \(O(\sqrt{\log N})\). Hence, the condition cannot hold asymptotically, and the MPF does not improve upon the optimized-order product formula when both \(T\) and \(1/\epsilon\) are polynomial in \(N\). However, a different conclusion holds at inverse-quasi-polynomial accuracy. Specifically, take
\begin{equation}\label{eqn:app-fast-bdg-quasipolynomial-parameters}
    T=N^\tau, \qquad \epsilon=\exp[-c(\log N)^2], \qquad c>0.
\end{equation}
Then we have \(R=c\log^2 N+(1+\tau)\log N\), and the MPF cost becomes
\begin{equation}\label{eqn:app-fast-bdg-quasipolynomial-mpf}
    G_{\mathrm{MPF}} = \mathcal{O}( N^{4/3+\tau}\log^9 N). 
\end{equation}
Moreover,
\[
    p_{\mathrm{opt}} = \sqrt{\frac{c}{\log a}}\,\log N+O(1)
\]
and \(2\sqrt{R\log a} = 2\log N\sqrt{c\log a}+O(1)\), so
\begin{equation}\label{eqn:app-fast-bdg-quasipolynomial-pf}
    G_{\mathrm{PF,opt}} = \mathcal{O}\!\left( N^{1+\tau+2\sqrt{c\log a}}\log N \right).
\end{equation}
The MPF has a smaller power of \(N\) when
\begin{equation}
    \frac43+\tau < 1+\tau+2\sqrt{c\log a},
\end{equation}
or equivalently,
$c>\frac{1}{36\log a}.$
As discussed earlier, the MPF scaling is simultaneously smaller than the post-Trotter scaling \(N^2\) when \(\tau<\frac23\). Both conditions are strict because the logarithmic factors disfavor the MPF when equality holds. For a more concrete example, consider the choice $\tau=\frac12$ and $\tau=\frac12$
gives
\begin{equation}
G_{\mathrm{MPF}} = \mathcal{O}(N^{11/6}\log^9N),
    \qquad G_{\mathrm{PF,opt}} = \mathcal{O}(N^{19/10}\log N),
    \qquad G_{\mathrm{post}} = \mathcal{O}(N^2).
\end{equation}
Thus \(G_{\mathrm{MPF}}\) is asymptotically smallest. 

\subsection{Kitaev's Honeycomb Model}\label{app:Kitaev}

\begin{proof}[Proof of \cref{kitaev-majorana}]
 For each \(j\in\Lambda\), introduce four Majorana operators $b_j^x, b_j^y, b_j^z$ and $c_j$.  These operators act on an enlarged local \(4\)-dimensional Hilbert space.  Define the operators
\begin{equation}\label{dj-ops}
D_j := b_j^x b_j^y b_j^z c_j
\end{equation}
for all \(j\in\Lambda\). Note that $D_j^\dagger=D_j$ and $D_j^2=I$.\footnote{
Indeed, \(D_j\) is a product of four distinct self-adjoint Majorana operators.  Hence, $D_j^\dagger=c_j b_j^z b_j^y b_j^x $. Reordering \(c_j b_j^z b_j^y b_j^x\) back to \(b_j^x b_j^y b_j^z c_j\) requires \(6\) transpositions, and each transposition gives a minus sign.  Hence, $D_j^\dagger = (-1)^6 b_j^x b_j^y b_j^z c_j = D_j $. 
Similarly, we have $D_j^2=(b_j^x b_j^y b_j^z c_j )
 (b_j^x b_j^y b_j^z c_j ) = (-1)^6
(b_j^x)^2(b_j^y)^2(b_j^z)^2c_j^2 = I$. 
}
Hence, the eigenvalues of \(D_j\) are \(\pm1\). Therefore, the constraint $D_j := b_j^x b_j^y b_j^z c_j = I$ cuts the $4$-dimensional local Majorana Hilbert space down to a $2$-dimensional Hilbert space. This follows from the simple observation that the \(\pm 1\)-eigenspaces of \(D_j\) have the same dimension. Indeed, \(D_j\) anti-commutes with each individual Majorana operator at site \(j\). For example, \(D_j b_j^x = - b_j^x D_j\). Thus, \(b_j^x\) maps the \(+1\)-eigenspace of \(D_j\) bijectively onto the \(-1\)-eigenspace, and conversely. Thus, the constraint \(D_j=I\) defines \(2\)-dimensional subspace, which can be identified with the local  Hilbert space corresponding to one spin-\(\frac12\) particle. Define
\begin{equation}\label{eqn:kitaev-spin-majorana-representation}
\widetilde\sigma_j^\alpha
:=
i b_j^\alpha c_j,
\qquad
\alpha=x,y,z
\end{equation}
on the ambient $4$-dimensional Hilbert space. These operators preserve the \(2\)-dimensional \(+1\)-eigenspace of \(D_j\). Indeed, \(D_j\) anti-commutes with each individual Majorana operator, and therefore commutes with every even product of Majorana operators.  Since \(\widetilde\sigma_j^\alpha\) is a product of two Majoranas, we have
\begin{equation}
[D_j,\widetilde\sigma_j^\alpha]=0.
\end{equation}
Hence, \(\widetilde\sigma_j^\alpha\)'s restrict to an operator on the \(2\)-dimensional \(+1\)-eigenspace of \(D_j\). On this subspace, the operators \(\widetilde\sigma_j^x,\widetilde\sigma_j^y,\widetilde\sigma_j^z\) satisfy the Pauli algebra. First note that $(\widetilde\sigma_j^\alpha)^2=(i b_j^\alpha c_j)^2=I$ for $\alpha=x,y,z$. Moreover, using the Majorana anti-commutation relations, we have
\begin{equation}
\widetilde\sigma_j^x\widetilde\sigma_j^y
=
(i b_j^x c_j)(i b_j^y c_j)
=
b_j^x b_j^y.
\end{equation}
On the \(+1\)-eigenspace of \(D_j\), we have \(b_j^x b_j^y b_j^z c_j=I\). Multiplying by \(c_j b_j^z\) and using the Majorana anti-commutation relations gives $b_j^x b_j^y=-b_j^z c_j.$
Since $i\widetilde\sigma_j^z=b_j^z c_j$, we have 
\begin{equation}
\widetilde\sigma_j^x\widetilde\sigma_j^y=i\widetilde\sigma_j^z.
\end{equation}
Similarly, we have
\begin{equation}
\widetilde\sigma_j^y\widetilde\sigma_j^z
=
i\widetilde\sigma_j^x,
\qquad
\widetilde\sigma_j^z\widetilde\sigma_j^x
=
i\widetilde\sigma_j^y.
\end{equation}
Thus, on the \(+1\)-eigenspace of \(D_j\), the operators \(\widetilde\sigma_j^\alpha=i b_j^\alpha c_j\) realize the usual Pauli matrices acting on a single spin-\(\frac12\). In this sense, the usual spin operators \(\sigma_j^\alpha\) are represented on the \(+1\)-eigenspace of \(D_j\) by the operators \(\widetilde\sigma_j^\alpha=i b_j^\alpha c_j\). For \(\{j,k\} \in E_\alpha\), define $u_{jk}:=i b_j^\alpha b_k^\alpha$. 
Note that $u_{jk}^\dagger=u_{jk}$ and $u_{jk}^2=I$. 
Moreover, using \(\widetilde\sigma_j^\alpha=i b_j^\alpha c_j\), we have
\begin{align}
\widetilde\sigma_j^\alpha \widetilde\sigma_k^\alpha= (i b_j^\alpha c_j)(i b_k^\alpha c_k)= -b_j^\alpha c_j b_k^\alpha c_k= b_j^\alpha b_k^\alpha c_j c_k= -u_{jk}  i c_j c_k .
\end{align}
Hence, \cref{eqn:kitaev-honeycomb-spin-hamiltonian} can be written on the enlarged Majorana Hilbert space as
 \begin{equation}\label{eqn:kitaev-honeycomb-enlarged-hamiltonian}
\widetilde H=\sum_{\alpha\in\{x,y,z\}}J_\alpha\sum_{\{j,k\}\in E_\alpha} u_{jk} i c_j c_k
=-\sum_{\alpha\in\{x,y,z\}}J_\alpha\sum_{\{j,k\}\in E_\alpha}  b_j^\alpha b_k^\alpha c_j c_k.
\end{equation}
Thus, Kitaev's honeycomb model is a quartic interacting Hamiltonian on the Majorana Hilbert space. Its special structure is that the operators \(u_{jk}\) are conserved. Consequently, the Hamiltonian becomes quadratic after restriction to any fixed sector, which is the key input for the commutator bound below.
\end{proof}

\begin{proof}[Proof of \cref{lem:kitaev-sectorwise-commutator-bound}]
Fix a sector \(u\).  We first prove the bound inside this sector.  Fix once and for all an ordering of the Majorana labels, and write $\Gamma_{rs}:=ic_rc_s$ for $r<s$. Every quadratic Majorana operator has a canonical expansion $X=\sum_{r<s}x_{rs}\Gamma_{rs}$. Define the coefficient norm by $\|X\|_{\#}:=\sum_{r<s}|x_{rs}|$.  Since \(\|\Gamma_{rs}\|=1\), we have $\|X\| \leq \|X\|_{\#}$. In the fixed sector \(u\), $\widetilde H_\alpha^{(u)}$ is given by
\begin{equation}
\widetilde H_\alpha^{(u)}=J_\alpha\sum_{\{j,k\}\in E_\alpha}u_{jk}\Gamma_{jk},
\end{equation}
where \(u_{jk}\in\{\pm1\}\).  We claim that, for every \(r<s\),
\begin{equation}\label{eqn:kitaev-single-gamma-comm-bound}
\|[\widetilde H_\alpha^{(u)},\Gamma_{rs}]\|_{\#}
\le
4|J_\alpha|.
\end{equation}
By the Majorana commutation relation \cref{eqn:maj-comm}, the commutator \([\Gamma_{jk},\Gamma_{rs}]\) vanishes unless the two unordered pairs \(\{j,k\}\) and \(\{r,s\}\) share exactly one endpoint.  When the commutator is non-zero, it is a scalar multiple of another quadratic Majorana operator with coefficient of absolute value \(2\).  Since \(E_\alpha\) is an independent edge set, the fixed pair \(\{r,s\}\) can share an endpoint with at most two edges of \(E_\alpha\).  Therefore at most two terms in the sum defining \(\widetilde H_\alpha^{(u)}\) contribute, and each contributes coefficient norm at most \(2|J_\alpha|\).  This proves \cref{eqn:kitaev-single-gamma-comm-bound}. Now let \(X=\sum_{r<s}x_{rs}\Gamma_{rs}\) be any quadratic Majorana operator.  We have
\begin{align}
\|[\widetilde H_\alpha^{(u)},X]\|_{\#}=\left\| \sum_{r<s}x_{rs}[\widetilde H_\alpha^{(u)},\Gamma_{rs}] \right\|_{\#} \le \sum_{r<s}|x_{rs}| \|[\widetilde H_\alpha^{(u)},\Gamma_{rs}]\|_{\#}\le 4|J_\alpha| \sum_{r<s}|x_{rs}|= 4|J_\alpha|\|X\|_{\#}.
\end{align}
Thus commutation with \(\widetilde H_\alpha^{(u)}\) increases the coefficient norm by at most \(4|J_\alpha|\). Since \(|u_{jk}|=1\), we have
\begin{equation}
\|\widetilde H_\alpha^{(u)}\|_{\#}=|J_\alpha|
\sum_{\{j,k\}\in E_\alpha}|u_{jk}|=|J_\alpha||E_\alpha|\le n|J_\alpha|.
\end{equation}
In particular, we have $\|\widetilde H_{\gamma_q}^{(u)}\|_{\#} \le n|J_{\gamma_q}|$. Each commutator with a \(\widetilde H_{\gamma_r}^{(u)}\) contributes a factor at most \(4|J_{\gamma_r}|\).  Hence, we have
\begin{equation}
\| [\widetilde H_{\gamma_1}^{(u)},[\cdots,[\widetilde H_{\gamma_{q-1}}^{(u)},\widetilde H_{\gamma_q}^{(u)}]\cdots]] \| \leq   \| [\widetilde H_{\gamma_1}^{(u)},[\cdots,[\widetilde H_{\gamma_{q-1}}^{(u)},\widetilde H_{\gamma_q}^{(u)}]\cdots]] \|_{\#}
\le n4^{q-1} \prod_{r=1}^{q}|J_{\gamma_r}|.
\end{equation}
This proves \cref{eqn:kitaev-sector-word-bound}.  Using \cref{lift-sector-comms}, for every word \(\gamma_1,\ldots,\gamma_q\), we have
\begin{align}
\| [\widetilde H_{\gamma_1},[\cdots,[\widetilde H_{\gamma_{q-1}},\widetilde H_{\gamma_q}]\cdots]] \|=  \max_u  \| [\widetilde H_{\gamma_1}^{(u)},[\cdots,[\widetilde H_{\gamma_{q-1}}^{(u)},\widetilde H_{\gamma_q}^{(u)}]\cdots]]  \|  \le n4^{q-1}
\prod_{r=1}^{q}|J_{\gamma_r}|.
\end{align}
The original spin Hamiltonian \(H\) is obtained by restricting \(\widetilde H\) to the physical spin subspace, namely the subspace obtained by the joint \(+1\)-eigenspaces of the operators \(D_j\)'s. Since taking this restriction cannot increase the operator norm, the same estimate applies to the corresponding nested commutator for the original spin Hamiltonian. This completes the proof.
\end{proof}

\end{document}